\documentclass{article}

\usepackage{PRIMEarxiv}

\usepackage[square,numbers,sort&compress,comma]{natbib}

\usepackage{amsmath}
\usepackage{amssymb}
\usepackage{caption}
\usepackage{graphicx}
\usepackage{latexsym}
\usepackage{times}
\usepackage[pagewise]{lineno}
\usepackage{hyperref}
\usepackage{amsmath,bm}
\usepackage[percent]{overpic}
\usepackage{float}
\usepackage{stfloats} 
\usepackage{xcolor}
\usepackage{placeins}

\title{Entropy-Constrained Machine Learning with Residual Data Augmentation for Modeling Chemical Kinetics}

\author{
  Okezzi Ukorigho, Opeoluwa Owoyele%
  \thanks{Corresponding author: \texttt{oowoyele@lsu.edu}} \\[3pt]
  Department of Mechanical and Industrial Engineering \\
  Louisiana State University, Baton Rouge, LA 70810, USA
}
        
\begin{document}
\maketitle

\begin{abstract}
We present a physics-constrained machine learning framework for accelerating the direct numerical simulation (DNS) of turbulent reacting flows. The model replaces the direct evaluation of detailed chemical source terms with a surrogate that predicts reaction rates from a reduced thermochemical state. To improve physical consistency, the second law of thermodynamics is incorporated as a training constraint by enforcing non-negative entropy generation, which restricts the evolution of the thermochemical state to physically admissible directions and improves stability during time integration. The approach is demonstrated on DNS of a two-dimensional planar lean premixed methane-air flame interacting with a turbulent flow field. The model reproduces detailed-chemistry results with high fidelity while achieving more than an order-of-magnitude reduction in computational cost. Furthermore, a residual-based synthetic data augmentation strategy enables parametric exploration by constructing new training data from the original dataset, allowing accurate simulation at new inlet \textcolor{black}{conditions} without additional detailed-chemistry CFD runs. These results demonstrate that thermodynamically constrained machine learning can provide reliable and computationally efficient surrogates for detailed chemistry in high-fidelity combustion simulations.
\end{abstract}

\keywords{turbulent reacting flows \and chemical kinetics \and direct numerical simulation \and entropy-constrained machine learning \and second-law constraints \and data augmentation}

\section{Introduction\label{sec:introduction}} 

Direct Numerical Simulation (DNS) of reacting flows remains an important tool for improving our understanding of turbulence–chemistry interactions, \textcolor{black}{but its application to practical fuels and geometries remains limited by the high cost of resolving detailed reacting-flow physics} \cite{chen2009terascale,lu2009toward,babkovskaia2011high}. As a result, several approaches have been developed to accelerate chemistry calculations. These include skeletal and reduced mechanism generation techniques \cite{lu2008criterion,pepiot2008efficient,gao2016global}, as well as manifold-based approaches such as flamelet-generated manifolds \cite{van2000modelling, van2016state}, intrinsic low-dimensional manifolds \cite{maas1992simplifying,maas1992implementation}, and data-driven manifold reduction methods that project the thermochemical state space onto lower-dimensional manifolds \cite{sutherland2009combustion, coussement2012kernel, owoyele2017toward, kumar2023acceleration}. Over the past decade, machine learning (ML) has gained traction as a tool for accelerating chemical source-term evaluations. Early work by Christo et al.\ \cite{christo1996integrated} demonstrated neural-network-based prediction of chemical source terms, followed by Blasco et al.\ \cite{blasco1998modelling}, who modeled the temporal evolution of chemically reacting systems. Subsequent studies have expanded these ideas, including micro-mixing–based data generation strategies for ML training \cite{wan2020chemistry}, Arrhenius- and law-of-mass-action-inspired neural network architectures \cite{ji2021autonomous}, neural ordinary differential equations for kinetics integration \cite{owoyele2022chemnode, kumar2025physics2}, and more recently, operator-learning approaches for accelerating combustion chemistry \cite{kumar2024combustion, weng2025extended}.

Concurrently, a growing body of work has sought to incorporate physical constraints into ML-based source-term prediction, enforcing conservation of mass or chemical elements either as soft constraints through loss-function regularization \cite{doppel2024robust, zhang2024crk, wang2025enforcing, kumar2025physics} or as hard constraints through post-prediction correction layers \cite{readshaw2023incorporation, doppel2025formation}. These efforts are valuable because purely data-driven models often do not automatically encode such laws, despite their fundamental importance and their foundation in centuries of accumulated scientific knowledge. This work is positioned within the same broader effort, which seeks to promote the satisfaction of physical laws in ML-based chemistry models. Existing conservation constraints targeting mass conservation primarily enforce consistency for local thermochemical states, but they do not address the dynamical behavior of the system, specifically, the directions in composition space along which the system is allowed to evolve.

In this context, the present work introduces a ML framework in which the second law of thermodynamics is embedded as a training constraint by enforcing non-negative entropy generation. This promotes a thermodynamic consistency that goes beyond elemental conservation. The primary scientific contribution of this paper is the demonstration, for the first time, that (1) unconstrained data-driven ML models for chemical source-term prediction can produce second-law violations in practice, and (2) enforcing such thermodynamic constraints significantly improves the reliability and stability of the predicted dynamical evolution of the thermochemical state. The constraint is incorporated within a hybrid ML regression framework that combines nonlinear neural-network representations with kernel-based interpolation \cite{chandrasekhar2026deep}. In addition, we introduce a residual-based synthetic data augmentation strategy that allows for parametric exploration of operating conditions without repeated detailed chemistry simulations at new conditions. The methodology is validated using a two-dimensional statistically planar lean premixed methane-air flame subjected to isotropic turbulence, demonstrating that the proposed physically constrained ML model reproduces detailed-chemistry DNS with substantial computational acceleration.

\section{Model Formulation and \textcolor{black}{Setup} \label{sec:model}}

\subsection{Hybrid Neural Network-Kernel Regression Framework\label{subsec:framework}} 

The chemical source terms are approximated using a hybrid model that combines the expressive power of neural networks with the interpolative robustness of kernel regression. Full details are provided in \cite{chandrasekhar2026deep}; only a brief summary is given here. We represent the local thermochemical state using a reduced variable vector, $\bm{\psi} = [T,\; Y_1,\; \ldots,\; Y_{n_r}]^T$, constructed from temperature and a subset of \(n_r\) species mass fractions. A feed-forward neural network, \(\mathcal{N}_{\theta}\), maps \(\bm{\psi}\) to a latent feature vector \(\mathbf{z} \in \mathbb{R}^{n_z \times 1}\), so that $\mathbf{z} = \mathcal{N}_{\theta}(\bm{\psi})$. These latent features are then used within a kernel regression model to predict species source terms. For a given species \(i\), let the latent vectors of \(n_p\) query points be collected in $\mathbf{Z}^{\{q\}} = [\mathbf{z}_1^{\{q\}}, \mathbf{z}_2^{\{q\}}, \ldots, \mathbf{z}_{n_p}^{\{q\}}]$, and let the latent vectors of \(n_a\) anchor points be collected in $\mathbf{Z}^{\{a\}} = [\mathbf{z}_1^{\{a\}}, \mathbf{z}_2^{\{a\}}, \ldots, \mathbf{z}_{n_a}^{\{a\}}]$.
If \(\dot{\boldsymbol{\omega}}_i^{\{a\}} \in \mathbb{R}^{n_a \times 1}\) denotes the ground-truth source terms of species \(i\) at the anchor points, then the predicted source terms at the query points are given by

\begin{equation}
    \hat{\dot{\boldsymbol{\omega}}}_i =
    K(\mathbf{Z}^{\{q\}}, \mathbf{Z}^{\{a\}})
    \Big(
    K(\mathbf{Z}^{\{a\}}, \mathbf{Z}^{\{a\}}) + \lambda I
    \Big)^{-1}
    \dot{\boldsymbol{\omega}}_i^{\{a\}},
\end{equation}
where \(\lambda\) is a regularization parameter, \textcolor{black}{set to $\lambda = 10^{-9}$ to preserve matrix conditioning while keeping the kernel regressor close to the anchor values}, and \(K\) is the radial basis function (RBF) kernel. For two latent vectors \(\mathbf{z}_j\) and \(\mathbf{z}_k\), selected as columns from some latent matrix $\mathbf Z$, the kernel entry is defined as

\begin{equation}
    K_{jk} = \exp\!\left(
    -\frac{\|\mathbf{z}_j - \mathbf{z}_k\|^2}{2\theta^2}
    \right),
\end{equation}
where \(\theta\) is the kernel bandwidth. The kernel regression layer acts as a ground-truth-aware interpolator that stabilizes predictions near the boundaries of the sampled thermochemical space. At any anchor point \(\mathbf{z}^{\{a\}}_j\), the predicted source term recovers the corresponding ground-truth value \(\dot{\omega}_{i,j}^{\{a\}}\) up to the regularization tolerance set by \(\lambda\), since the prediction is constructed directly from the anchor data rather than from a purely parametric approximation. In addition, the prediction variance increases monotonically with distance from the nearest anchor point in latent space, providing a signal for detecting failure modes and influencing the selection of anchor points. The neural network component provides a nonlinear mapping that captures complex relationships in the thermochemical state space.

\subsection{Thermodynamic Constraint \textcolor{black}{Formulation} \label{subsec:entropy}} 

A well-known limitation of purely data-driven models is that they may
violate fundamental physical principles. In chemical kinetics
modeling, ML-based source-term predictors are typically trained to
match pointwise targets, with limited regard for the admissible
dynamical evolution of the thermochemical state. However, the second
law of thermodynamics imposes a natural constraint on this evolution, and this provides a physically grounded criterion that can be incorporated
during training. For a chemically reacting multicomponent mixture, the entropy generation rate due to chemical reactions can be derived from the Gibbs relation at constant temperature and pressure (derivation in Supplement, Section~S1):
\begin{equation}
   \sigma = -\frac{1}{T}\sum_{i=1}^{N_s}\frac{\mu_i}{W_i}\,\dot{\omega}_i,
   \label{eq:entropy}
\end{equation}

where \(W_i\) is the molecular weight of species \(i\), and
\(\dot{\omega}_i\) is the net mass production rate per unit volume
\((\mathrm{kg\,m^{-3}\,s^{-1}})\). Thus, the second law requires $\sigma \geq 0$ as the species evolve at a given CFD grid point at a given instant. Physically, this
condition implies that the chemically reacting system evolves only
along thermodynamically admissible directions in composition space. Any
model prediction for which \(\sigma < 0\) corresponds to negative
entropy generation and is therefore thermodynamically inadmissible. To ensure that the learned reaction rates remain consistent
with irreversible thermodynamics, we add a penalty term to the training loss to suppress model parameters that lead to spontaneous local entropy reduction during the temporal evolution: \textcolor{black}{The entropy-penalty weight is set as $\lambda_\sigma = \lambda_0/\sigma_{\rm ref}^2$, where $\lambda_0 = 10^{-2}$ is a fixed coefficient and $\sigma_{\rm ref}$ is the median of $|\sigma_{\rm DNS}|$ within each subdomain. The factor $1/\sigma_{\rm ref}^2$ normalizes the entropy penalty by a characteristic magnitude of the entropy generation rate within the subdomain, so that the entropy-penalty contribution remains comparable in magnitude to the data MSE term during training.}

\begin{equation}
\begin{aligned}
    \mathcal{L} = \frac{1}{N}\bigg(
    & \sum_{n=1}^{N}\|\dot{\boldsymbol{\omega}}_n^{\mathrm{DNS}}
      - \dot{\boldsymbol{\omega}}_n^{\mathrm{pred}}\|^2
    + \lambda_{\sigma}\sum_{n=1}^{N}\big[\max(0,-\sigma_n)\big]^2
    \bigg)
    \label{eq:loss}
\end{aligned}
\end{equation}

\begin{figure}[h!]
\centering
\includegraphics[width=0.4\columnwidth]{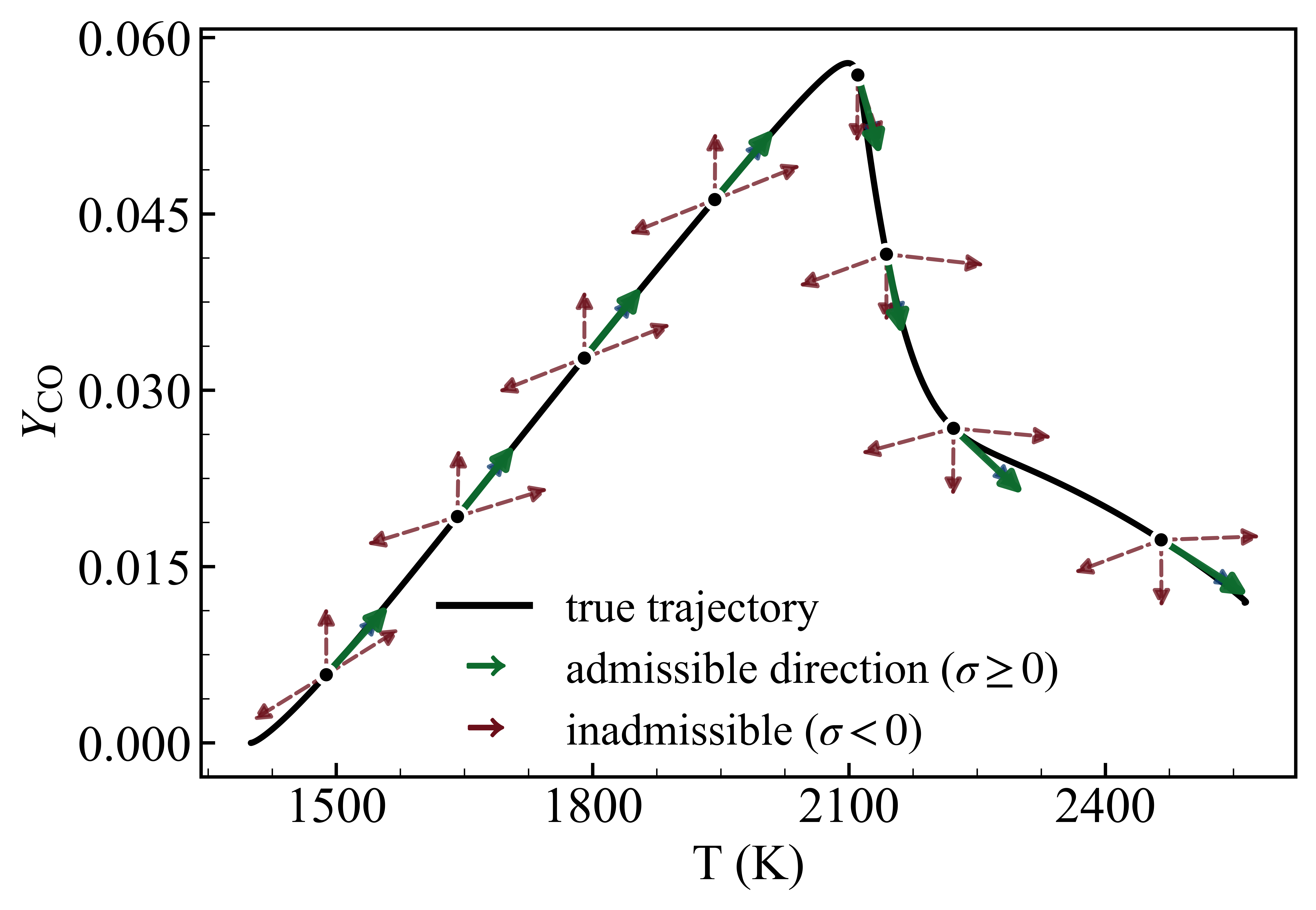}
\caption{\footnotesize System evolution trajectory for a zero-dimensional auto-ignition trajectory in $(T,\,Y_{\mathrm{CO}})$ space, indicating thermodynamically admissible directions ($\sigma \geq 0$) and evolution directions that violate the second law ($\sigma < 0$).} 
\label{fig:direction_field}
\end{figure}

This regularization promotes learned reaction rates that remain consistent with irreversible thermodynamics. \textcolor{black}{We note, however, that inclusion of the entropy generation in the loss can increase conventional training errors. Specifically, this term led to a twofold increase in MSE as well as a decrease in $R^2$ from 0.97 to 0.94 during training for the baseline case. This suggests that in our case, ML model parameters that minimize error also spontaneously destroy entropy. As shown in the results section, despite the worse \emph{a priori} regression metrics, the constrained model produces substantially improved \emph{a posteriori} stability and predictive accuracy during coupled time integration.}
The role of the entropy constraint is illustrated in Fig.~\ref{fig:direction_field}, which shows a zero-dimensional constant-pressure autoignition trajectory in $(T,\,Y_{\mathrm{CO}})$ state space for a lean methane--air mixture ($\phi = 0.7$, $T_0 = 1400$~K). The figure shows the true evolution, which respects the second law, as well as some directions that violate it, obtained via source term perturbations. At each point along the trajectory, the second law restricts the admissible evolution of the system to directions that produce non-negative entropy ($\sigma > 0$, green arrows); all other directions ($\sigma < 0$, red arrows) correspond to thermodynamically infeasible pathways that would require a spontaneous decrease in entropy.  An unconstrained surrogate has no mechanism to prevent evolution in physically inadmissible directions, and therefore may begin to diverge from the true solution, accumulating errors at each integration step until failure occurs.  The penalty term in Eq.~(\ref{eq:loss}) is meant to confine the learned source terms to the admissible directions and prevents such error accumulations that leads to the divergent behavior, as will be shown in Section ~\ref{subsec:constraint_effect}.

\subsection{Synthetic Data Augmentation for Generalization\label{sec:aug}} \addvspace{1pt}

Another challenge when developing data-driven models is generalization to unseen conditions. Here, we introduce an approach that systematically modifies existing data to represent new regions of parameter space. In practice, this approach may allow parametric studies to be conducted without running additional CFD simulations by transforming the available data to emulate new operating conditions. To achieve such model robustness under unseen states, we introduce a synthetic data augmentation strategy based on residual synthesis of the normalized temperature. The temperature observations are first normalized as
\begin{equation}
    T^\star = \frac{T - T_{in}}{T_{ex} - T_{in}},
    \label{eq:augmentation_norm}
\end{equation}
where $T_{in}$ and $T_{ex}$ denote the inlet and exit flame temperatures, respectively. This mapping transforms the temperature at all points in the domain to $T^\star \in [0,1]$, and allows for comparison across different initial conditions. Residuals are then computed between the DNS thermochemical vector $\bm{\psi}_{\mathrm{DNS}}(T^\star)$ from the wrinkled flame and a reference laminar flame profile $\bm{\psi}_{\mathrm{lam}}(T^\star)$ at the same normalized temperature:

\begin{equation}
    \mathcal{R}(T^\star) = \bm{\psi}_{\mathrm{DNS}}(T^\star) - \bm{\psi}_{\mathrm{lam}}(T^\star).
\end{equation}

These residuals capture deviations from the laminar state induced by turbulent mixing and flame wrinkling. To generate synthetic data at a new inlet condition, a laminar reference profile, $\bm{\psi}_{\mathrm{lam}}(T^\star)$, is first computed for that condition. For each target state, $\bm{\psi}_{\mathrm{lam}}(T^\star_{\mathrm{new}})$, the $k$ nearest neighbors are identified from the full baseline DNS dataset using distance in the reduced thermochemical state space. The corresponding residual vectors are collected in the set $\{\mathcal{R}_j\}_{j=1}^k$, from which one residual is selected at random to define the sampled residual, $\tilde{\mathcal{R}}$. To account for unresolved local variability, a binning strategy is applied in normalized temperature space to estimate the standard deviation, $\sigma(T^\star)$, as a function of $T^\star$. A perturbation is then sampled from a Gaussian distribution as

\begin{equation}
\bm \varepsilon \sim \mathcal{N}\!\left(0,\alpha\ \cdot\bm\sigma^2(T^\star_{\mathrm{new}}) \bm I\right),
\end{equation}
where $\alpha$ is a scaling factor \textcolor{black}{chosen as $\alpha = 0.5$.} The synthetic state is finally constructed as

\begin{equation}
\bm{\psi}_{\mathrm{syn}}(T^\star_{\mathrm{new}})
=
\bm{\psi}_{\mathrm{lam}}(T^\star_{\mathrm{new}})
+
\tilde{\mathcal{R}}
+
\bm \varepsilon.
\end{equation}

\textcolor{black}{$\tilde{\mathcal{R}}$ is computed once from the baseline DNS and stored. For new operating conditions, starting from the corresponding laminar flame profile, this procedure generates synthetic samples that populate the thermochemical state space at the new condition without requiring a new DNS for training. As an initially laminar flame interacts with the imposed flow field, turbulence and unsteady transport disperse this trajectory and scatter the local thermochemical states away from the underlying laminar manifold. In the normalized-temperature coordinate $T^\star$, the laminar profile $\bm{\psi}_{\mathrm{lam}}(T^\star)$ provides a reference trajectory between the unburned and burned states, while the residual $\bm{\psi}_{\mathrm{DNS}}(T^\star)-\bm{\psi}_{\mathrm{lam}}(T^\star)$ represents the deviation caused by turbulence-induced wrinkling, mixing, and unsteady transport at matched $T^\star$. The underlying assumption is not that these deviations are identical across operating conditions, but that their statistical structure is sufficiently similar over the range considered. Under this assumption, residuals sampled from the baseline DNS can be used to enrich laminar profiles at nearby operating conditions, without requiring new DNS data for training. The procedure is therefore intended for nearby variations global thermodynamic parameters, and is not expected to extrapolate to large changes in pressure or turbulent flow field statistics.}


\subsection{DNS Configuration and Numerical Setup\label{sec:dns_setup}}
All simulations were performed using Pencil Code \cite{babkovskaia2011high}, a fully compressible DNS solver based on a high-order finite-difference formulation. The governing equations include mass, momentum, energy, and species transport, with chemical source terms either evaluated from the detailed mechanism or replaced by the physics-constrained hybrid ML surrogate. Temporal integration employs a third-order explicit Runge--Kutta scheme, while spatial derivatives use sixth-order central differencing to minimize numerical dissipation. The test configuration \textcolor{black}{for all cases} corresponds to an initially planar premixed methane-air flame within a decaying isotropic turbulent flow field. \textcolor{black}{The computational domain spans $L_x \times L_y = 1.0$~cm $\times$ 0.5~cm, discretized with $768 \times 384$ grid points yielding uniform spacing $\Delta x = \Delta y = 13$~\textmu m}. This resolution adequately captures the laminar flame thickness $\delta_F \approx 0.5$~mm and all reactive-diffusive scales. 
The left boundary supplies fresh reactants at 1~atm, while the right boundary applies non-reflecting outflow conditions. Periodic boundary conditions are imposed in the transverse direction. \textcolor{black}{The turbulence is prescribed using a 2D von K\'arm\'an--Pao kinetic energy spectrum \cite{hinze1975turbulence}, with turbulence intensity $u'=1.49~\mathrm{m/s}$, integral length scale $\ell_t=0.25~\mathrm{mm}$, Kolmogorov scale $\eta=2.22\times10^{-5}~\mathrm{m}$, turbulent kinetic energy $k=2.21~\mathrm{m^2/s^2}$, and dissipation rate $\varepsilon=1.68\times10^4~\mathrm{m^2/s^3}$.}
The base training simulation was performed at an inlet temperature of $T_{\rm in} = 500$~K and equivalence ratio $\phi = 0.7$; validation cases spanning $T_{\rm in} = 300$--600~K and $\phi = 0.7$-1.2 are described in Section~\ref{sec:results}.

The full chemistry reference uses a skeletal 30-species, 184-reaction methane oxidation mechanism \cite{lu2008criterion}, with transport properties computed using mixture-averaged diffusion coefficients. The hybrid ML model operates on a reduced seven-species state vector: $[T, Y_{\rm H_2}, Y_{\rm O_2}, Y_{\rm OH}, Y_{\rm H_2O}, Y_{\rm CH_4}, Y_{\rm CO}, Y_{\rm CO_2}]$. \textcolor{black}{The retained species span the major reactants and products (CH$_4$, O$_2$, H$_2$O, CO$_2$), key partial-oxidation intermediates (CO, H$_2$), and OH as a reaction-zone marker. The current work does not explicitly enforce mass conservation. However, together with N$_2$, which is transported as an inert diluent ($\dot{\omega}_{\rm N_2}=0$), the retained species account for at least $99.86\%$ of the total mixture mass at the initial condition, while remaining within $99.77\%$-$100.08\%$ of the full-chemistry mixture mass at the end of the baseline simulation time. The remaining minor species are not transported or recovered. The entropy generation term during training was computed based on this reduced set. The ML model directly predicts only the chemical source terms for the seven transported reactive species. The temperature source term in the energy equation is then computed from these species source terms and the corresponding species enthalpies.}
Simulations were executed in parallel on 10 compute nodes (48 cores per node). \textcolor{black}{The detailed chemistry was advanced using a timestep of $\Delta t = 5$ ns over 30,000 integration steps until the flame was sufficiently wrinkled, corresponding to an end time of 0.15 ms.}

\section{Results\label{sec:results}}

The performance of the proposed physics-constrained ML model is evaluated from three perspectives: reproduction of full-mechanism DNS when trained on baseline DNS data, prediction at a new operating condition using a model retrained on synthetically augmented data derived from the baseline case, and the impact of entropy-generation constraints on stability. The second assessment is particularly important because the augmented data do not perfectly represent the new condition, thereby introducing some discrepancy and requiring the model to accommodate some degree of extrapolation.

\subsection{Training methodology and anchor selection\label{subsec:training}} \addvspace{1pt}
The ML model takes as input $T$ and the reduced seven-species state vector and predicts the corresponding chemical source terms for each tracked species. The thermochemical state space is partitioned into four temperature-based subdomains, each trained by a separate model, reducing the data range that any single model must represent. \textcolor{black}{The subdomains are defined in terms of a normalized temperature $T^\star = (T - T_{\min})/(T_{\max} - T_{\min})$, with $\mathcal{D}_1$--$\mathcal{D}_4$ spanning $[0, 0.075]$, $[0.05, 0.75]$, $[0.7, 0.85]$, and $[0.8, 1]$. During training, a $5\%$ overlap between adjacent temperature subdomains was employed to promote consistency at the model boundaries. 
Four subdomains provided a good balance between improved local accuracy and the additional training cost associated with a larger number of subdomains.}  From the full $768 \times 384$ grid, representative training points are sampled uniformly per subdomain.  Prior to training, all input state vectors and output source terms are transformed by a fifth-root mapping $\tilde{\psi}_k = \mathrm{sgn}(\psi_k)\,|\psi_k|^{1/5}$ followed by z-score normalization using subdomain-specific means and standard deviations, thereby transforming the magnitude discrepancies in mass fractions and temperature into comparable scales for model training. Within each subdomain, $N_a = 25$ anchor points are selected to span the full range of the species source terms.

The neural network component $\mathcal{N}_{\theta}$ consists of two hidden layers with 32 neurons each, resulting a $8 \times 32 \times 32 \times 8$ architecture. Since the neural network is shallow, we use sigmoidal activations with variable slope parameters per layer \cite{chandrasekhar2026deep}. The output defines the latent feature space in which the Gaussian RBF kernel operates, with a learned length scale $\theta$ per species. All parameters: neural network weights and biases, activation slopes, and the RBF length scale are jointly optimized using an Adam optimization approach with a mini-batch size of 256\textcolor{black}{. The Adam parameters are $\beta_1 = 0.9$, $\beta_2 = 0.999$, and $\epsilon = 10^{-8}$. The initial learning rate is $0.05$, decayed by a factor of $0.5$ every $50$ epochs with a floor of $10^{-5}$. No warm-up or other scheduler was used}.

The original training dataset was generated at $T_{\rm in} = 500$~K from a CFD simulation with full chemistry. \textcolor{black}{For each new condition, the synthetic dataset size is not treated as a free parameter, but is instead fixed by the CFD grid used in the target case. Specifically, a laminar flame profile computed at the same inlet temperature, equivalence ratio, and pressure as the intended CFD condition, containing 768 points, is mapped across the 384 transverse grid levels, with residuals sampled independently at each level, leading to a dataset size of 294,912.} Results presented below correspond to five conditions: $T_{\rm in} = 300$, 400, and 600~K at $\phi = 0.7$, $T_{\rm in} = 300$~K at $\phi = 1.0$, and $T_{\rm in} = 500$~K at $\phi = 1.2$, chosen to probe the model's response to changes in both inlet temperature and mixture composition relative to the baseline condition.  These include lower and higher unburned-gas temperatures at fixed equivalence ratio, as well as stoichiometric and fuel-rich mixtures that alter the thermochemical manifold in composition space.  For each of these conditions, only a steady laminar flame solution was computed; the residual synthesis approach (Section~\ref{subsec:augmentation}) was then used to emulate the turbulent flame statistics at the new condition.  The 500~K data and synthetically generated data at each new condition were combined into an augmented training set, which was then used to train the respective ML models.

\subsection{Data augmentation results\label{subsec:augmentation}} 

\begin{figure*}[H]
    \centering
    \includegraphics[width=1.0\textwidth]{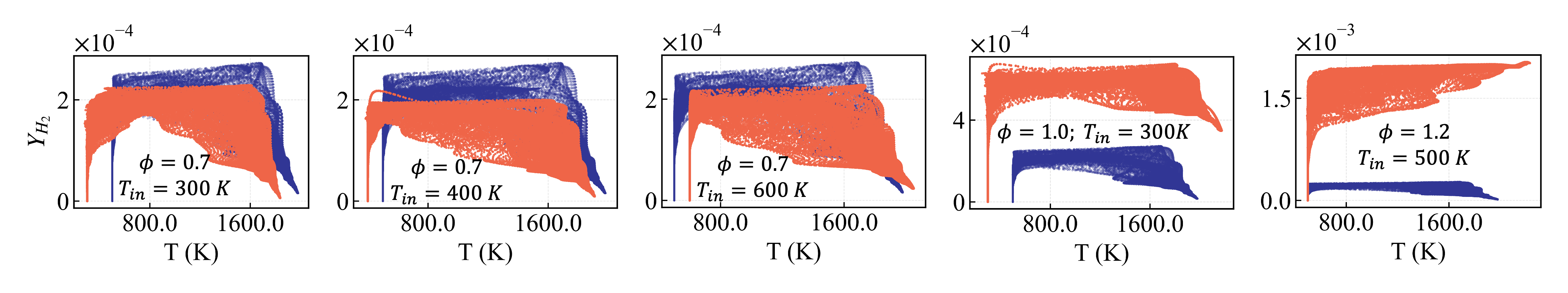}
    \caption{\footnotesize Distribution of H$_2$ mass fraction versus temperature \textcolor{black}{at the final time step} for the base training data ($T_{\rm in}=500$~K, $\phi=0.7$, dark blue) and each target condition (orange), showing the differences in the state distribution.}
    \label{fig:mismatch_all}

\end{figure*}

\begin{figure*}[H]
    
    \centering
    \includegraphics[width=1.0\textwidth]{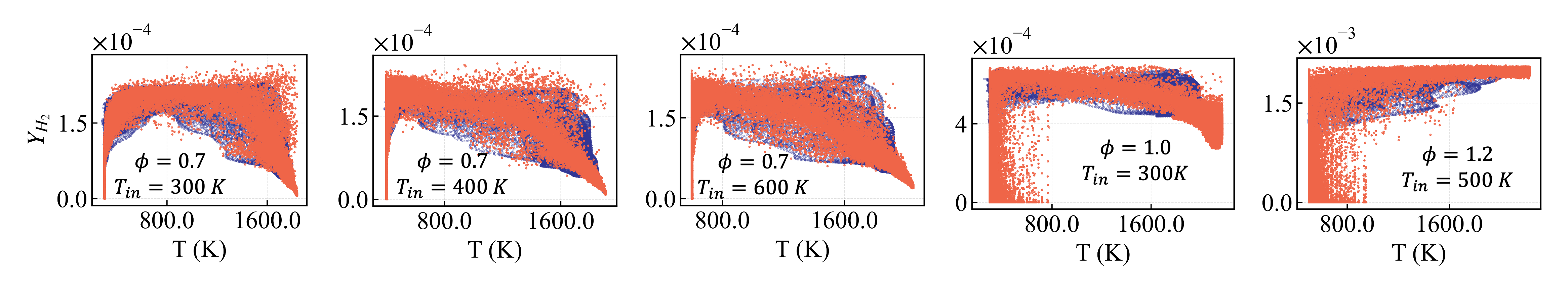}
    \caption{\footnotesize Residual-based synthetic data (orange) generated by the augmentation strategy of Section~\ref{sec:aug} overlaid on the baseline data (dark blue) for each target condition, showing improved overlap between the data distributions compared to Fig. \ref{fig:mismatch_all}.}
    \label{fig:synthetic_all}
    
\end{figure*}

Figure~\ref{fig:mismatch_all} compares the data obtained from the 500~K baseline training case \textcolor{black}{at the final time step} with each new inlet condition using a $Y_{\mathrm{H}_2}$--$T$ slice of the composition space as a representative example. In all cases, large regions of non-overlap are observed, indicating that a model trained solely on the baseline dataset would quickly encounter states outside the training distribution. In several cases, the entire state distribution of the new operating condition lies outside that of the baseline case, which would likely lead to failure for any purely data-driven ML model. In contrast, Fig.~\ref{fig:synthetic_all} shows that the proposed data-augmentation strategy substantially increases the overlap between the baseline dataset and the distributions associated with the new operating conditions. Nevertheless, the learning problem remains challenging. In some cases (see Fig.~\ref{fig:synthetic_all}), the augmented data are sparsely populated near the boundaries of the state space, which can affect model performance. In addition, small regions of non-overlap remain in parts of the composition space, particularly for richer conditions relative to the baseline case. As a result, some degree of extrapolative robustness is still required from the ML model.

\subsection{Predictive accuracy\label{subsec:validation}}

Prediction accuracy for a thermochemical scalar $\psi$ is quantified using the normalized mean absolute error ($\bar{\epsilon}$),
\begin{equation}
\bar{\epsilon}_n = \frac{\bar{\epsilon}}{\psi_{\max} - \psi_{\min}},
\label{eq:nmae}
\end{equation}
where $\bar{\epsilon}$ is the mean absolute error,

\begin{equation}
\bar{\epsilon} = \frac{1}{n_g} \sum_{k=1}^{n_g} 
\left| \psi^{\mathrm{ML}}_k - \psi^{\mathrm{detailed}}_k \right|,
\end{equation}

computed over the $n_g$ grid points in the domain. Here,
$\psi^{\mathrm{ML}}$ and $\psi^{\mathrm{detailed}}$ denote the machine-learning prediction and the detailed-chemistry reference solution, respectively, while $\psi_{\max}$ and $\psi_{\min}$ are the maximum and minimum values of the detailed-chemistry field over the domain.

\begin{figure}[h!]
\centering
\begin{tabular}{@{}c@{\hspace{1pt}}c@{\hspace{1pt}}c@{}}
\begin{overpic}[width=0.25\textwidth]{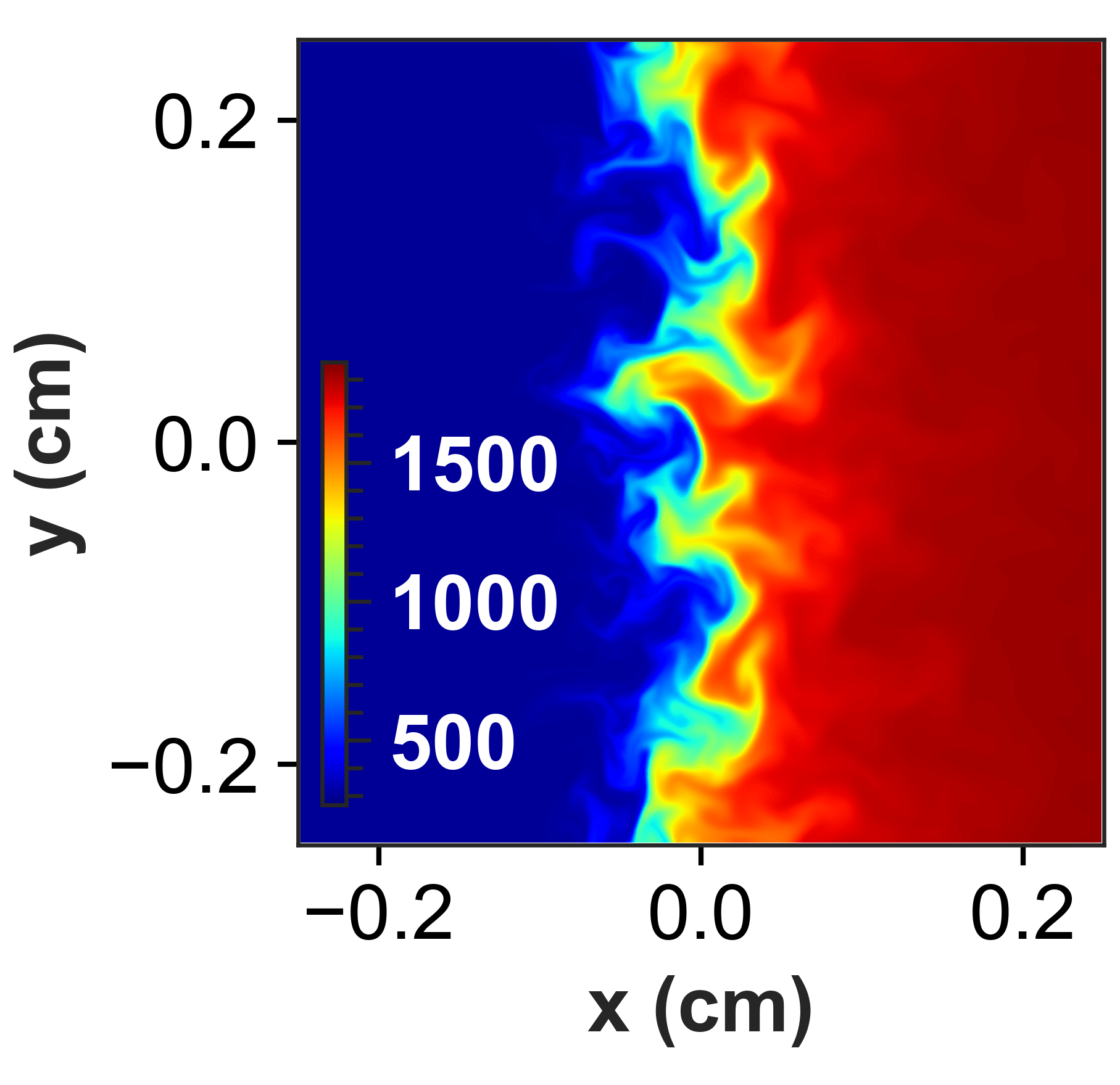}
\put(2,95){\footnotesize\textbf{(a)}}
\end{overpic} &
\begin{overpic}[width=0.25\textwidth]{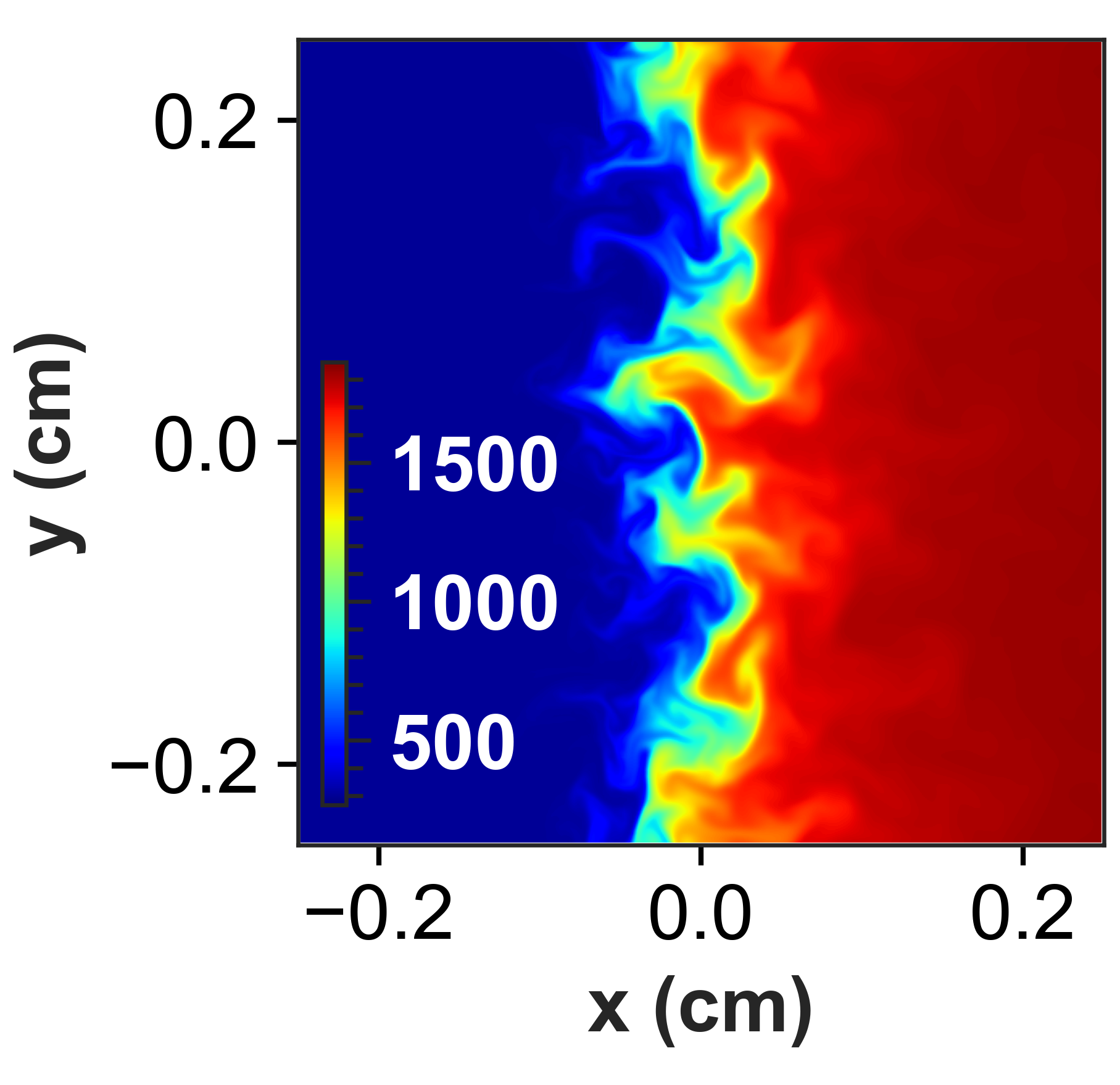}
\put(2,95){\footnotesize\textbf{(b)}}
\end{overpic} &
\begin{overpic}[width=0.25\textwidth]{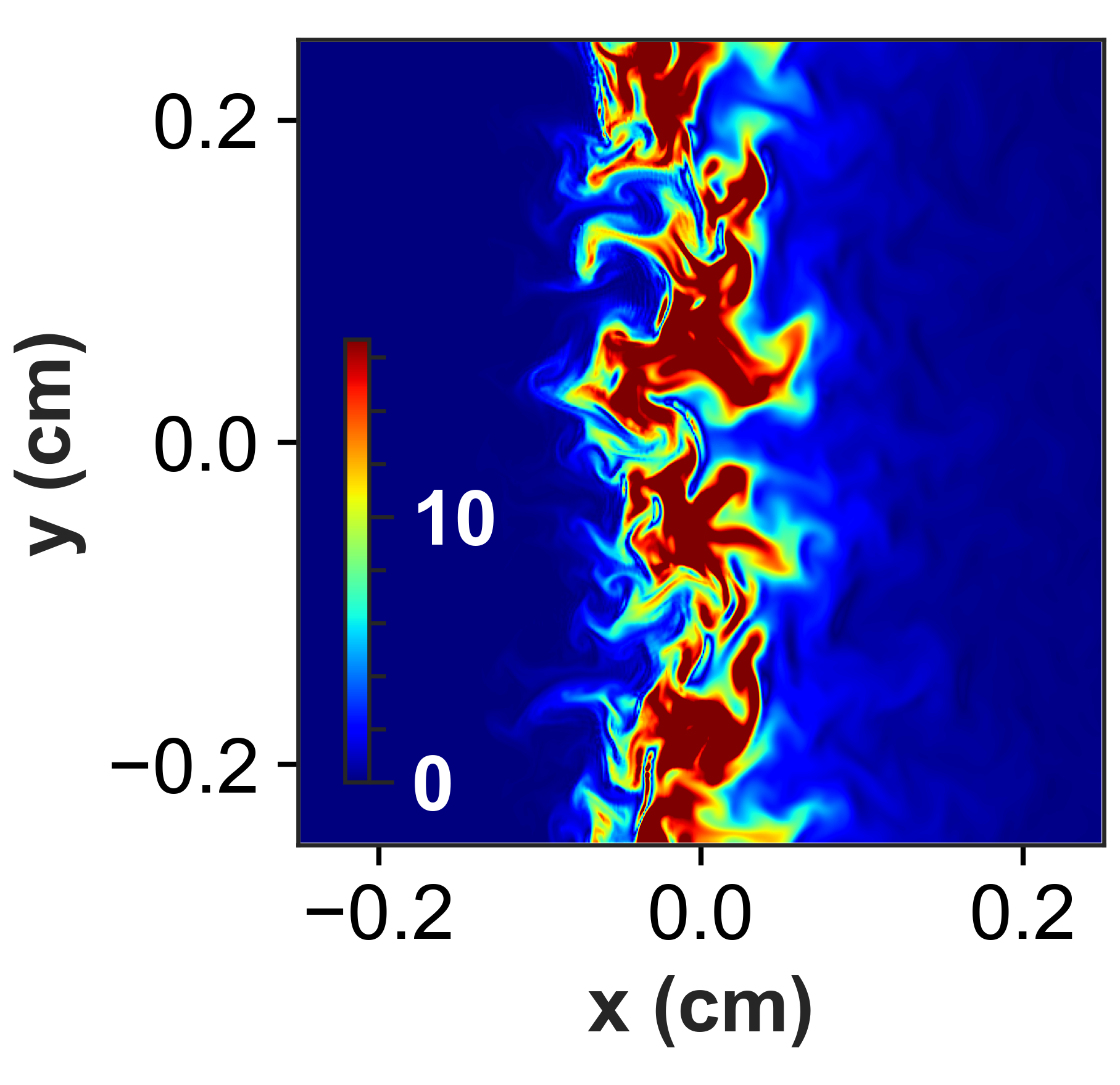}
\put(2,95){\footnotesize\textbf{(c)}}
\end{overpic} \\[-2pt]
\begin{overpic}[width=0.25\textwidth]{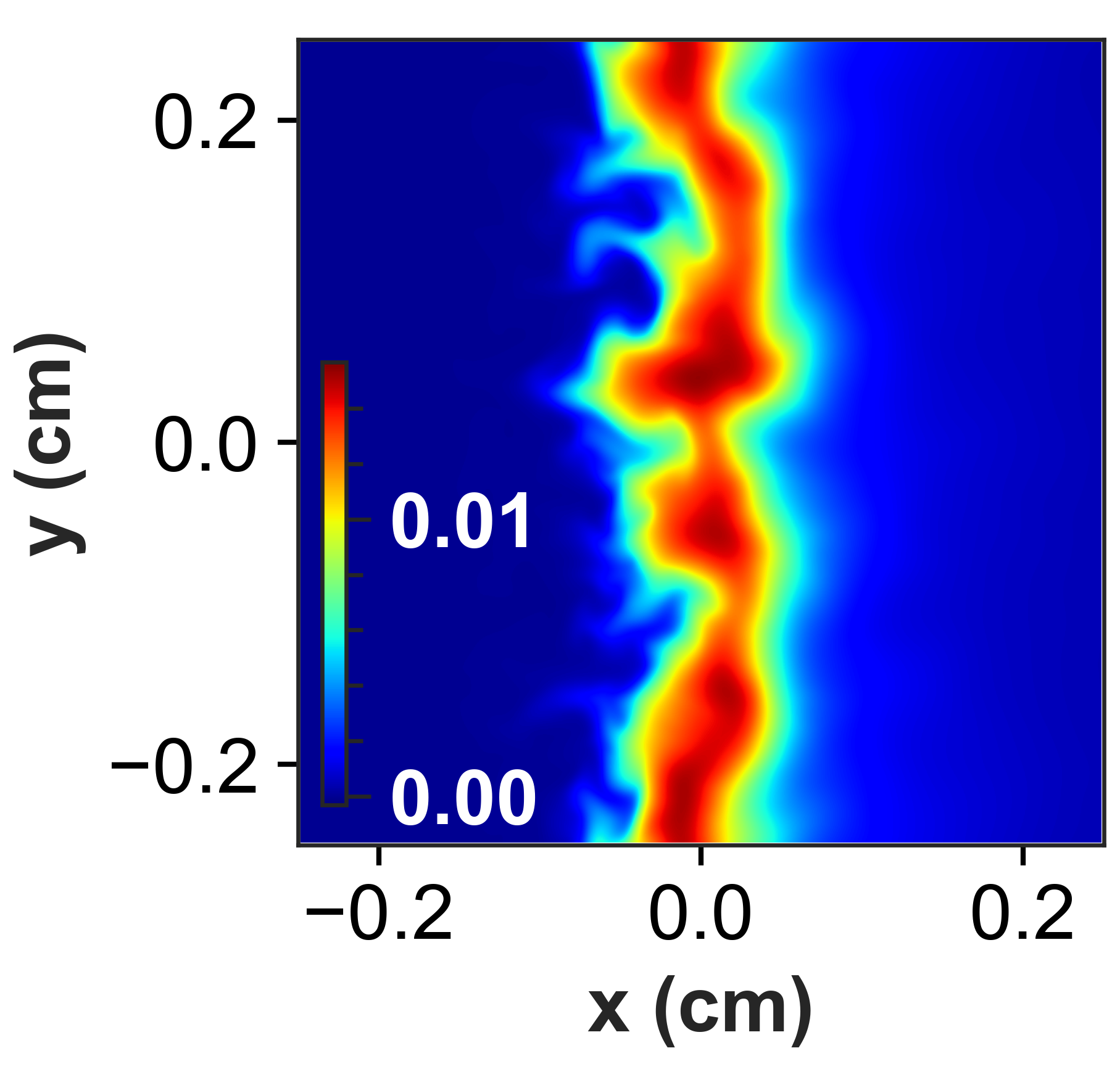}
\put(2,95){\footnotesize\textbf{(d)}}
\end{overpic} &
\begin{overpic}[width=0.25\textwidth]{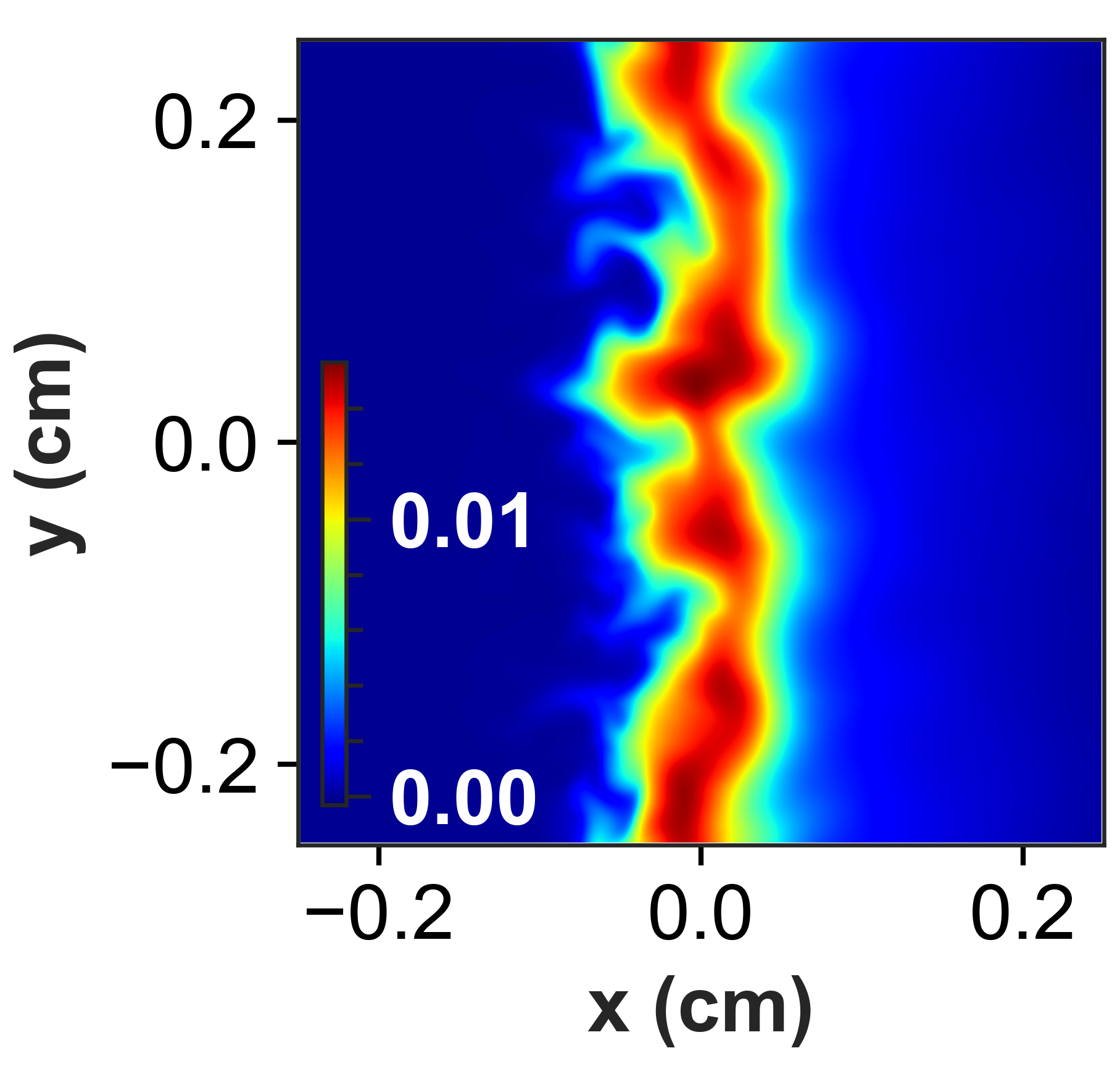}
\put(2,95){\footnotesize\textbf{(e)}}
\end{overpic} &
\begin{overpic}[width=0.25\textwidth]{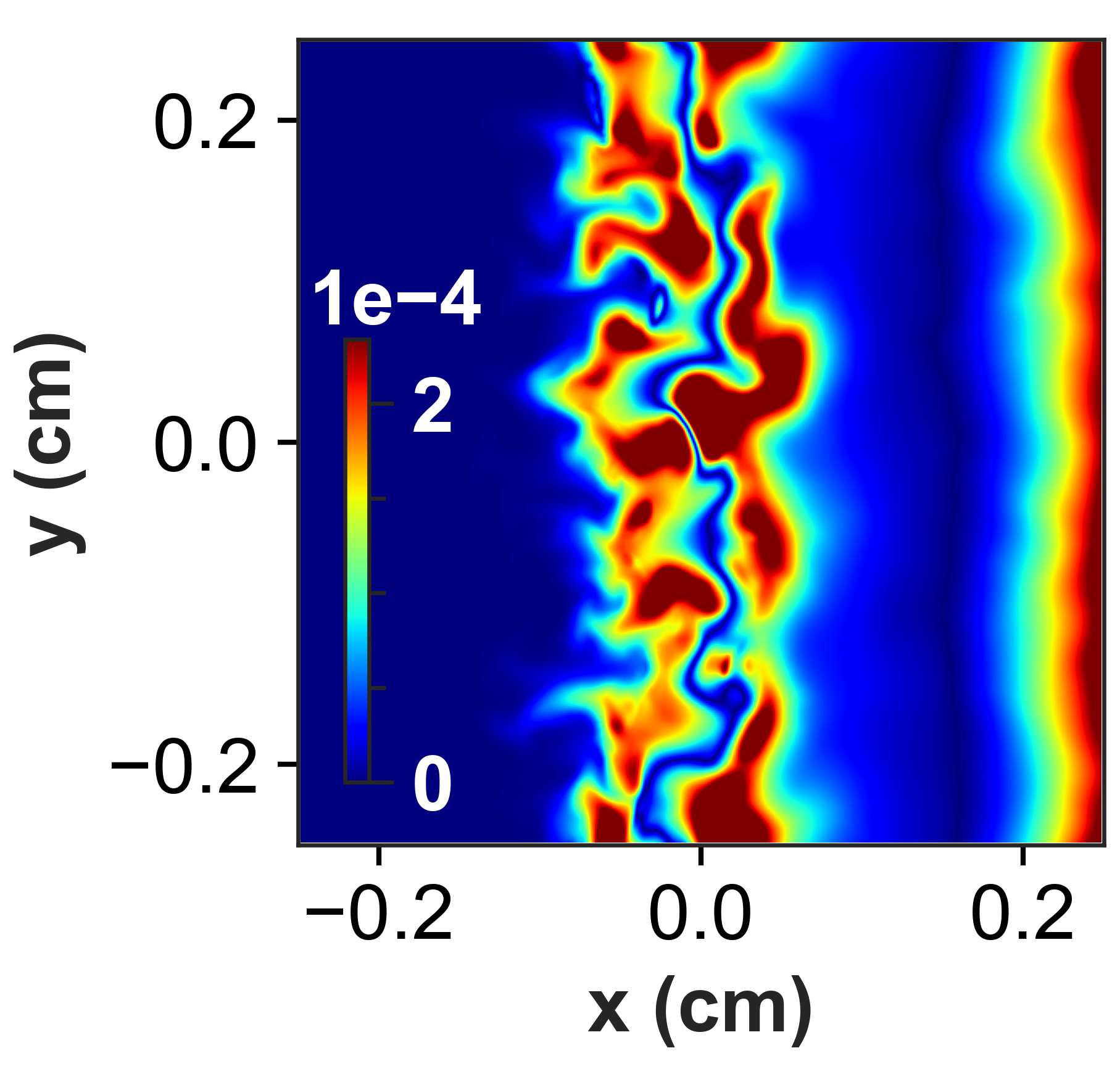}
\put(2,95){\footnotesize\textbf{(f)}}
\end{overpic}
\end{tabular}
\caption{\footnotesize Validation at $T_{\rm in} = 300$~K, $\phi = 0.7$: (a,d)~full 30-species mechanism contours; (b,e)~ML model contours; (c,f)~absolute error fields. Top row: temperature; bottom row: $Y_{\rm CO}$.}
\label{fig:validation}
\end{figure}

\begin{figure}[h!]
\centering
\begin{tabular}{@{}c@{\hspace{1pt}}c@{\hspace{1pt}}c@{}}
\begin{overpic}[width=0.25\textwidth]{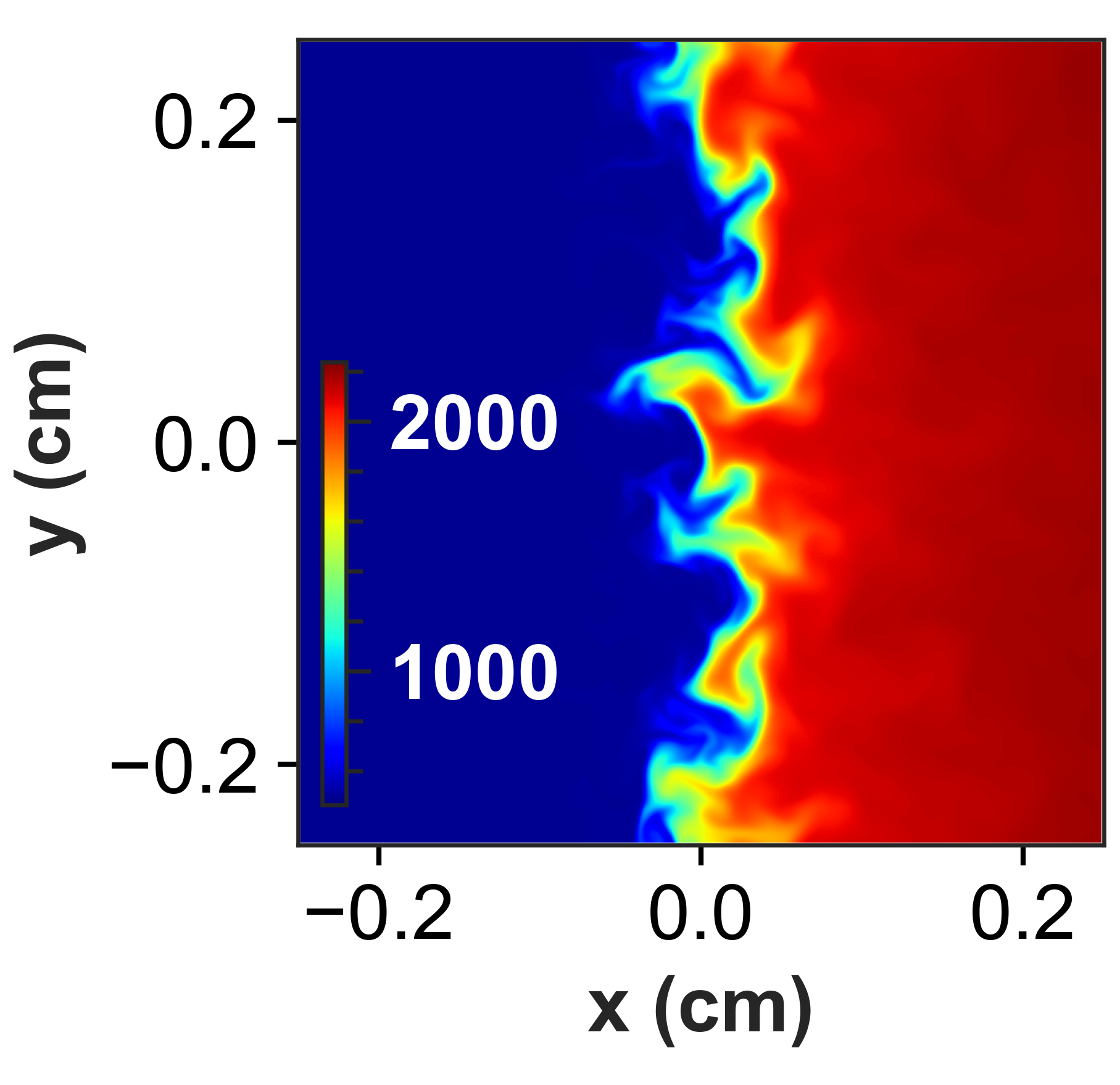}
\put(2,95){\footnotesize\textbf{(a)}}
\end{overpic} &
\begin{overpic}[width=0.25\textwidth]{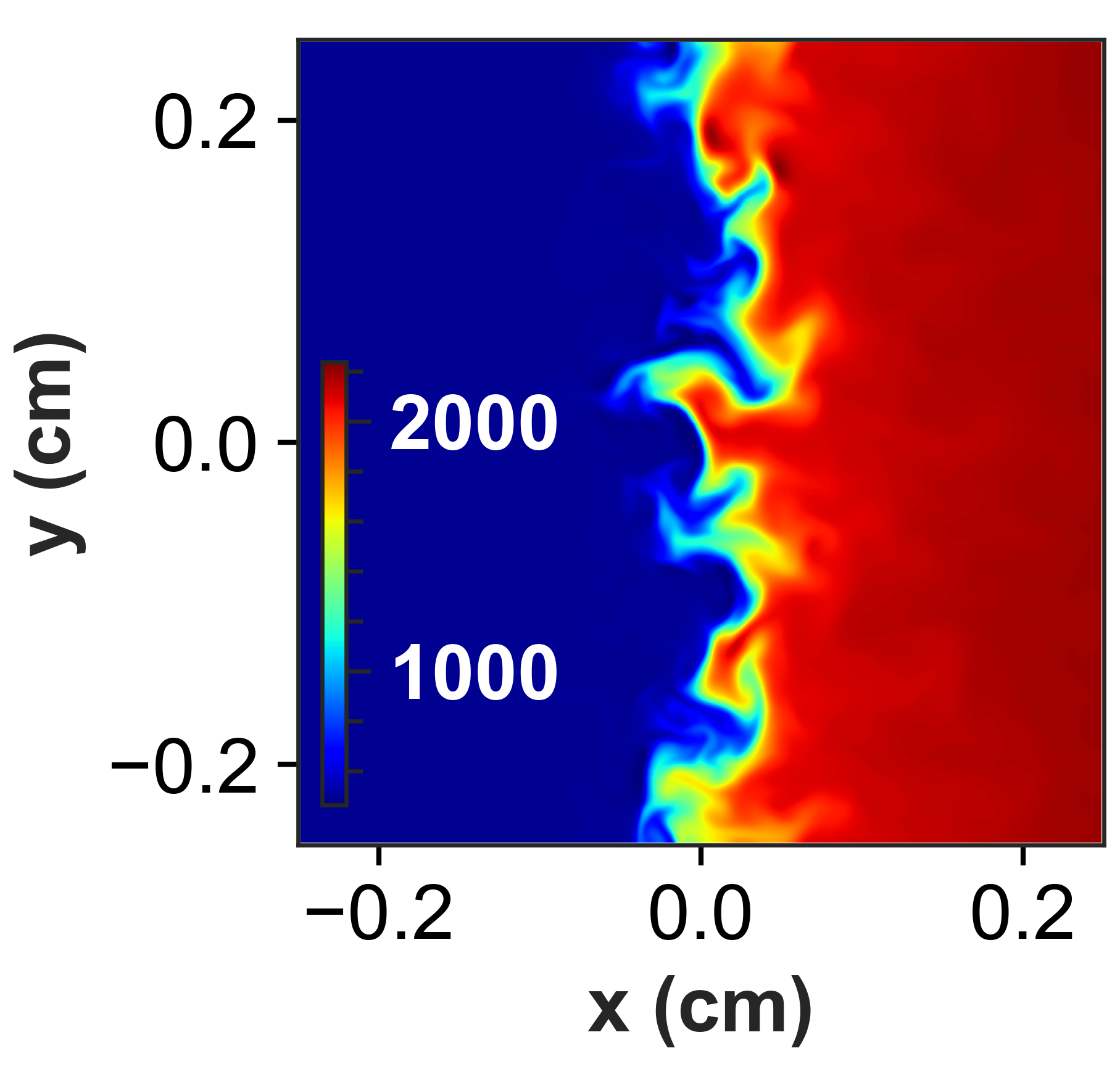}
\put(2,95){\footnotesize\textbf{(b)}}
\end{overpic} &
\begin{overpic}[width=0.25\textwidth]{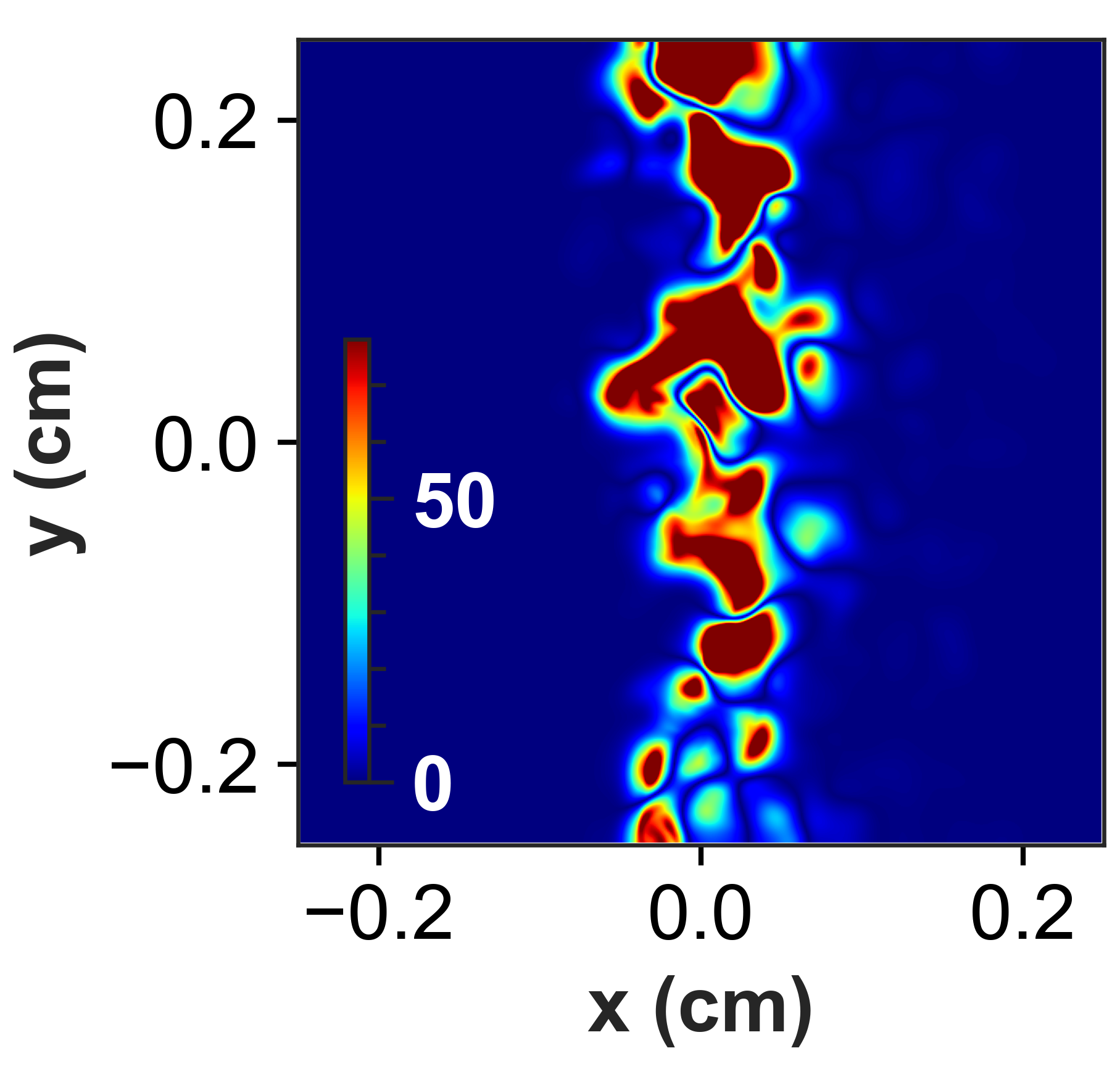}
\put(2,95){\footnotesize\textbf{(c)}}
\end{overpic} \\[-2pt]
\begin{overpic}[width=0.25\textwidth]{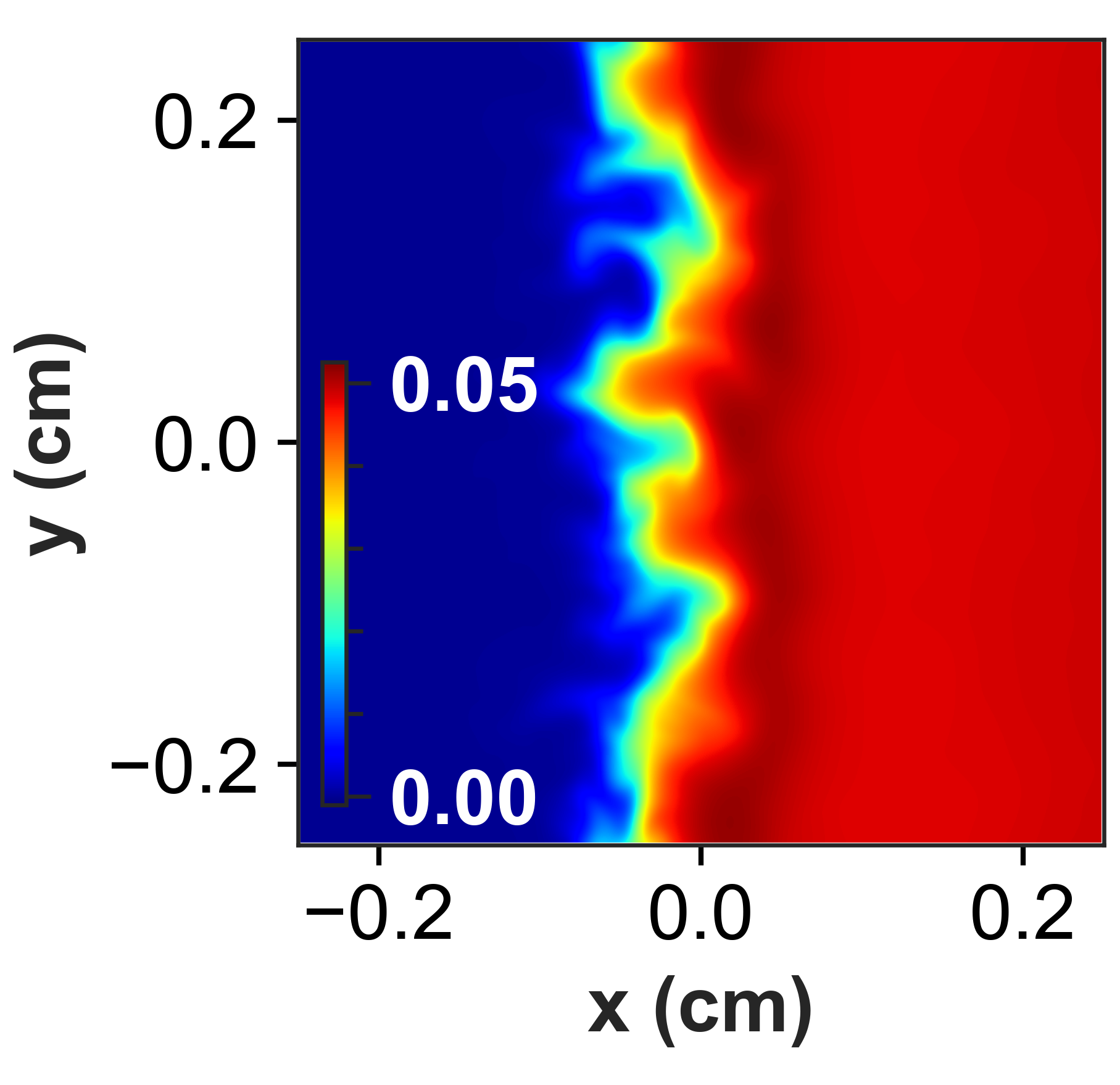}
\put(2,95){\footnotesize\textbf{(d)}}
\end{overpic} &
\begin{overpic}[width=0.25\textwidth]{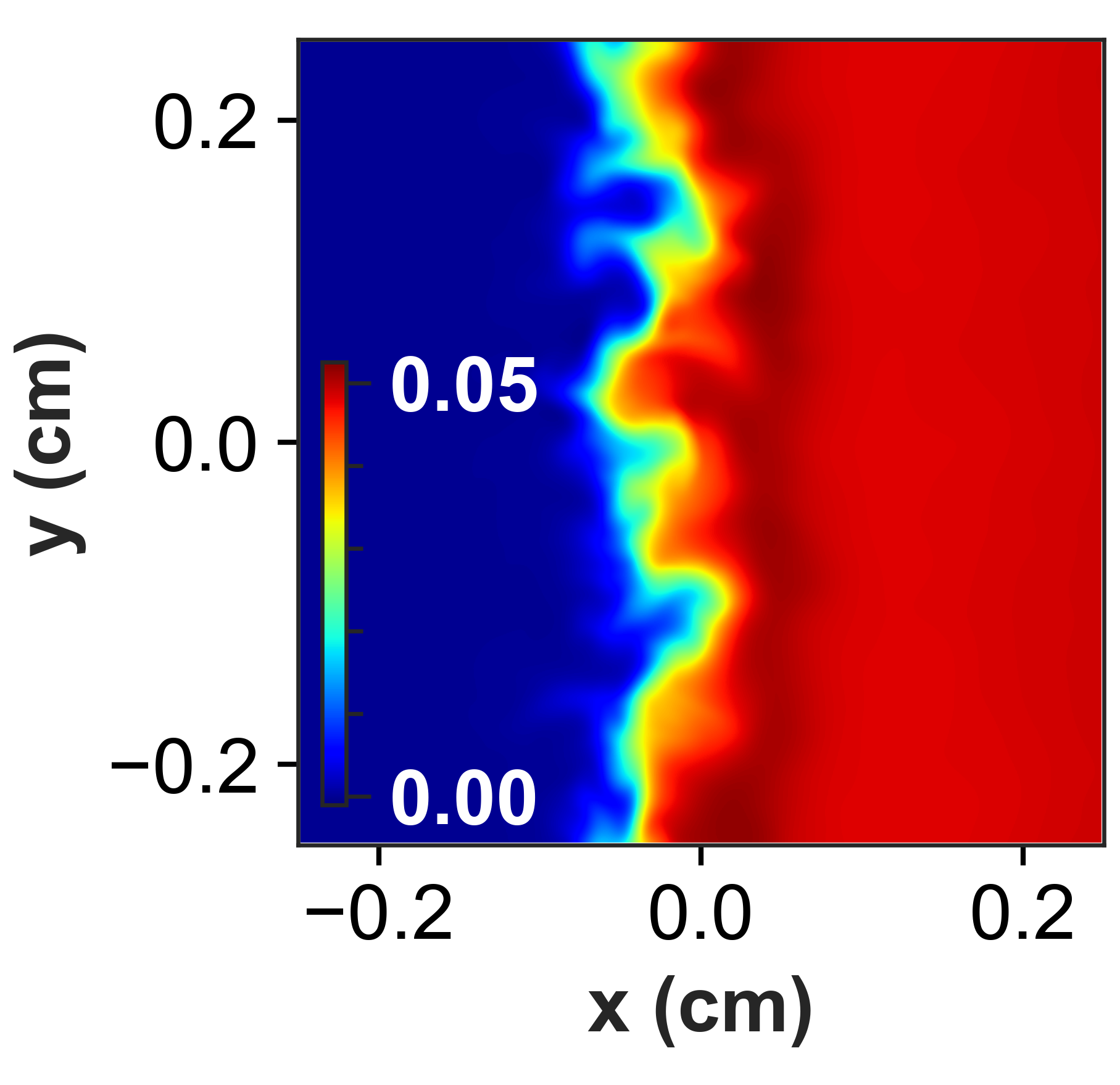}
\put(2,95){\footnotesize\textbf{(e)}}
\end{overpic} &
\begin{overpic}[width=0.25\textwidth]{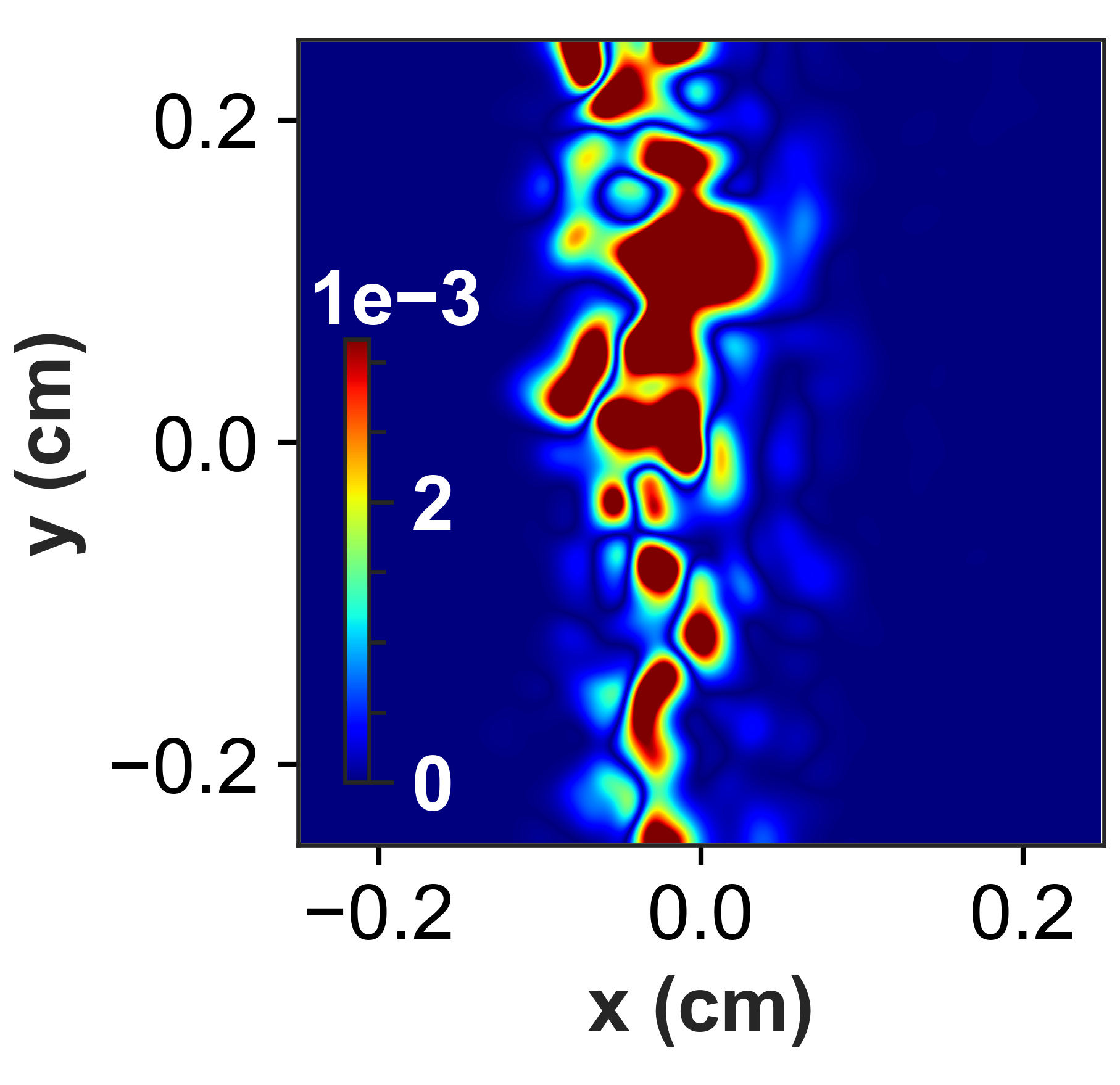}
\put(2,95){\footnotesize\textbf{(f)}}
\end{overpic}
\end{tabular}
\caption{\footnotesize Validation at $T_{\rm in} = 500$~K, $\phi = 1.2$: (a,d)~full mechanism; (b,e)~ML model; (c,f)~absolute error. Top: temperature; bottom: $Y_{\rm CO}$.}
\label{fig:val_500_1p2}
\end{figure}

\textcolor{black}{Figures~\ref{fig:validation} and \ref{fig:val_500_1p2}} compare simulation results from the reduced physics-constrained ML model against the full 30-species mechanism DNS for temperature and $Y_{\rm CO}$ \textcolor{black}{for two representative validation conditions: the case closest to the baseline ($T_{\rm in} = 300$~K, $\phi = 0.7$) and the most challenging case ($T_{\rm in} = 500$~K, $\phi = 1.2$). The remaining three conditions are quantified in Table~\ref{tab:extended_performance} with contours plots of $T$, CO and OH shown in the Supplemental Material.} \textcolor{black}{All comparisons in this section and the accompanying figures are at the final time step, where $t = t_{\rm end}$.} At the test case of $T_{\rm in} = 300$~K, $\phi = 0.7$, Fig.~\ref{fig:validation}, the two-dimensional contours demonstrate that the reduced model accurately reproduces the spatial flame structure for both fields, capturing the wrinkled flame front topology, the sharp transition from unburned to burned gases, and the detailed temperature distribution across the reaction zone. The absolute error fields (Fig.~\ref{fig:validation}c,f) provide a pointwise measure of prediction quality across the entire domain. For temperature, the value of the normalized mean absolute error, $\bar{\epsilon}_{T}$, is $1.80 \times 10^{-3}$, with the errors concentrated along the wrinkled flame front as expected. In this region, errors are amplified by large gradients and differences in fluid properties between the surrogate and detailed-chemistry solutions. In the unburned and post-flame regions the error is negligible. The CO error field exhibits a similar spatial pattern ($\bar{\epsilon}_{\text{CO}} = 4.25 \times 10^{-3}$), with the relatively large discrepancies localized to the reaction zone. \textcolor{black}{Errors also appear relatively higher toward the boundary, likely due to the diminished accuracy in the post-flame region}. 

\textcolor{black}{For the intermediate conditions ($T_{\rm in} = 400$~K and $600$~K at $\phi = 0.7$, and $T_{\rm in} = 300$~K at $\phi = 1.0$), the surrogate reproduces the wrinkled flame front topology and CO production-consumption structure with error magnitudes that scale with the departure from the baseline condition (Table~\ref{tab:extended_performance}).} The $\phi = 1.2$ case (Fig.~\ref{fig:val_500_1p2}) represents the most challenging condition, with the highest $\bar{\epsilon}_T$ of $7.45 \times 10^{-3}$ and $\bar{\epsilon}_{\text{CO}}$ of $8.89 \times 10^{-3}$; the surrogate nonetheless tracks the elevated post-flame CO levels and broader intermediate-species distributions characteristic of rich combustion. Similar agreement was observed for all tracked species (H$_2$, OH, H$_2$O, CH$_4$, CO$_2$) across all conditions.

\begin{figure}[h!]
\centering
\textcolor{black}{\includegraphics[width=0.4\columnwidth]{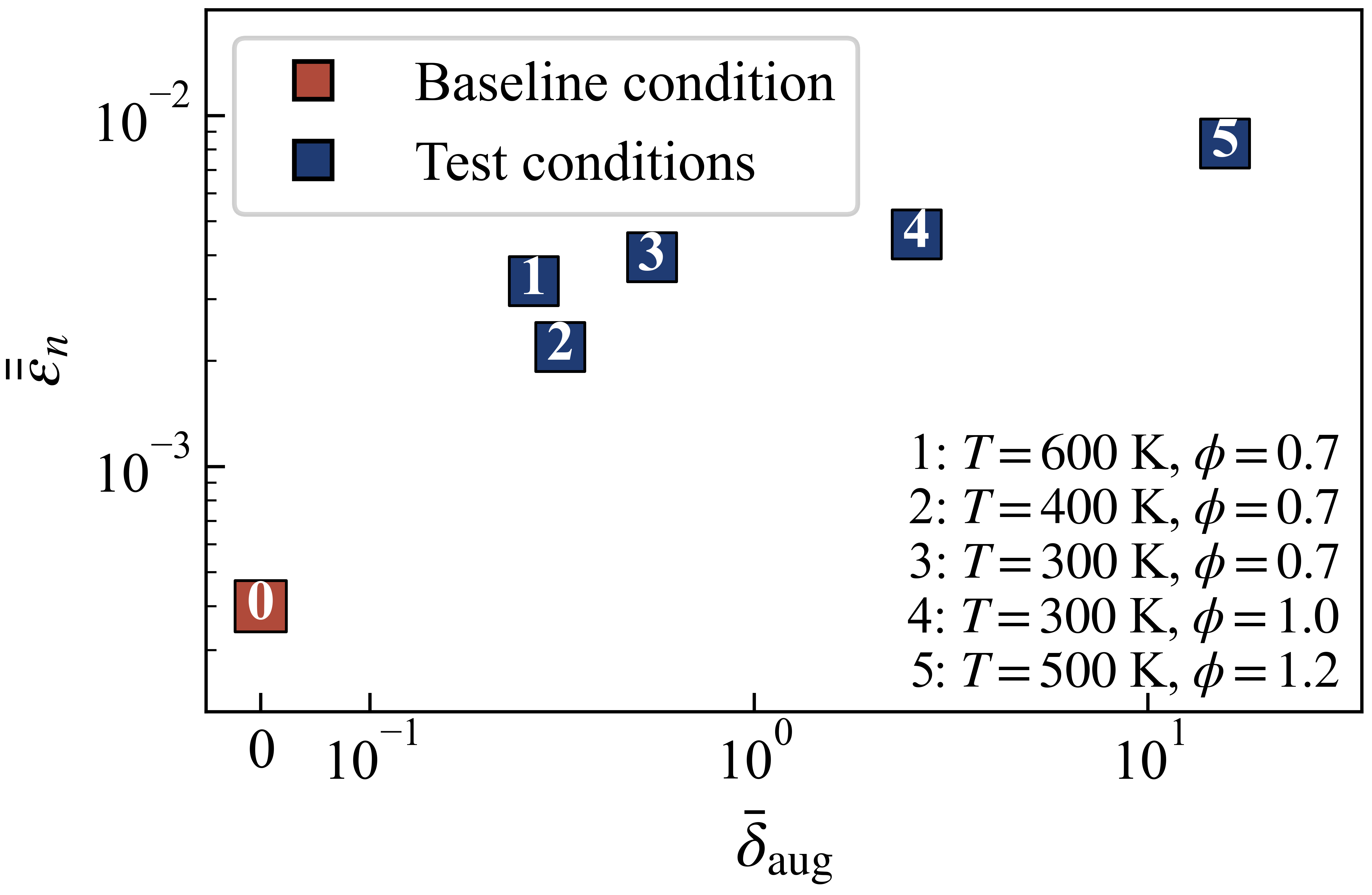}}
\caption{\footnotesize \textcolor{black}{Species-averaged normalized mean absolute error $\bar{\bar{\epsilon}}_n$ versus the mean nearest-neighbor distance $\bar{\delta}_{\mathrm{aug}}$.}}
\label{fig:error_vs_mismatch}
\end{figure}

\begin{table*}[h!]
\footnotesize
\caption{Normalized MAE ($\bar{\epsilon}_n$) across all validated conditions \textcolor{black}{at the final time step}.}
\centering
\begin{tabular}{@{}l c c c c c@{}}
\hline
 & \multicolumn{5}{c}{Condition ($T_{\rm in}$, $\phi$)} \\
\cline{2-6}
Species 
  & 300\,K  & 400\,K  & 600\,K  & 300\,K  & 500\,K \\
  & 0.7     & 0.7     & 0.7     & 1.0     & 1.2    \\
\hline
$T$              & $1.80 \times 10^{-3}$ & $2.17 \times 10^{-3}$ & $3.50 \times 10^{-3}$ & $6.13 \times 10^{-3}$ & $7.45 \times 10^{-3}$ \\
$Y_{\rm H_2}$    & $8.75 \times 10^{-3}$ & $3.75 \times 10^{-3}$ & $4.39 \times 10^{-3}$ & $6.28 \times 10^{-3}$ & $1.13 \times 10^{-2}$ \\
$Y_{\rm O_2}$    & $2.14 \times 10^{-3}$ & $1.73 \times 10^{-3}$ & $2.28 \times 10^{-3}$ & $2.06 \times 10^{-3}$ & $7.63 \times 10^{-3}$ \\
$Y_{\rm OH}$     & $9.12 \times 10^{-3}$ & $2.04 \times 10^{-3}$ & $2.97 \times 10^{-3}$ & $4.60 \times 10^{-3}$ & $7.07 \times 10^{-3}$ \\
$Y_{\rm H_2O}$   & $2.12 \times 10^{-3}$ & $1.67 \times 10^{-3}$ & $3.25 \times 10^{-3}$ & $6.35 \times 10^{-3}$ & $8.14 \times 10^{-3}$ \\
$Y_{\rm CH_4}$   & $1.65 \times 10^{-3}$ & $1.51 \times 10^{-3}$ & $2.97 \times 10^{-3}$ & $1.77 \times 10^{-3}$ & $8.18 \times 10^{-3}$ \\
$Y_{\rm CO}$     & $4.25 \times 10^{-3}$ & $3.45 \times 10^{-3}$ & $5.41 \times 10^{-3}$ & $7.38 \times 10^{-3}$ & $8.89 \times 10^{-3}$ \\
$Y_{\rm CO_2}$   & $1.77 \times 10^{-3}$ & $1.21 \times 10^{-3}$ & $2.21 \times 10^{-3}$ & $2.07 \times 10^{-3}$ & $7.96 \times 10^{-3}$ \\
\hline
\end{tabular}
\label{tab:extended_performance}

\end{table*}
Table~\ref{tab:extended_performance} consolidates the values of $\bar{\epsilon}$ for temperature and all tracked species across the five validation conditions. The test cases with a slightly perturbed inlet temperature: $T_{\rm in} = 300$~K, $\phi = 0.7$ and $T_{\rm in} = 400$~K, $\phi = 0.7$ yield the lowest errors. On the other hand, the stoichiometric ($\phi = 1.0$) and rich ($\phi = 1.2$) cases exhibit progressively higher errors, since shifts in equivalence ratio alter both the equilibrium composition and the thermochemical manifold structure, representing a greater departure from the lean condition at which the DNS training data was generated. \textcolor{black}{To quantify this departure, we scale the datasets to zero mean and unit variance, compute the mean nearest-neighbor distance $\bar{\delta}_{\mathrm{aug}}$ between each target-condition state and the augmented training set, then plot it against the species-averaged error $\bar{\bar{\epsilon}}_n$ in Fig.~\ref{fig:error_vs_mismatch}. The trend is broadly monotonic: lean cases cluster at moderate $\bar{\delta}_{\mathrm{aug}}$ with $\bar{\bar{\epsilon}}_n \approx 2$--$4 \times 10^{-3}$, while the stoichiometric and rich cases lie at larger $\bar{\delta}_{\mathrm{aug}}$ and reach $\sim 9 \times 10^{-3}$ at $\phi = 1.2$. This identifies the residual training--target mismatch as the dominant source of degradation, and suggests that beyond some threshold of $\bar{\delta}_{\mathrm{aug}}$ the surrogate would require retraining on a baseline DNS closer to the target regime.} Nevertheless, in the table, across all  conditions considered, the value of $\bar{\epsilon}$ for every tracked species remains below $1.2 \times 10^{-2}$, indicating that the combined augmentation and entropy-constraint strategy extends the applicability of the surrogate beyond a single operating point.

\subsection{Effect of the thermodynamic constraint\label{subsec:constraint_effect}} 

\begin{figure}[h!]
\centering
\includegraphics[width=0.4\columnwidth]{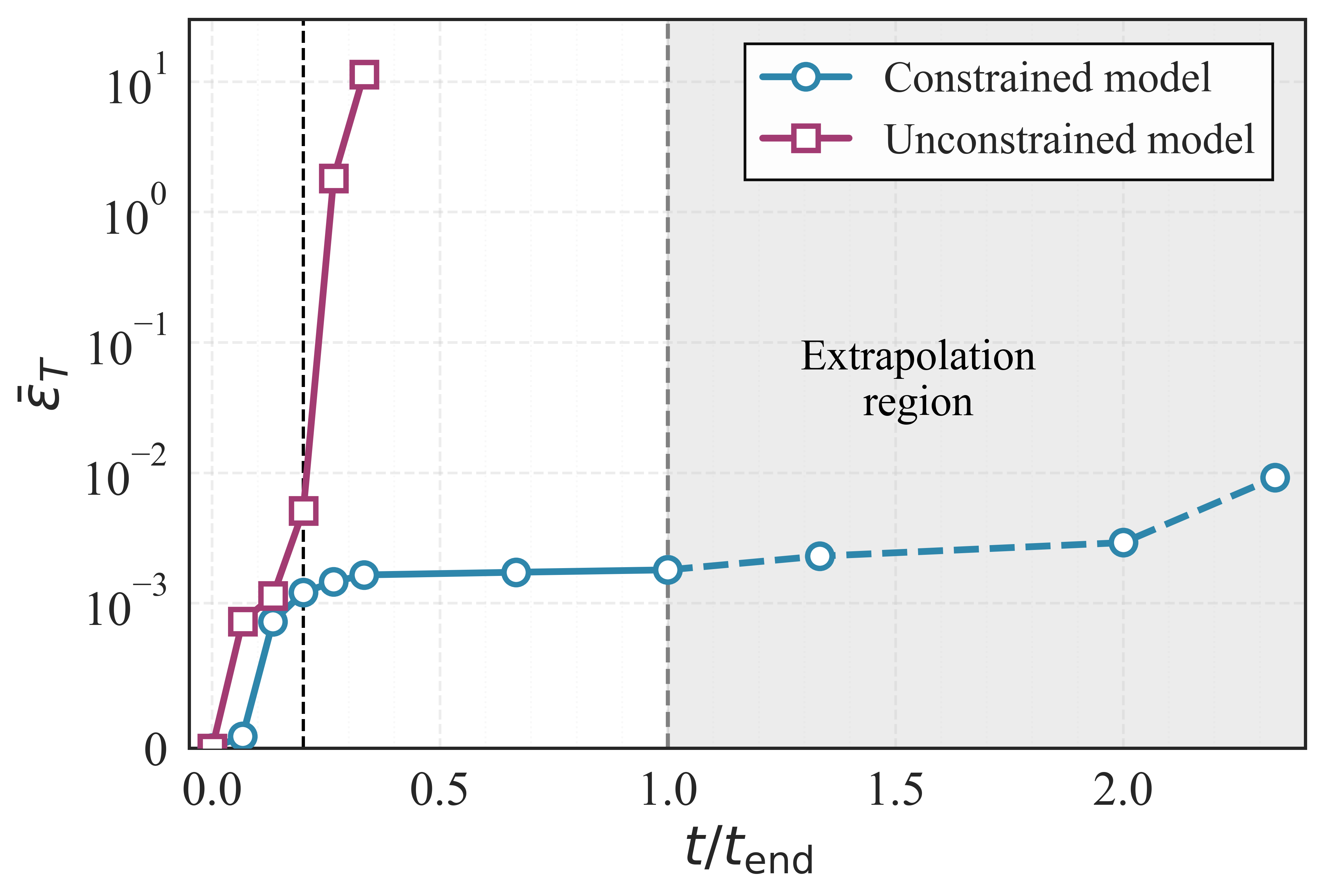}
\caption{\textcolor{black}{\footnotesize Temperature error $\bar{\epsilon}_T$ versus normalized time for the constrained and unconstrained models at $\phi = 0.7$, $T_{\rm in}=300$~K. The black dashed line marks the onset of rapid divergence in the unconstrained model.}}
\label{fig:error_comparison}
\end{figure}

Fig.~\ref{fig:error_comparison} compares the normalized relative errors for a standard ML model without thermodynamic constraints for the case with $\phi = 0.7$ and \textcolor{black}{$T_{\rm in} = 300~\mathrm{K}$} . For the unconstrained model, the error grows rapidly and eventually diverges before $t_{\rm end}$, with the temperature becoming unbounded and simulation failing due to numerical divergence. This divergence is associated with violations of the second law, where the entropy generation becomes negative in parts of the domain. In contrast, the entropy-constrained model maintains a bounded error throughout the simulation and exhibits substantially more stable behavior relative to the unconstrained case. 

\textcolor{black}{To probe behavior beyond the training horizon, the constrained simulation was continued beyond $t_{\rm end}$ (shaded region in Fig.~\ref{fig:error_comparison}). The error remained bounded and below $2\times10^{-3}$ throughout the training interval ($t \le t_{\rm end}$). Beyond the training horizon, the error grew slowly, reaching approximately $3\times10^{-3}$ at $t = 2t_{\rm end}$, indicating that the constraint sustains stability over a meaningful extrapolation window even as the solution departs from the training distribution. Past this point, error accumulation accelerated more noticeably, reaching approximately $9\times10^{-3}$ by $t \approx 2.3t_{\rm end}$. The simulation eventually diverged near $t \approx 3t_{\rm end}$. These results suggest that the constraint substantially delays instability and maintains physically consistent evolution well beyond the training horizon, although eventual divergence is expected as the trajectory drifts sufficiently far from the training manifold.}

\begin{figure}[h!]
\centering
\begin{tabular}{@{}c@{\hspace{0pt}}c@{}}
\includegraphics[width=0.4\columnwidth]{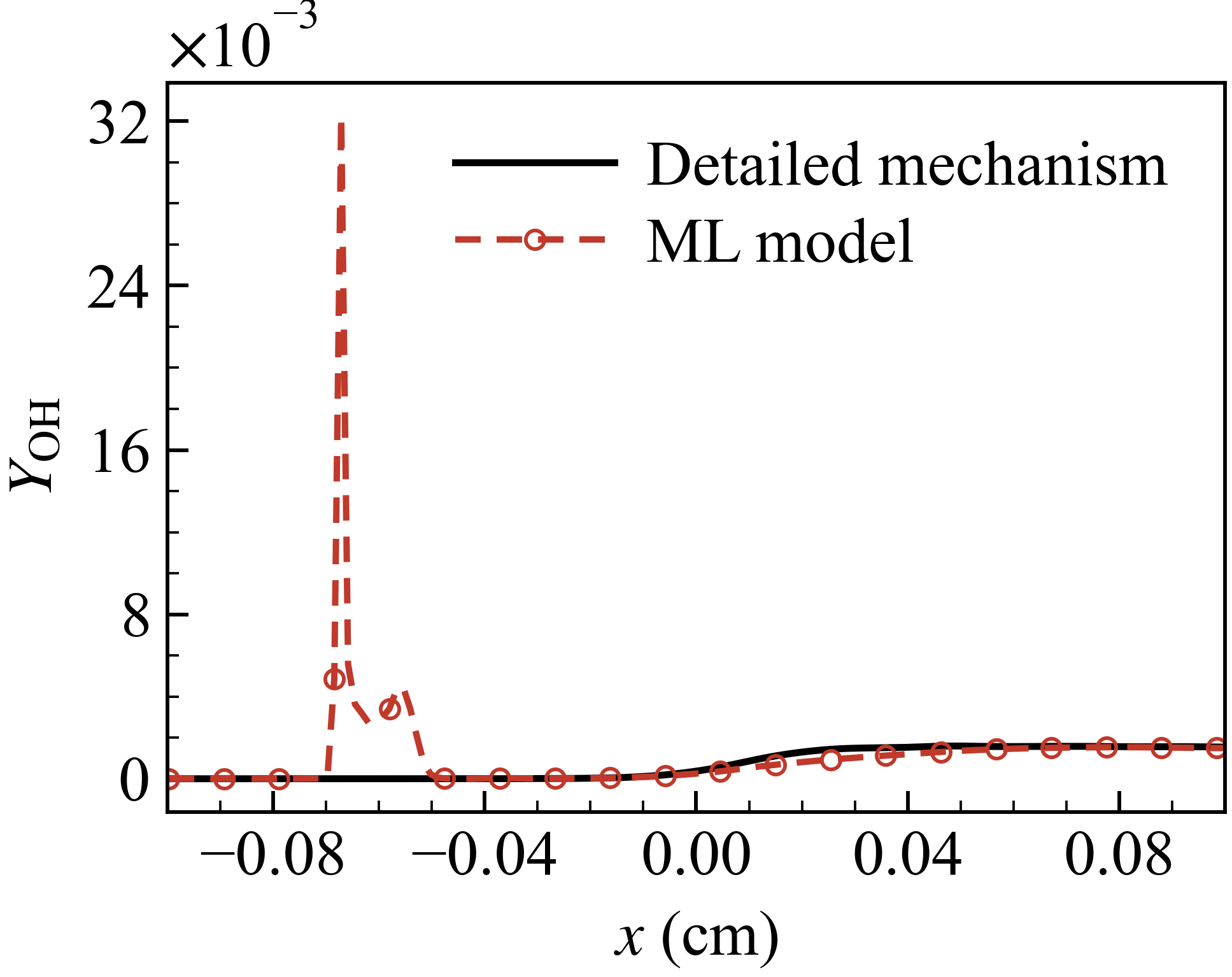} &
\textcolor{black}{\includegraphics[width=0.4\columnwidth]{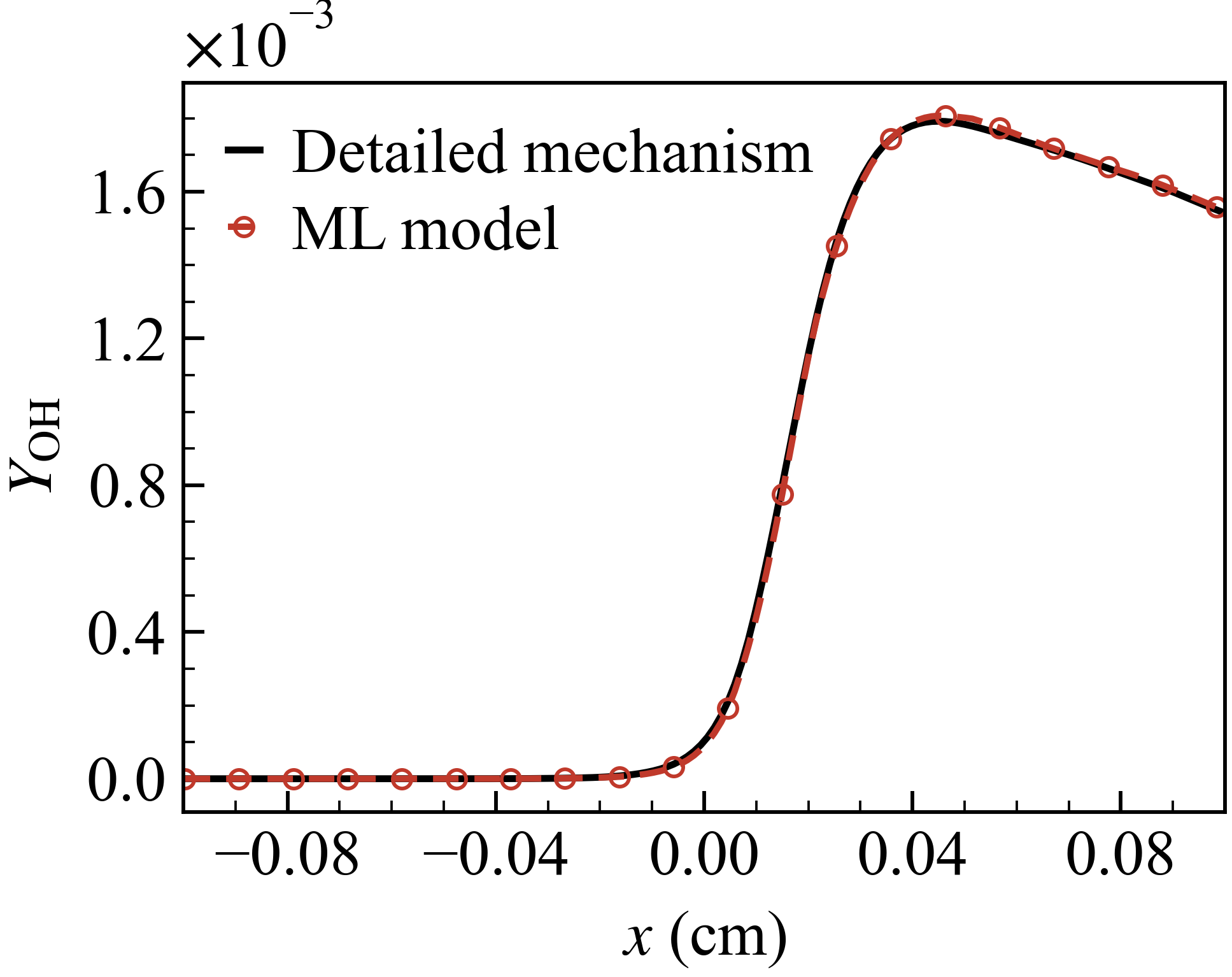}} \\
\footnotesize (a) Unconstrained ML & \textcolor{black}{\footnotesize (b) Constrained ML}
\end{tabular}
\caption{\footnotesize $Y_{\mathrm{OH}}$ profiles along the horizontal line through the global OH maximum for (a) unconstrained model \textcolor{black}{; (b) constrained model.}}
\label{fig:oh_constraint}
\end{figure}

The physical origin of this divergence is illustrated in Fig.~\ref{fig:oh_constraint}, which compares $Y_{\mathrm{OH}}$ profiles extracted along the horizontal line passing through the global OH maximum shortly before the unconstrained model diverges and numerically fails. The detailed-chemistry DNS (solid line) exhibits the expected distribution: negligible OH in the unburned mixture, a bounded rise through the flame front near $x \approx -0.065$~cm, and a gradual approach to post-flame equilibrium levels on the order of $1 \times 10^{-3}$.  The unconstrained surrogate (dashed line) develops a pronounced spurious spike at the flame front, where $Y_{\mathrm{OH}}$ exceeds $3 \times 10^{-2}$, which represents an order of magnitude above the physical peak.  This localized overshoot indicates that the unconstrained model predicts reaction rates that drive the thermochemical state along non-physical trajectories, generating radical concentrations in excess of what the local chemical-potential landscape permits, which is a sustained consequence of unconstrained source-term prediction. \textcolor{black}{Fig.~\ref{fig:oh_constraint}b shows that the constrained model evaluated at the same physical time is in good agreement with the DNS profile.}

\begin{figure}[h!]
\centering
\includegraphics[width=\columnwidth]{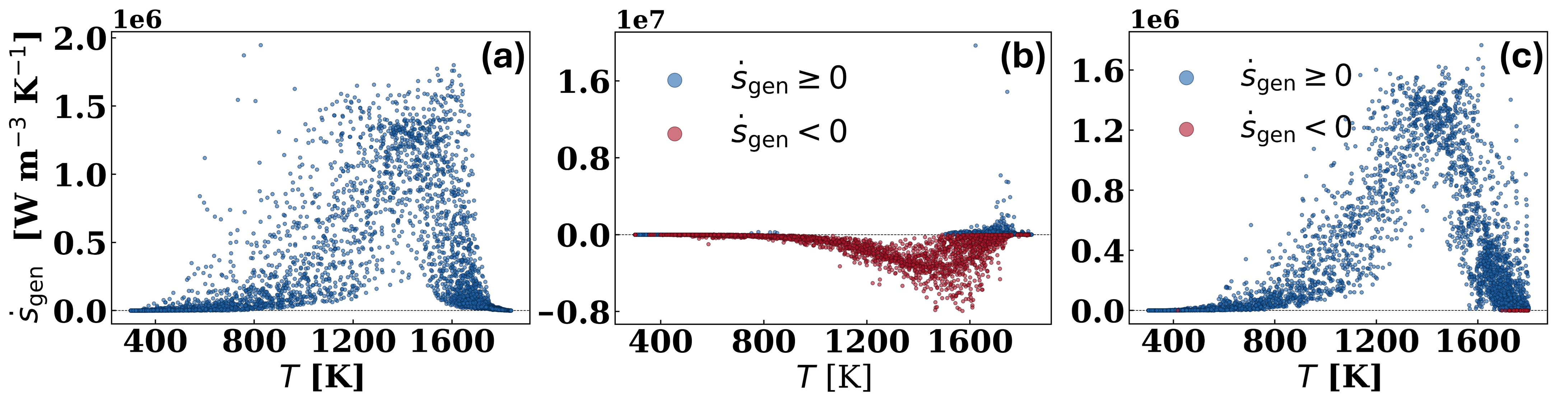}
\caption{\footnotesize Local entropy generation rate vs $T$: (a)~detailed-chemistry DNS; (b)~unconstrained ML model \textcolor{black}{; (c)~constrained ML model}}
\label{fig:entropy_scatter}
\end{figure}

Figure~\ref{fig:entropy_scatter} provides a direct diagnostic of second-law compliance by plotting the local entropy generation rate $\dot{s}_{\rm gen}$ against temperature for every grid point in the domain, as obtained at $t = 2/3 \ t_{end}$. For the detailed-chemistry DNS (Fig.~\ref{fig:entropy_scatter}a), the entropy generation is non-negative everywhere: $\dot{s}_{\rm gen}$ remains near zero in the unburned and fully equilibrated post-flame regions and rises through the reaction zone, reaching peak values of approximately $2 \times 10^{6}$~W\,m$^{-3}$\,K$^{-1}$ in the 1000--1700~K range where irreversible chemical reactions are most pronounced. 

The unconstrained ML model (Fig.~\ref{fig:entropy_scatter}b) presents a qualitatively different picture.  A large population of points exhibits negative entropy generation (represented with red markers), concentrated in the 750-1750~K temperature range. The magnitudes of these violations are substantial, reaching $-0.5 \times 10^{7}$~W\,m$^{-3}$\,K$^{-1}$, which is comparable in magnitude to the largest positive entropy generation rates. \textcolor{black}{In contrast, the constrained ML model (Fig.~\ref{fig:entropy_scatter}c), evaluated on its own simulation state at the same instant, recovers a distribution closely resembling the detailed-chemistry case: nearly all points satisfy $\dot{s}_{\rm gen} \geq 0$, while the remaining small negative values are approximately four orders of magnitude smaller than the peak entropy-generation rates. These residual violations arise from the soft-constraint formulation and the difficulty of eliminating extremely small violations during gradient-based optimization, where the associated penalty gradients become very weak.} Taken together with the results of Section \ref{subsec:validation}, these show that (1) unconstrained ML models for chemical kinetics, beyond mass conservation concerns, cannot be assumed to automatically satisfy second law constraints, and that (2) these second law constraints are important in promoting physically valid evolutions of the chemical state during a \textit{posteriori} validations. In this case, the widespread second-law violations (as seen in Fig.~\ref{fig:entropy_scatter}b) drive the thermochemical state toward non-physical compositions, producing the spurious species concentrations seen in Fig.~\ref{fig:oh_constraint} and the error divergence observed in Fig.~\ref{fig:error_comparison}.

\subsection{Computational performance\label{subsec:performance}} 
The machine-learning model benefits from the reduced stiffness of the learned thermochemical source terms, allowing the CFD solver to operate with a significantly larger time step ($\Delta t \approx 10^{-8}$~s) while maintaining acceptably low errors. \textcolor{black}{This reduction in temporal stiffness arises because several short-lived minor species are omitted from the transported state vector, yielding a reduced system with a smaller range of chemical time scales.} The full-chemistry simulation requires 3.125 hours of wall-clock time, whereas the machine-learning model completes in approximately 12 minutes. \textcolor{black}{The speedup arises from the larger allowable time step, the reduced set of transported variables which lowers transport-related costs, and the replacement of expensive chemical kinetics evaluations with a more efficient ML surrogate.} 
Furthermore, the reported speedups represent a lower bound. Machine-learning models are particularly amenable to additional optimization through vectorized kernels, hardware-accelerated linear algebra, and GPU acceleration-techniques that were not exploited in the present study.

\subsection{Conclusion}
In this study, we introduced an approach for incorporating second-law consistency into machine-learning models for chemical kinetics. The results show improved stability and accuracy compared with an unconstrained model. We also demonstrated that a residual-based data augmentation strategy can extend DNS data generated for a baseline condition to new inlet conditions while maintaining acceptable predictive accuracy. More broadly, the methodology presented in this study represents a shift in the perspective of physics-constrained learning from enforcing consistency of predicted states to governing the dynamical directions produced by the model. The work therefore establishes a foundation for thermodynamically informed machine-learning models for reacting-flow simulations.

\section*{Acknowledgments}
This research was supported by the National Science Foundation under Grant No. NSF-2201297. Portions of this research were conducted with high performance computational resources provided by the Louisiana Optical Network Infrastructure ( http://www.loni.org) and Louisiana State University (http://www.hpc.lsu.edu).

\bibliographystyle{unsrt}
\bibliography{references}

\FloatBarrier
\clearpage
\appendix
\section*{Supplementary Material}
\setcounter{section}{0}
\renewcommand{\thesection}{S\arabic{section}}
\renewcommand{\theequation}{S\arabic{equation}}
\renewcommand{\thefigure}{S\arabic{figure}}
\renewcommand{\thetable}{S\arabic{table}}
\setcounter{equation}{0}
\setcounter{figure}{0}
\setcounter{table}{0}

\section{Derivation of the entropy generation rate}
For a chemically reacting multicomponent mixture, the Gibbs free energy satisfies
\vspace{-5pt}
\begin{equation}
    dG = -S\,dT + V\,dp + \sum_{i=1}^{N_s} \mu_i\,dn_i,
    \label{eq:gibbs_change}
    \vspace{-2pt}
\end{equation}
where \(S\) is the mixture entropy, \(T\) is the temperature, \(V\) is
the volume, \(p\) is the pressure, and \(\mu_i\) and \(n_i\) are the
chemical potential and number of moles of species \(i\),
respectively. Since our interest here is in the entropy production
associated with chemical reactions, we consider the change in Gibbs
free energy at constant temperature and pressure. Equation
\eqref{eq:gibbs_change} then reduces to
\begin{equation}
    \left(dG\right)_{T,p} = \sum_{i=1}^{N_s} \mu_i\,dn_i.
\end{equation}
For convenience, the \((T,p)\) subscript is dropped in what follows. For an irreversible process at constant temperature and pressure, the
change in Gibbs free energy is related to the entropy generated by $dG = -T\,dS_{\mathrm{gen}}$, so that
\vspace{-5pt}
\begin{equation}
    \frac{dG}{dt}
    =
    \sum_{i=1}^{N_s} \mu_i \frac{dn_i}{dt}
    =
    -T\,\frac{dS_{\mathrm{gen}}}{dt}
    \leq 0.
    \vspace{-5pt}
\end{equation}
It follows that the entropy generation rate due to chemical reactions is
\vspace{-5pt}
\begin{equation}
    \frac{dS_{\mathrm{gen}}}{dt}
    =
    -\frac{1}{T}
    \sum_{i=1}^{N_s} \mu_i \frac{dn_i}{dt}
    \geq 0.
    \vspace{-5pt}
\end{equation}

Therefore, the local entropy generation rate can be expressed as
\vspace{-5pt}
\begin{equation}
\begin{gathered}
    \sigma
    =
    \frac{1}{V}\frac{dS_{\mathrm{gen}}}{dt}
    =
    -\frac{1}{T}\sum_{i=1}^{N_s}
    \mu_i \frac{1}{V}\frac{dn_i}{dt}
    =\\
    -\frac{1}{T}\sum_{i=1}^{N_s}
    \frac{\mu_i}{W_i}\,\dot{\omega}_i,
    \label{eq:entropy}
    \vspace{-5pt}
\end{gathered}
\end{equation}

where \(W_i\) is the molecular weight of species \(i\), and
\(\dot{\omega}_i\) is the net mass production rate per unit volume
\((\mathrm{kg\,m^{-3}\,s^{-1}})\).

\section{Validation contours for intermediate operating conditions\label{sec:intermediate_cases}}
\addvspace{1pt}

The main manuscript reports two-dimensional validation contours for the case
closest to the baseline ($T_{\rm in} = 300$~K, $\phi = 0.7$) and the most
challenging case ($T_{\rm in} = 500$~K, $\phi = 1.2$). For completeness, this
section provides the corresponding contours for the three intermediate
conditions. Quantitative errors for these conditions are reported in
Table~1 of the main text.

\begin{figure*}[t]
\centering
\begin{tabular}{@{}c@{\hspace{1pt}}c@{\hspace{1pt}}c@{}}
\begin{overpic}[width=0.25\textwidth]{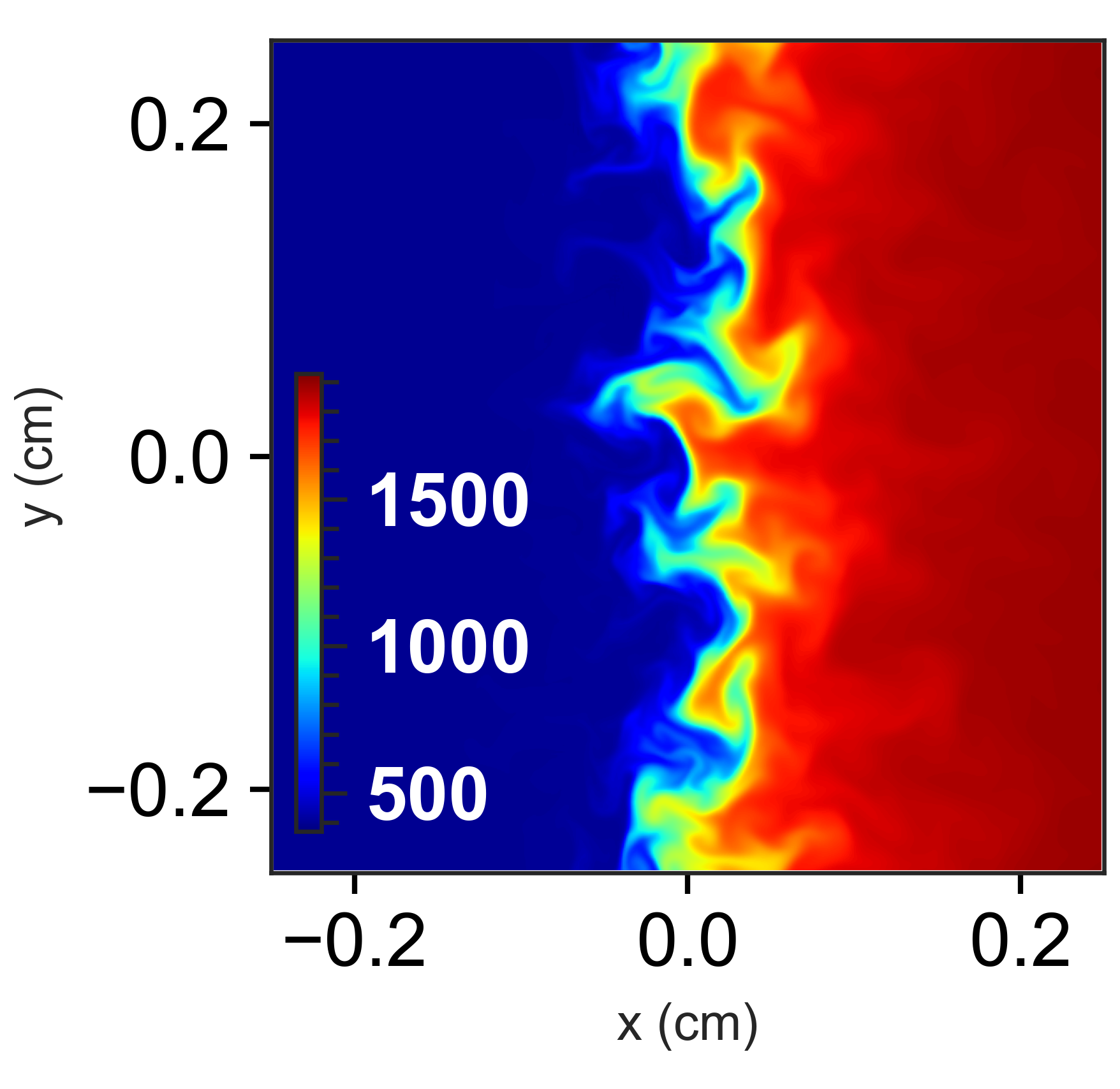}
\put(2,95){\footnotesize\textbf{(a)}}
\end{overpic} &
\begin{overpic}[width=0.25\textwidth]{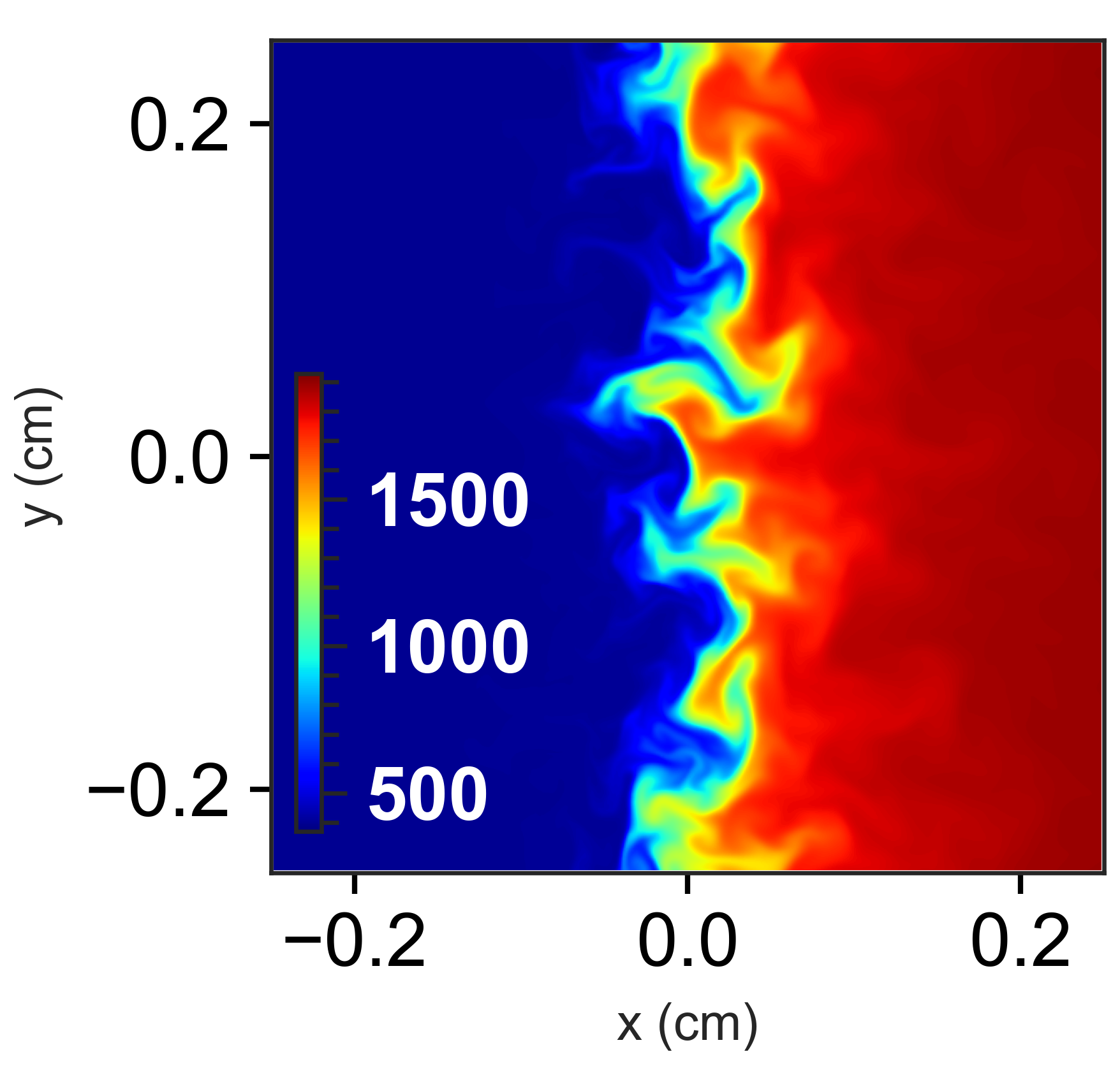}
\put(2,95){\footnotesize\textbf{(b)}}
\end{overpic} &
\begin{overpic}[width=0.25\textwidth]{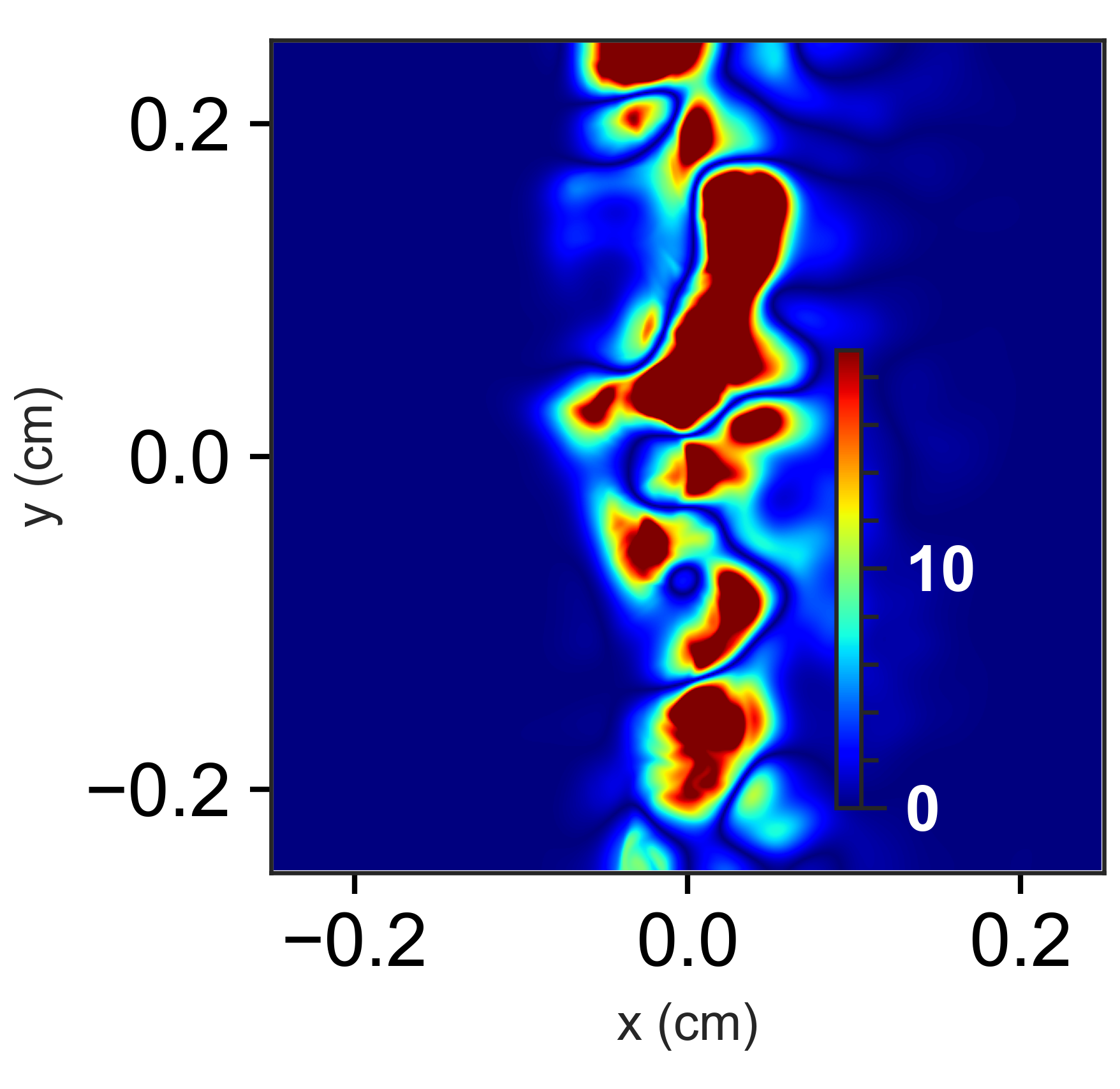}
\put(2,95){\footnotesize\textbf{(c)}}
\end{overpic} \\[-2pt]
\begin{overpic}[width=0.25\textwidth]{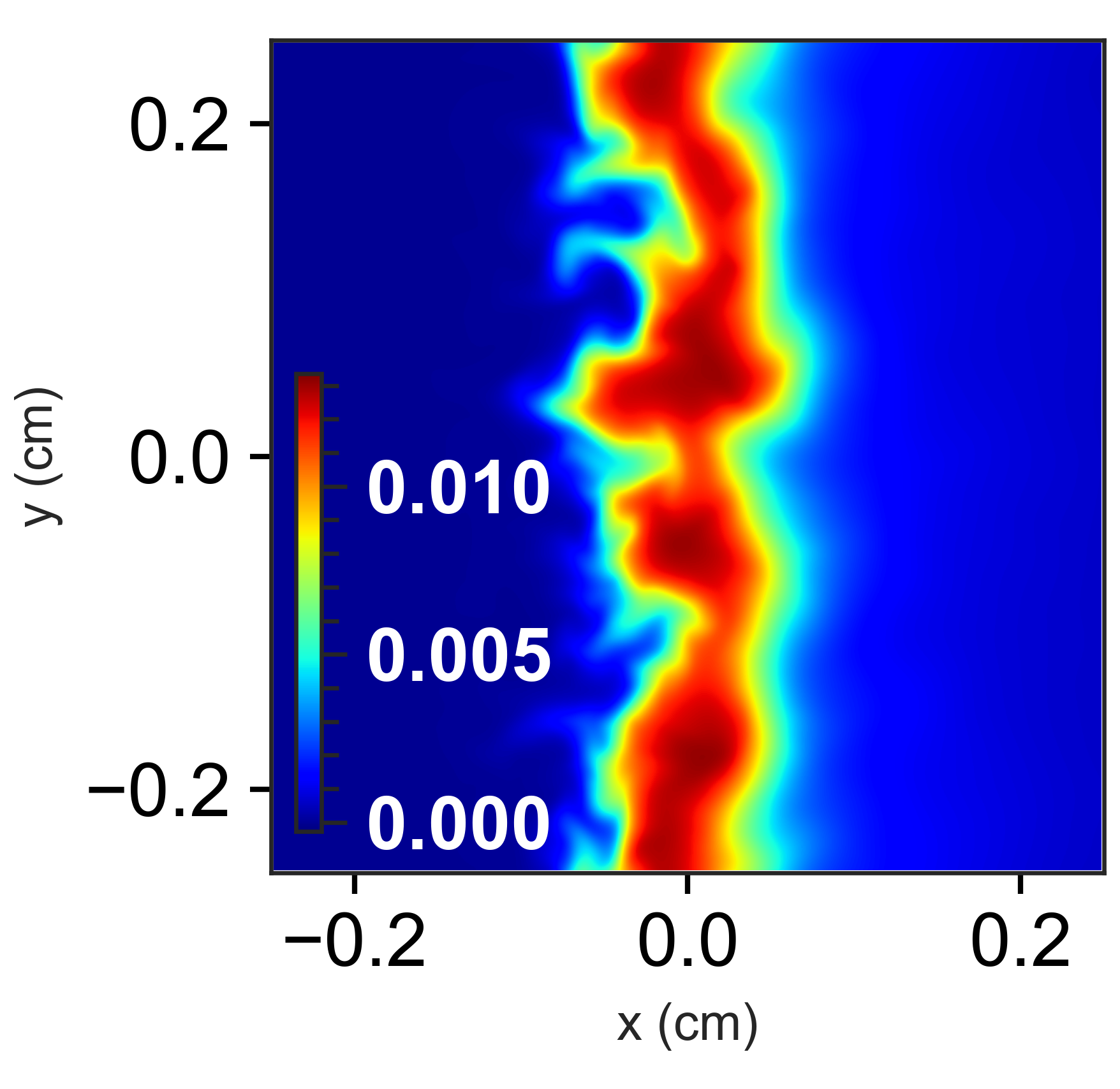}
\put(2,95){\footnotesize\textbf{(d)}}
\end{overpic} &
\begin{overpic}[width=0.25\textwidth]{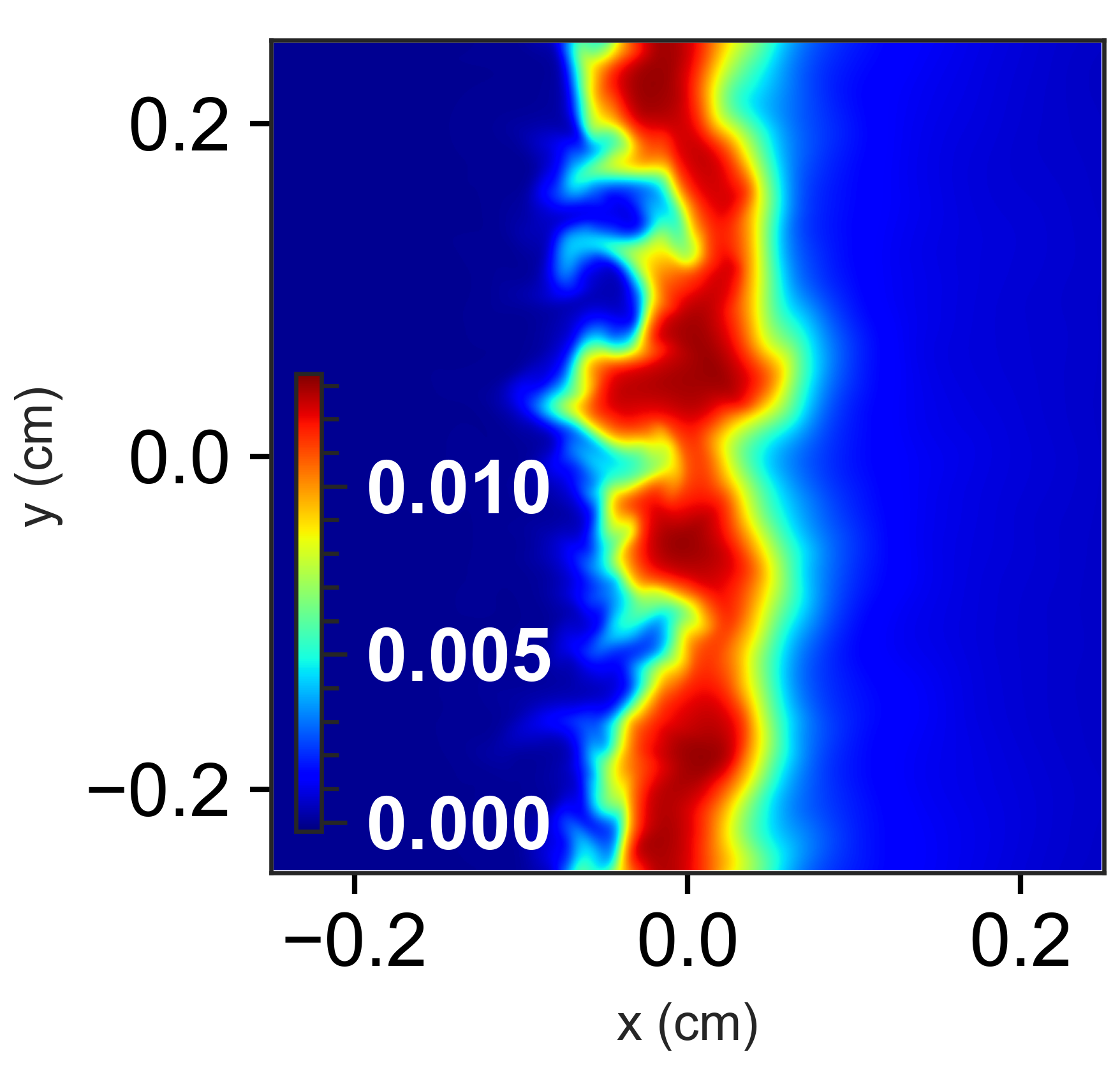}
\put(2,95){\footnotesize\textbf{(e)}}
\end{overpic} &
\begin{overpic}[width=0.25\textwidth]{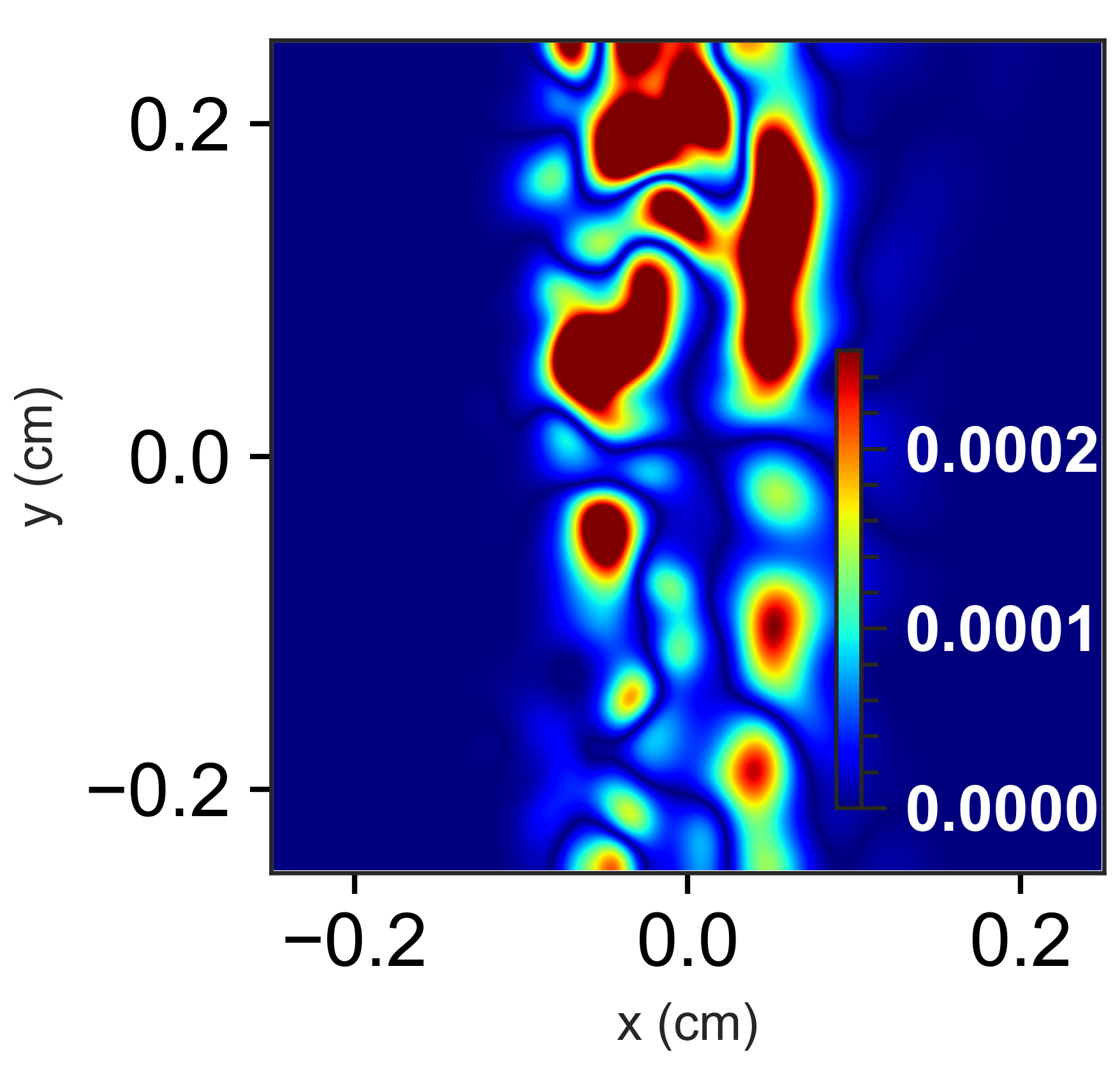}
\put(2,95){\footnotesize\textbf{(f)}}
\end{overpic}
\end{tabular}
\caption{\footnotesize Validation at $T_{\rm in} = 400$~K, $\phi = 0.7$:
(a,d)~full mechanism; (b,e)~ML model; (c,f)~absolute error.
Top: temperature; bottom: $Y_{\rm CO}$.}
\label{fig:S_val_400_p7}
\end{figure*}

\begin{figure*}[t]
\centering
\begin{tabular}{@{}c@{\hspace{1pt}}c@{\hspace{1pt}}c@{}}
\begin{overpic}[width=0.25\textwidth]{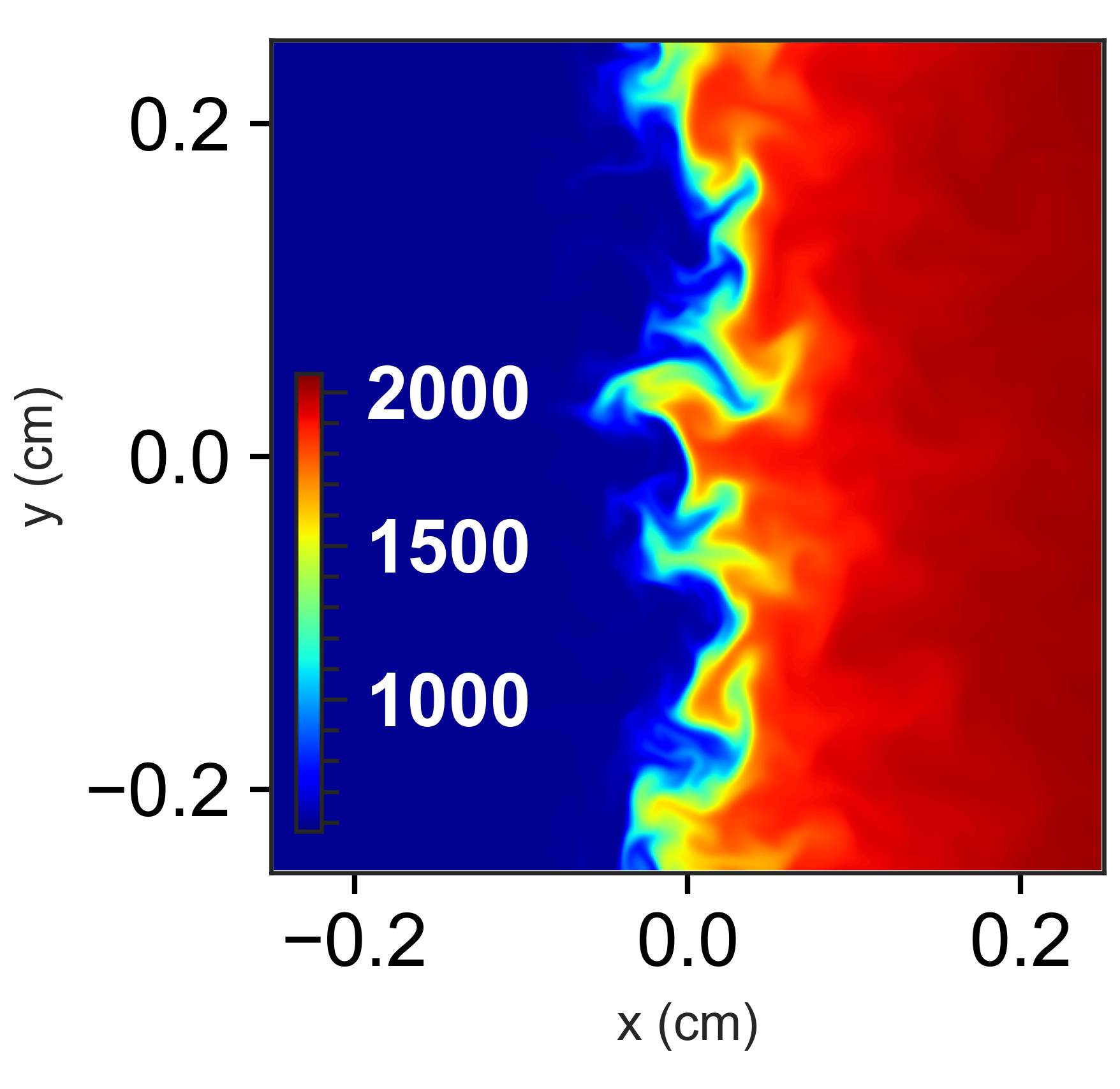}
\put(2,95){\footnotesize\textbf{(a)}}
\end{overpic} &
\begin{overpic}[width=0.25\textwidth]{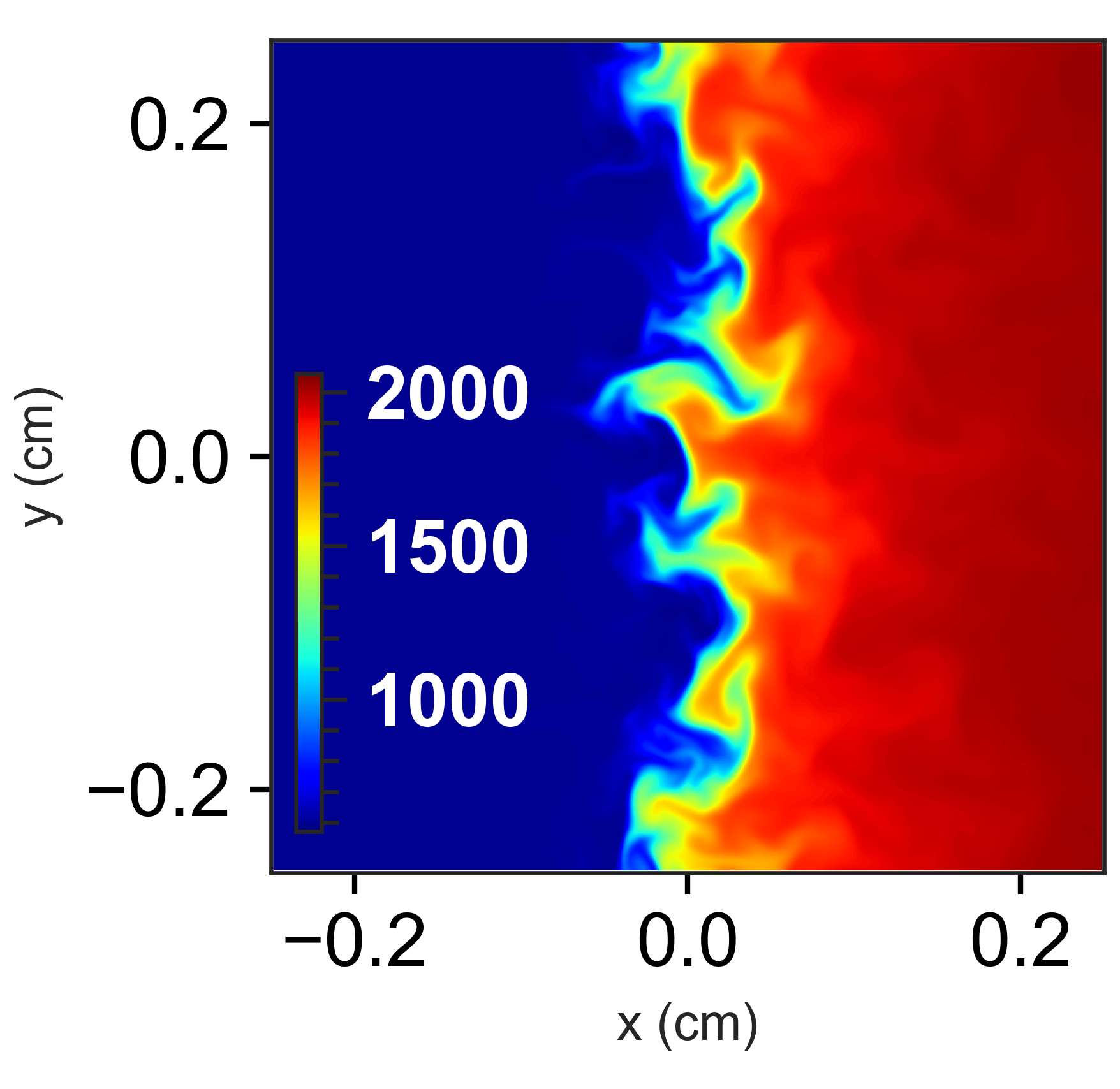}
\put(2,95){\footnotesize\textbf{(b)}}
\end{overpic} &
\begin{overpic}[width=0.25\textwidth]{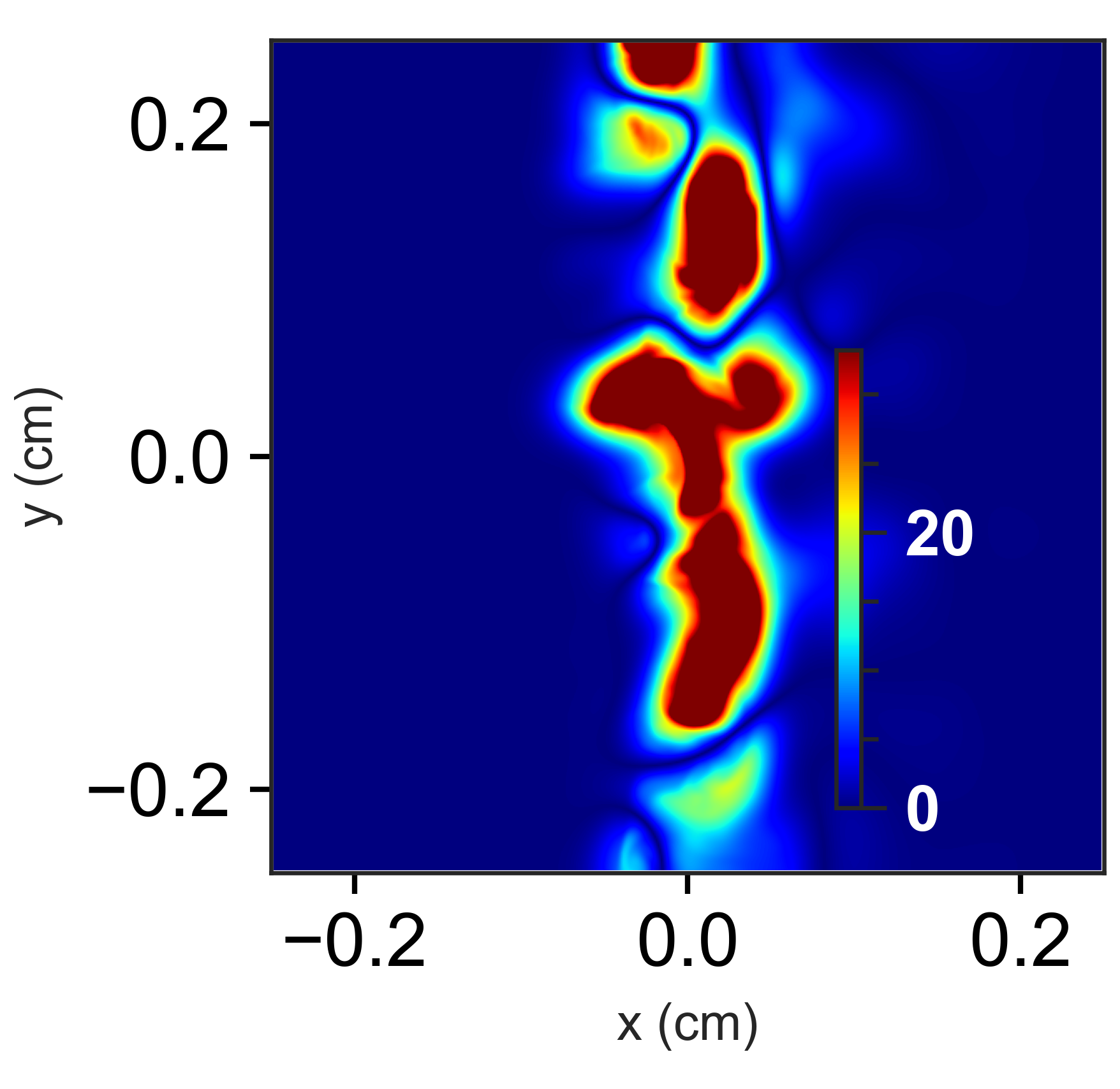}
\put(2,95){\footnotesize\textbf{(c)}}
\end{overpic} \\[-2pt]
\begin{overpic}[width=0.25\textwidth]{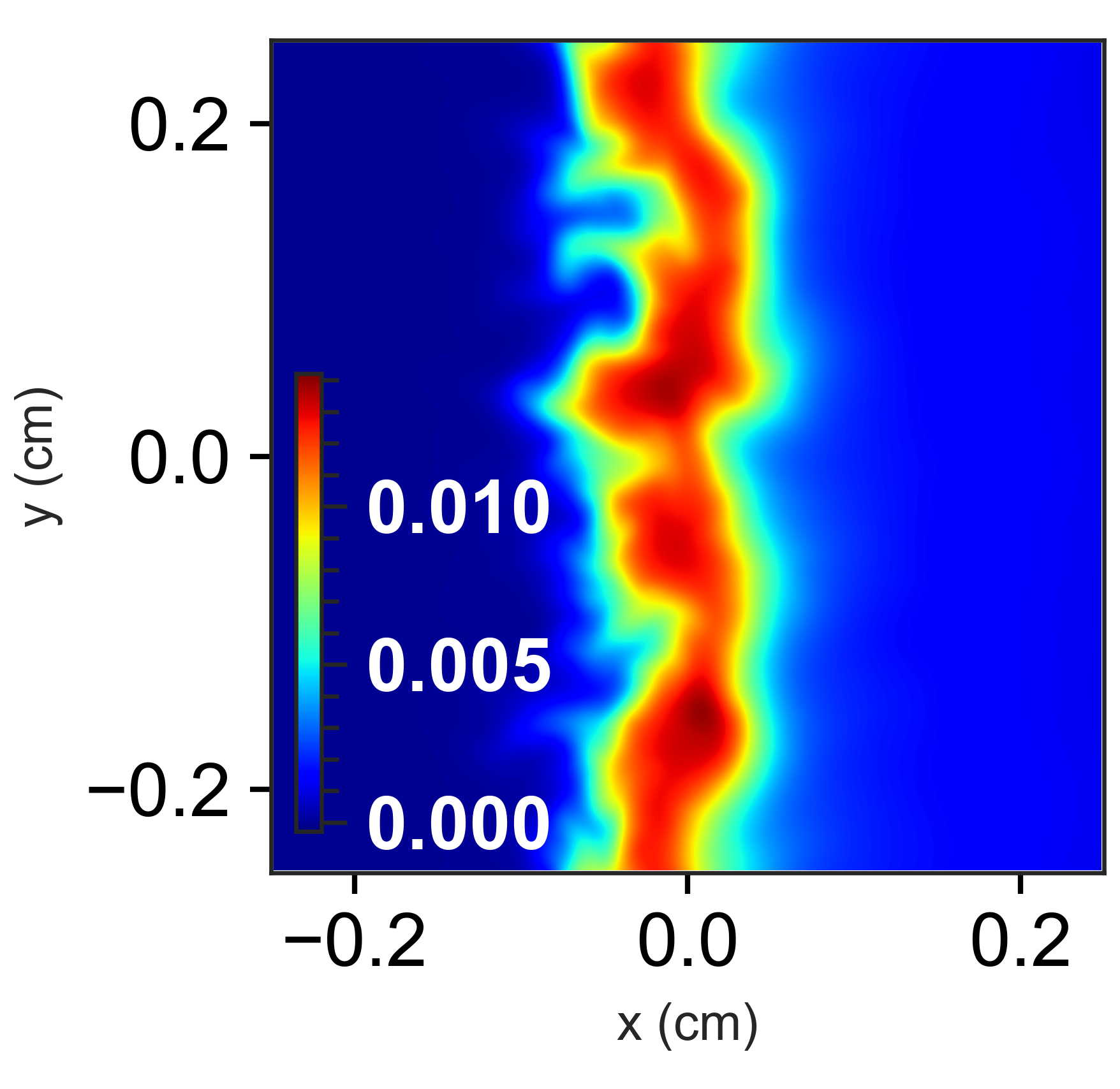}
\put(2,95){\footnotesize\textbf{(d)}}
\end{overpic} &
\begin{overpic}[width=0.25\textwidth]{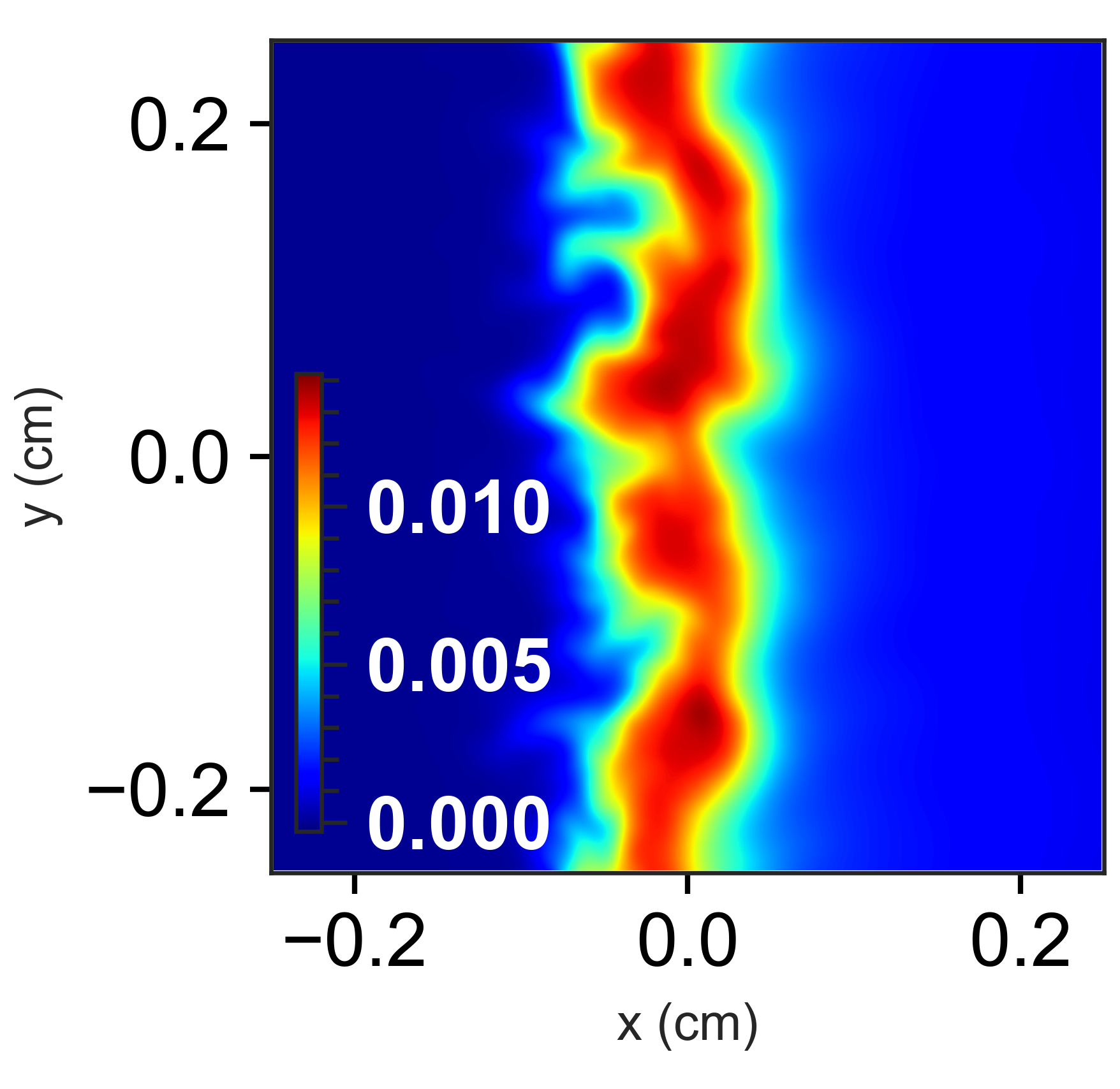}
\put(2,95){\footnotesize\textbf{(e)}}
\end{overpic} &
\begin{overpic}[width=0.25\textwidth]{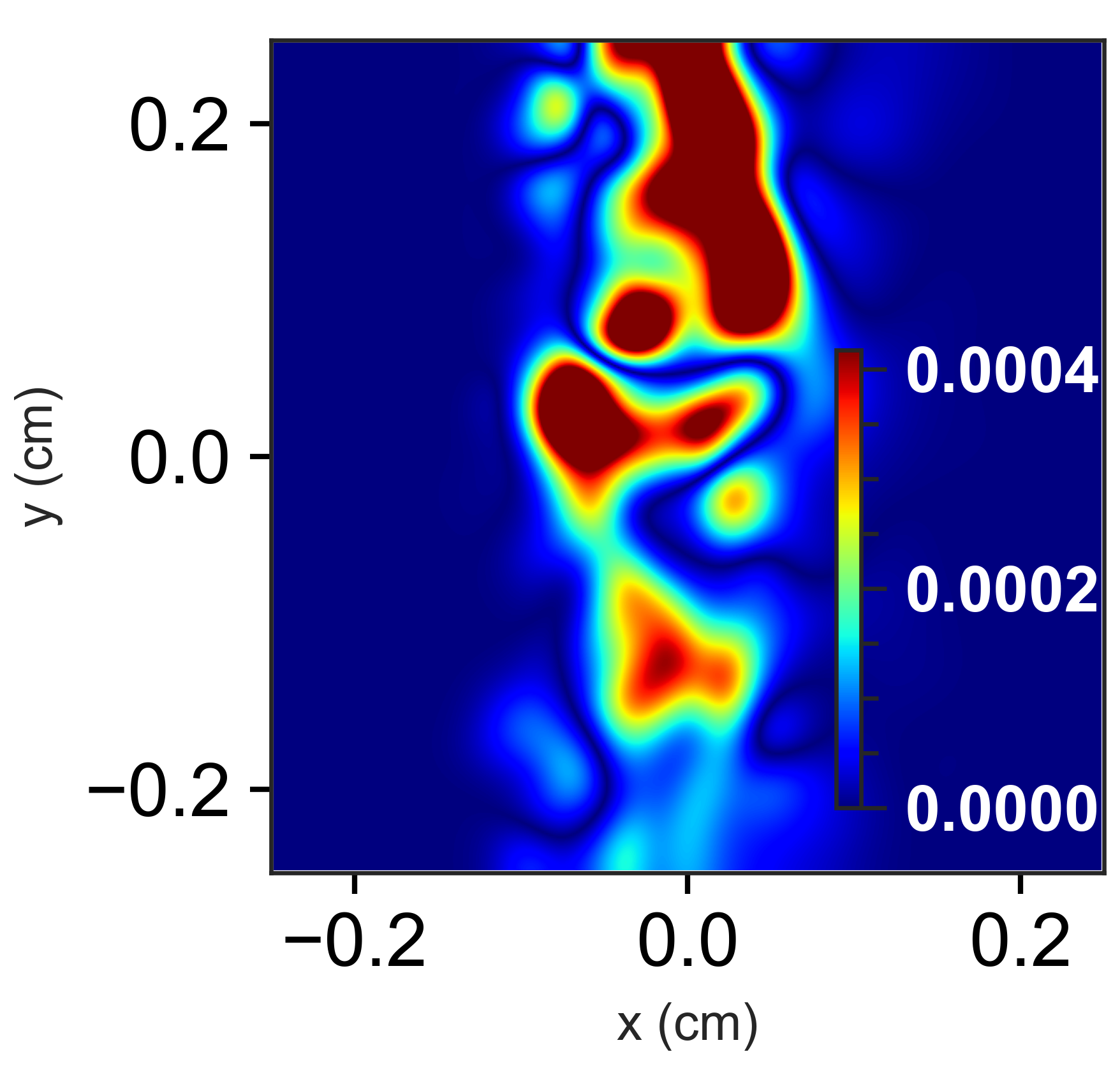}
\put(2,95){\footnotesize\textbf{(f)}}
\end{overpic}
\end{tabular}
\caption{\footnotesize Validation at $T_{\rm in} = 600$~K, $\phi = 0.7$:
(a,d)~full mechanism; (b,e)~ML model; (c,f)~absolute error.
Top: temperature; bottom: $Y_{\rm CO}$.}
\label{fig:S_val_600_p7}
\end{figure*}

\begin{figure*}[t]
\centering
\begin{tabular}{@{}c@{\hspace{1pt}}c@{\hspace{1pt}}c@{}}
\begin{overpic}[width=0.25\textwidth]{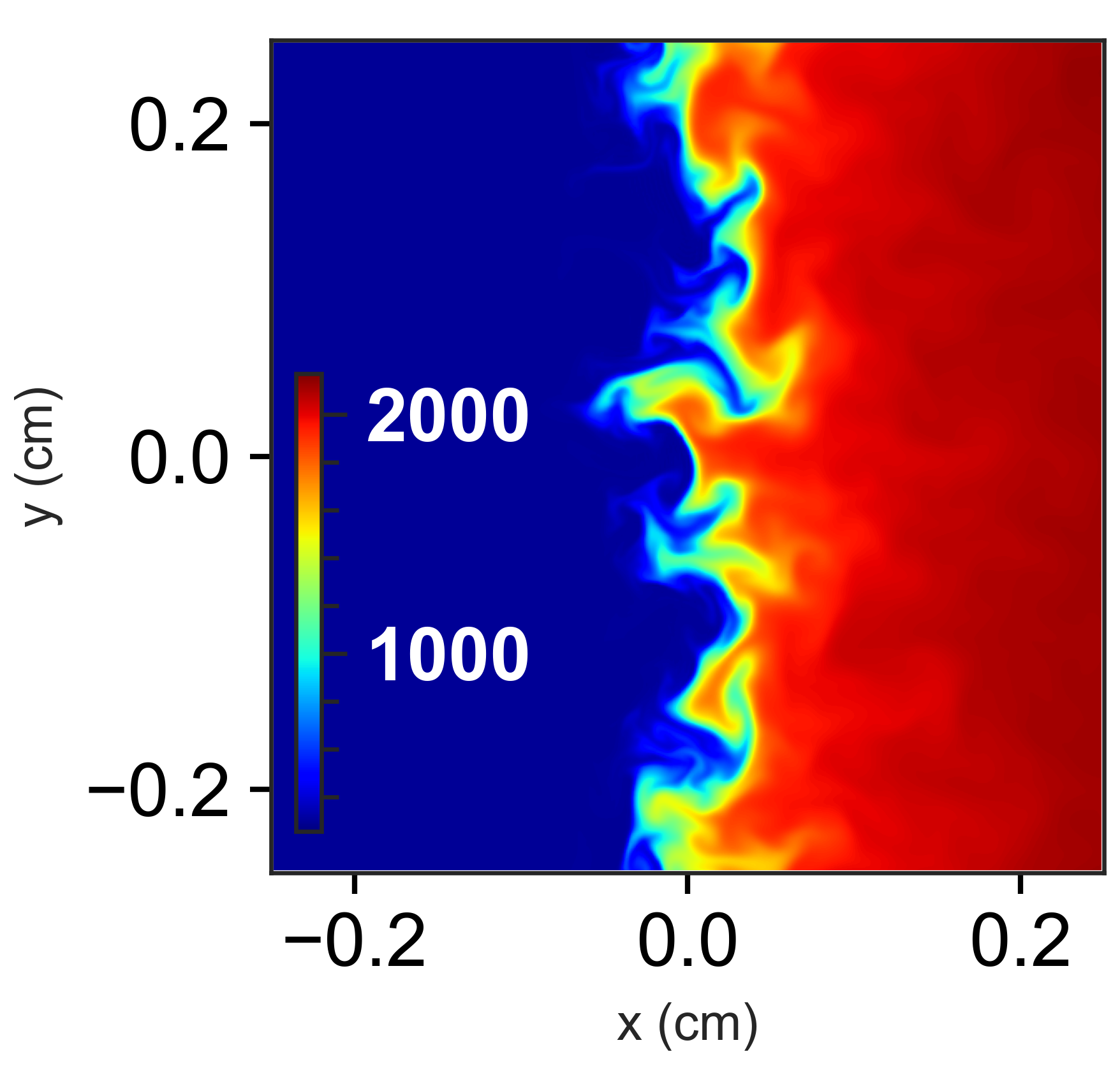}
\put(2,95){\footnotesize\textbf{(a)}}
\end{overpic} &
\begin{overpic}[width=0.25\textwidth]{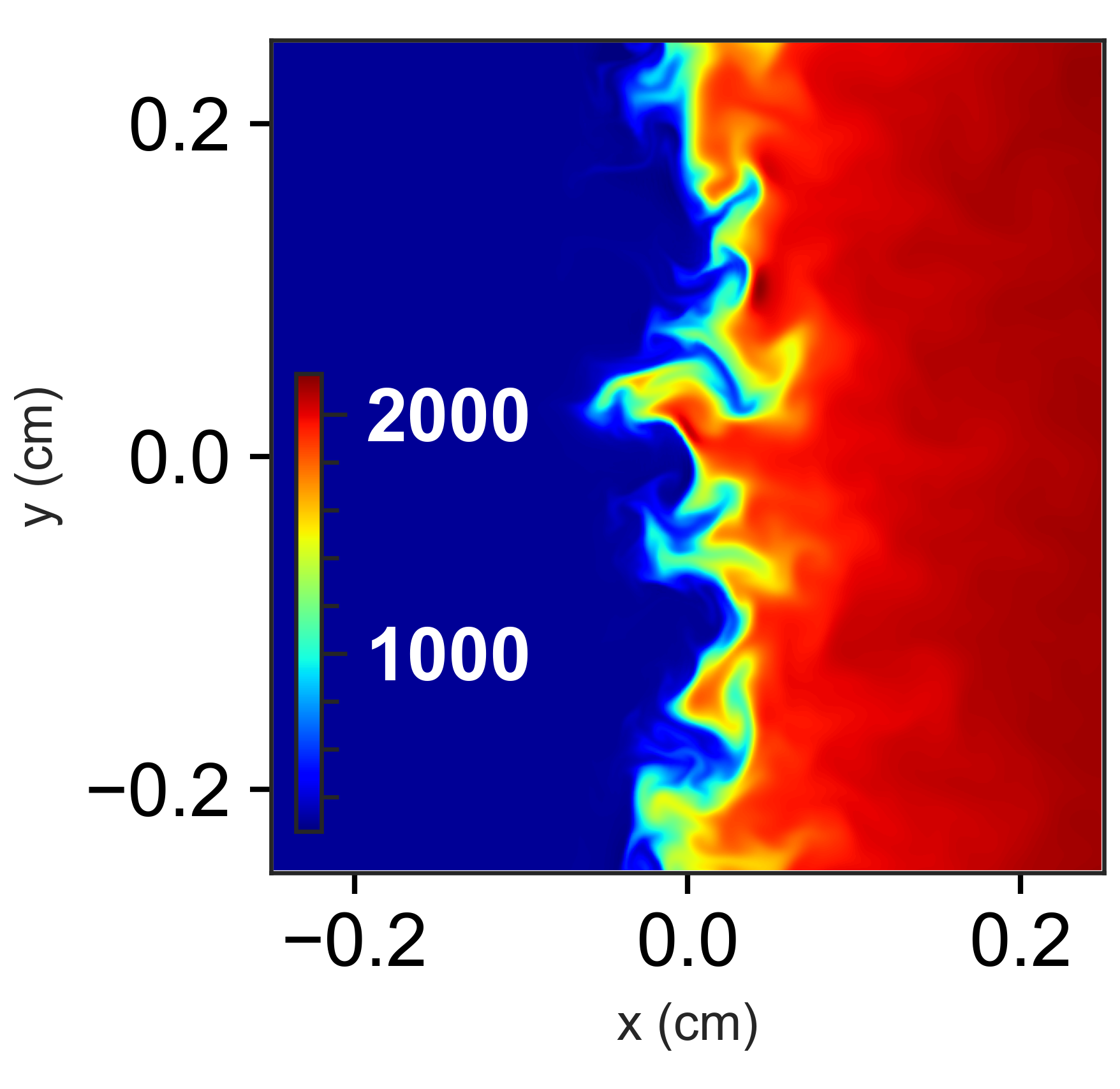}
\put(2,95){\footnotesize\textbf{(b)}}
\end{overpic} &
\begin{overpic}[width=0.25\textwidth]{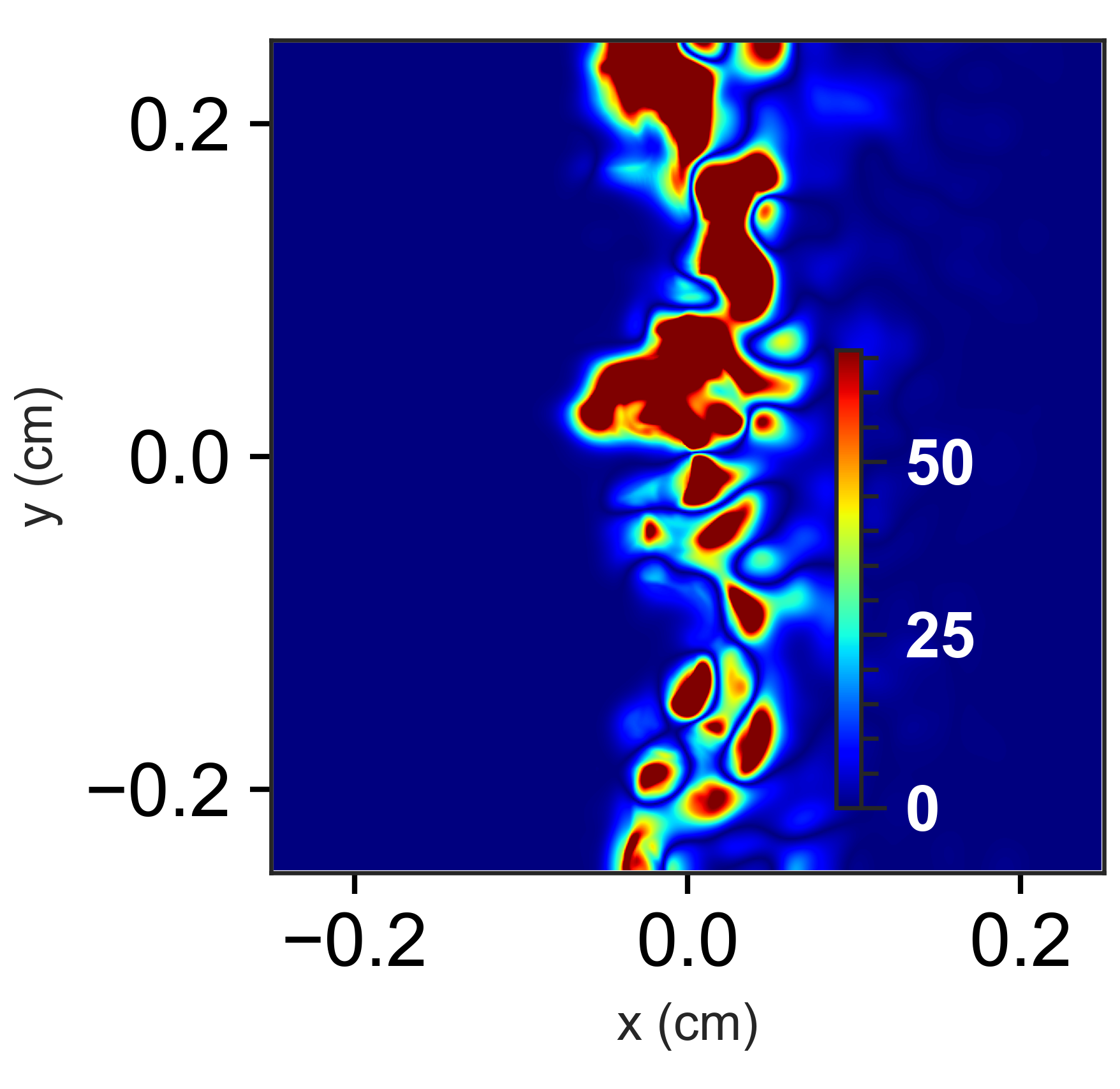}
\put(2,95){\footnotesize\textbf{(c)}}
\end{overpic} \\[-2pt]
\begin{overpic}[width=0.25\textwidth]{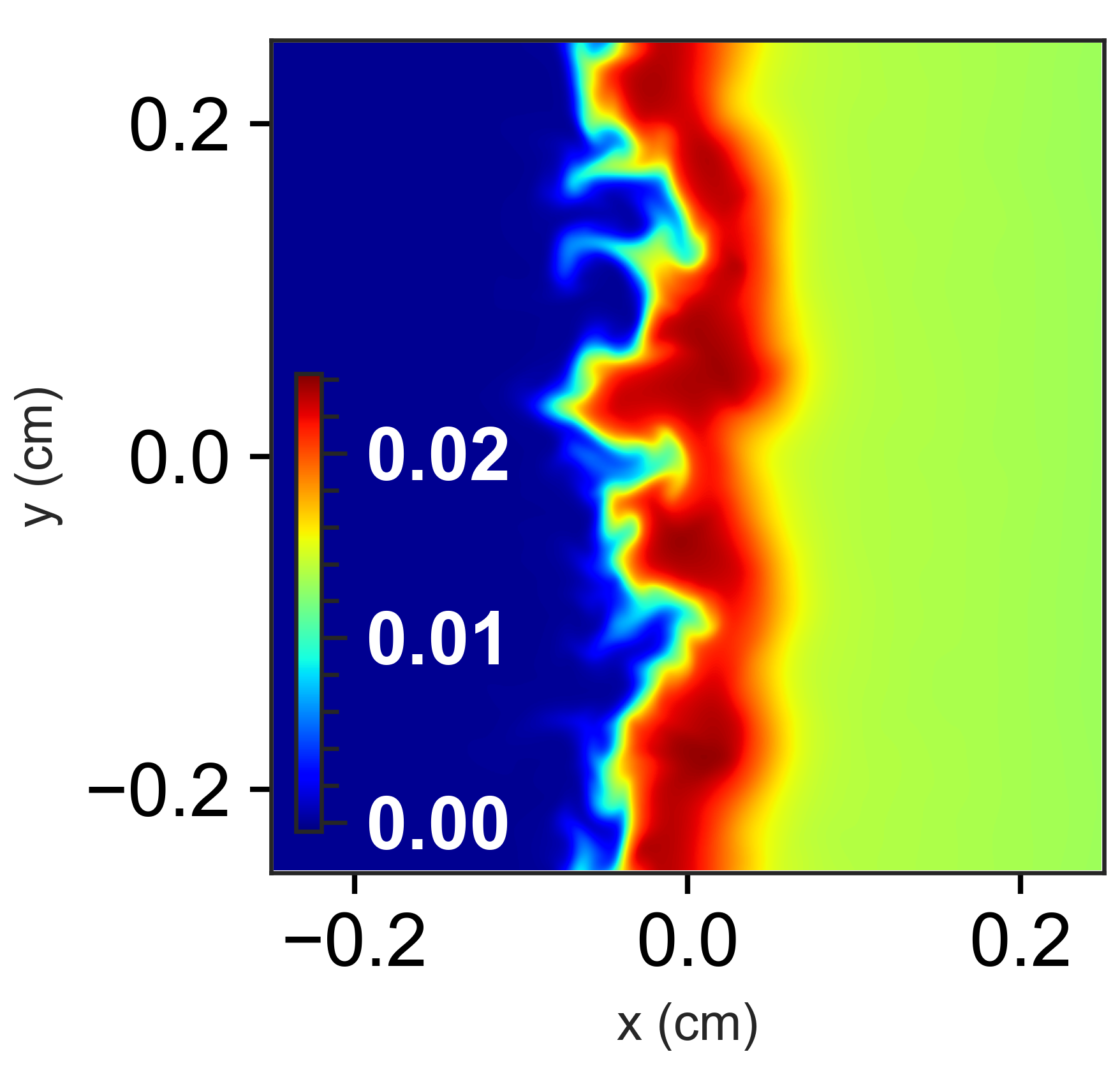}
\put(2,95){\footnotesize\textbf{(d)}}
\end{overpic} &
\begin{overpic}[width=0.25\textwidth]{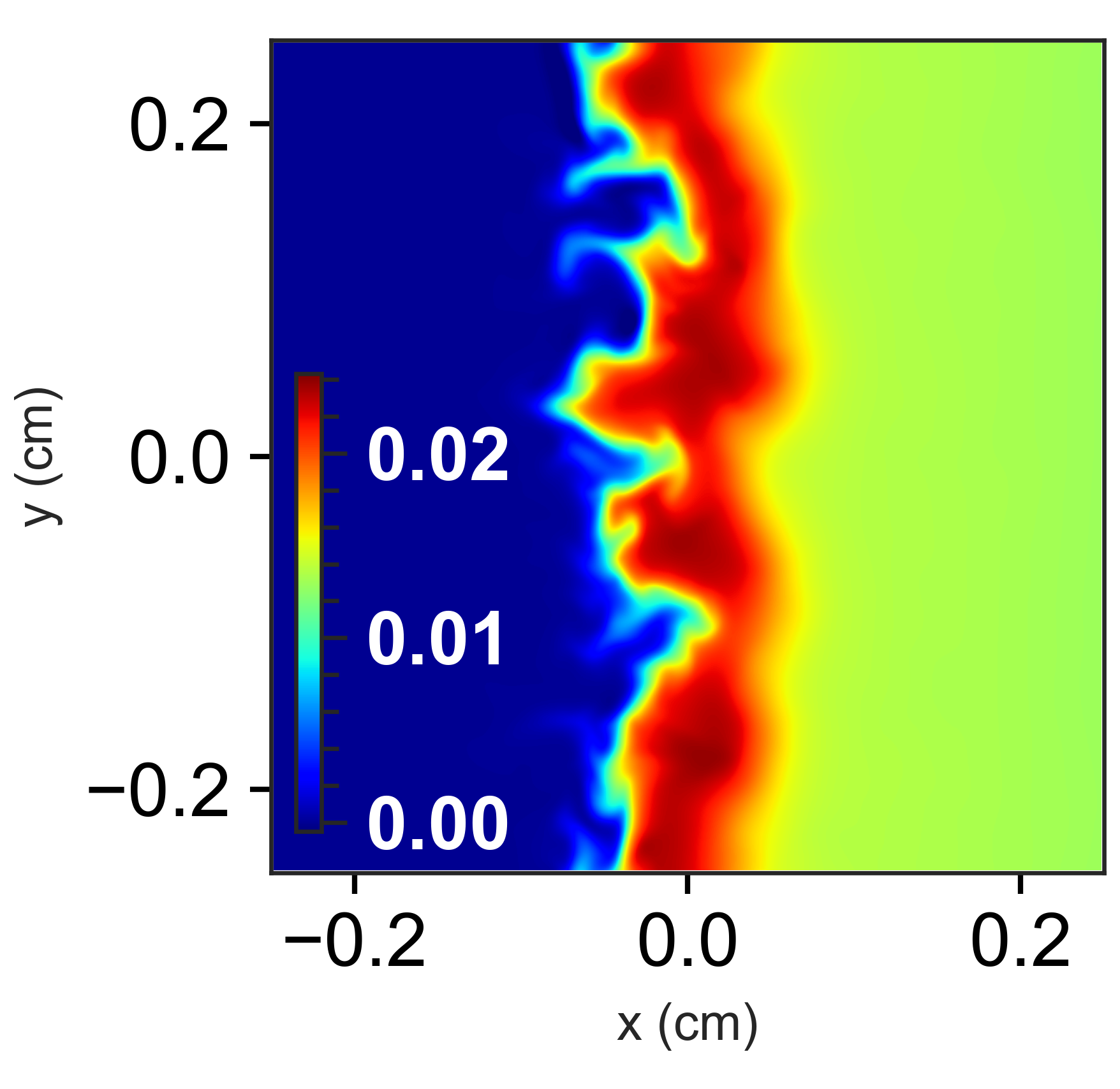}
\put(2,95){\footnotesize\textbf{(e)}}
\end{overpic} &
\begin{overpic}[width=0.25\textwidth]{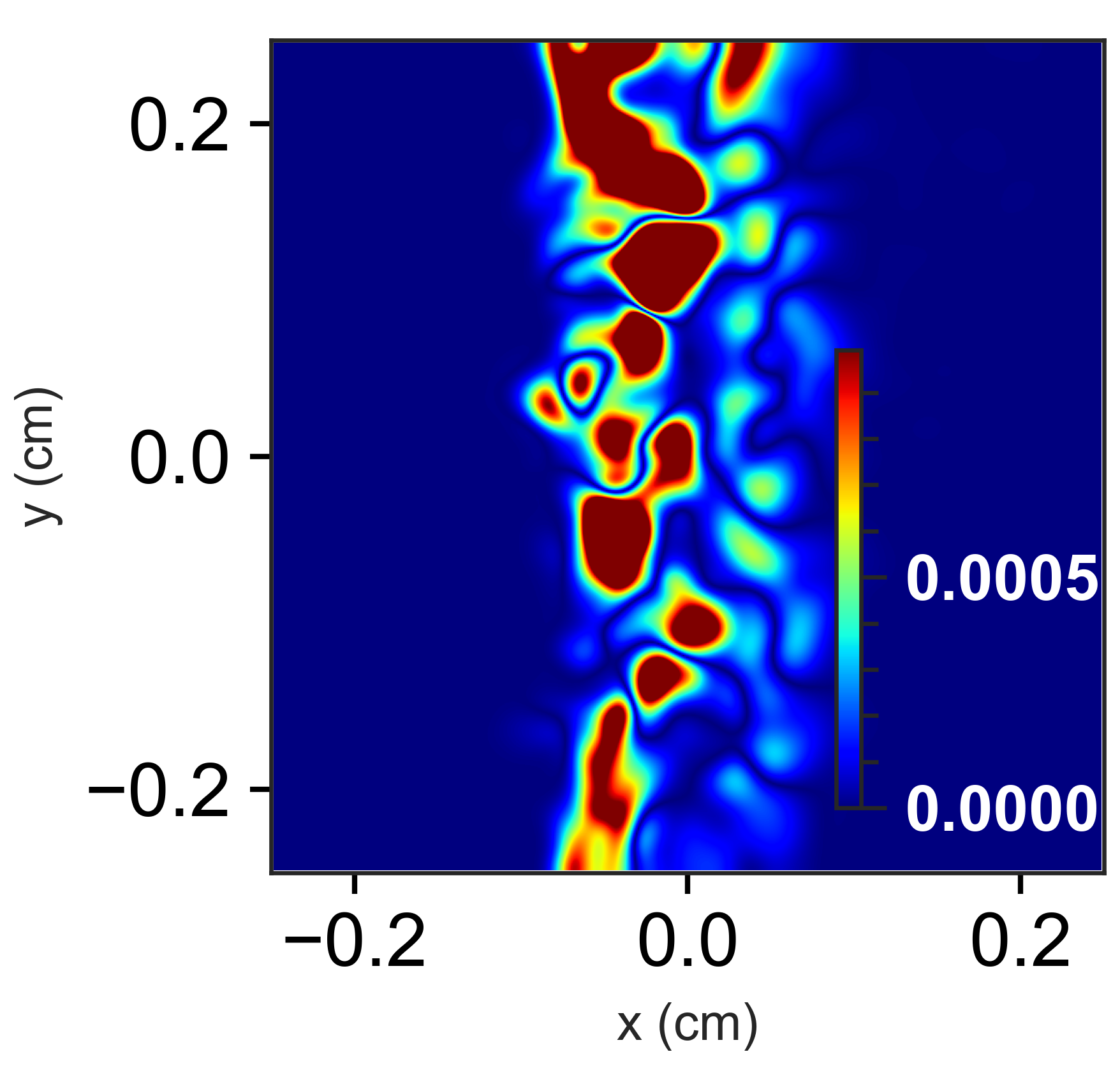}
\put(2,95){\footnotesize\textbf{(f)}}
\end{overpic}
\end{tabular}
\caption{\footnotesize Validation at $T_{\rm in} = 300$~K, $\phi = 1.0$:
(a,d)~full mechanism; (b,e)~ML model; (c,f)~absolute error.
Top: temperature; bottom: $Y_{\rm CO}$.}
\label{fig:S_val_300_1p0}
\end{figure*}

\section{$Y_{\mathrm{OH}}$ contours for all validation conditions\label{sec:OH_contours}}
\addvspace{1pt}

This section presents two-dimensional $Y_{\rm OH}$ contour fields for all
five validation conditions, arranged in the same layout used for the
temperature and $Y_{\rm CO}$ contours in the main text: full 30-species
mechanism, ML model, and absolute error. Quantitative errors for $Y_{\rm OH}$
are reported in Table~1 of the main text.

\begin{figure*}[t]
\centering
\begin{tabular}{@{}c@{\hspace{1pt}}c@{\hspace{1pt}}c@{}}
\begin{overpic}[width=0.25\textwidth]{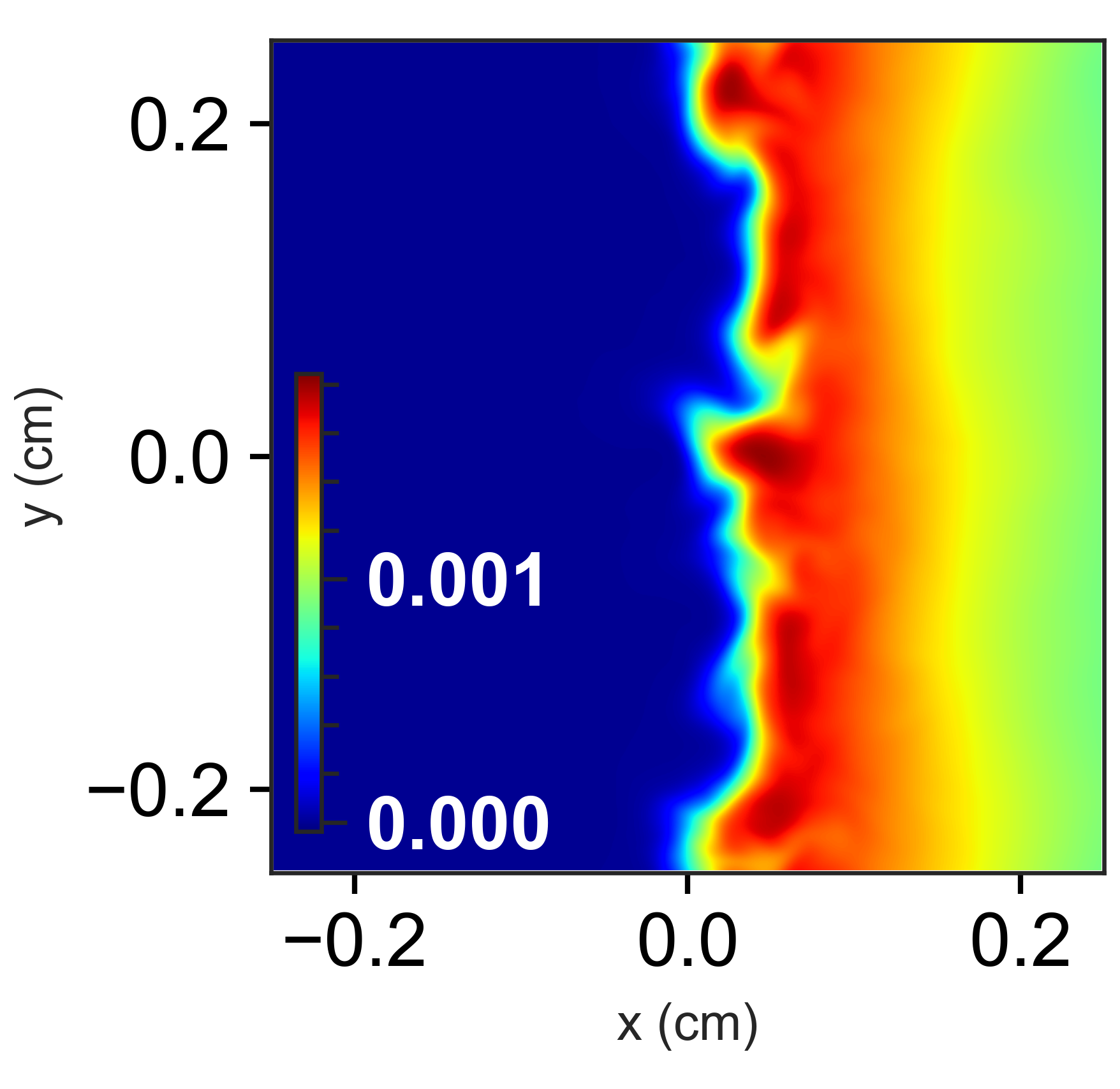}
\put(2,95){\footnotesize\textbf{(a)}}
\end{overpic} &
\begin{overpic}[width=0.25\textwidth]{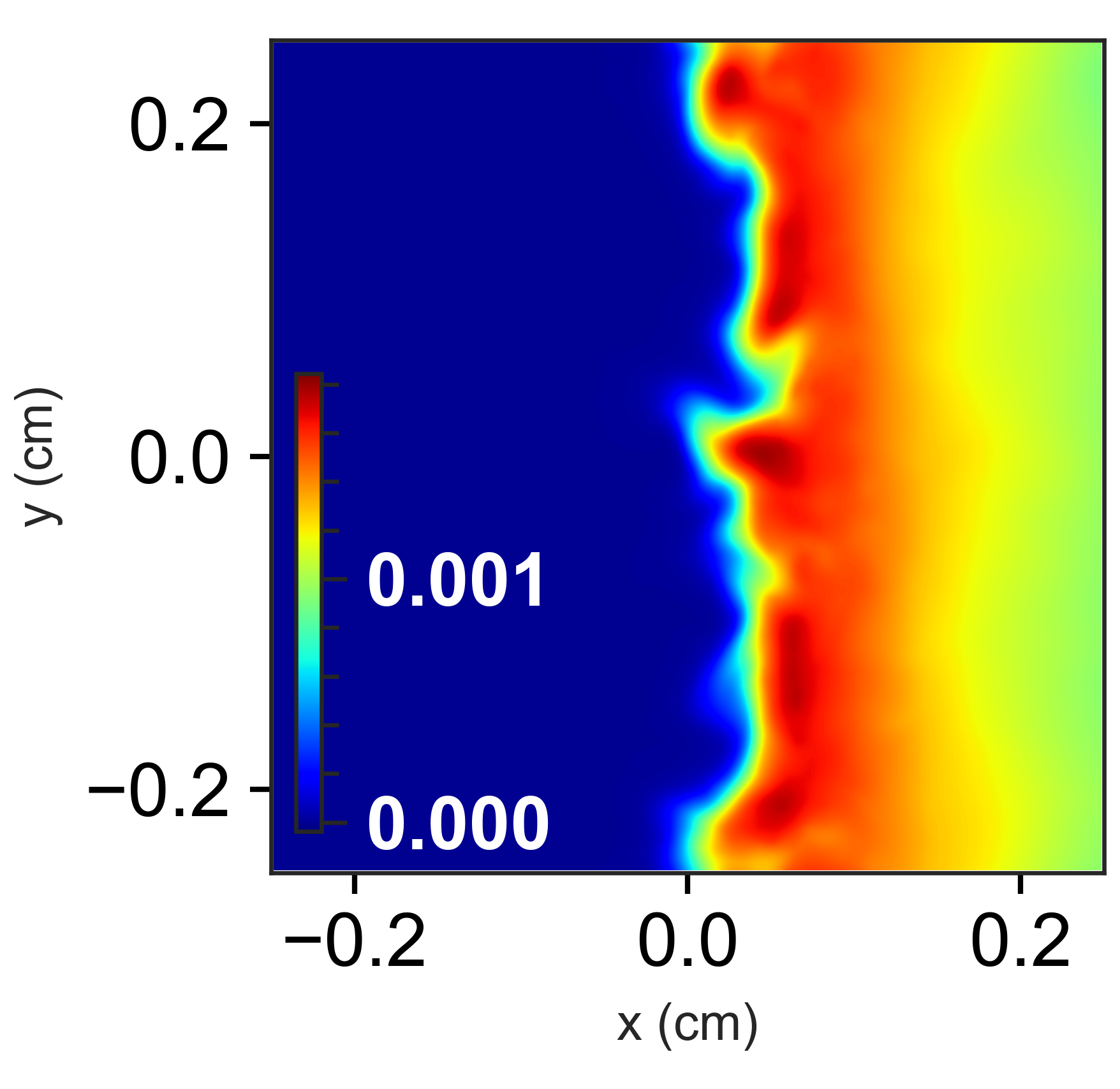}
\put(2,95){\footnotesize\textbf{(b)}}
\end{overpic} &
\begin{overpic}[width=0.25\textwidth]{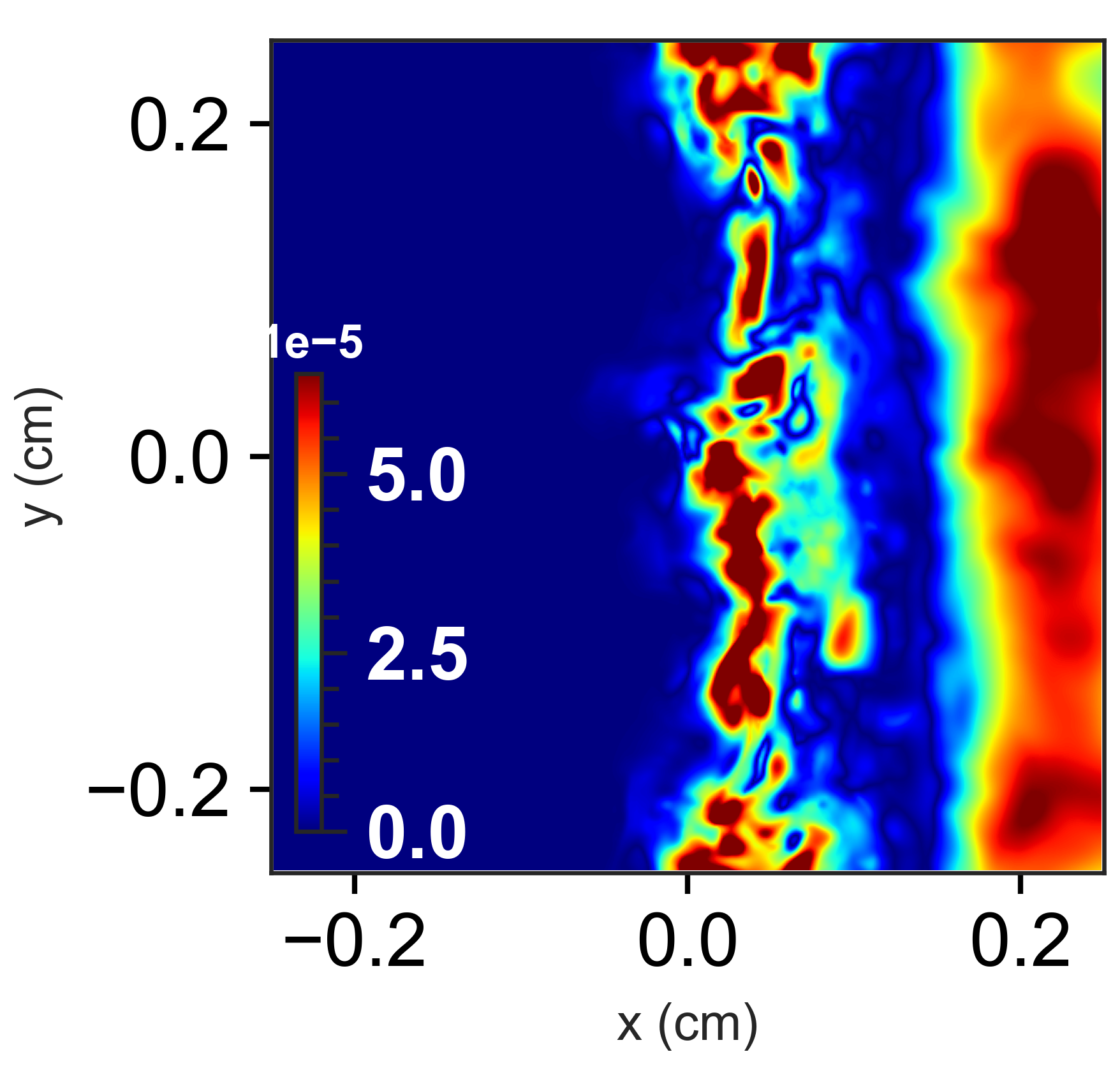}
\put(2,95){\footnotesize\textbf{(c)}}
\end{overpic}
\end{tabular}
\caption{\footnotesize $Y_{\rm OH}$ validation at $T_{\rm in} = 300$~K,
$\phi = 0.7$: (a)~full mechanism; (b)~ML model; (c)~absolute error.}
\label{fig:S_OH_300_p7}
\end{figure*}

\begin{figure*}[t]
\centering
\begin{tabular}{@{}c@{\hspace{1pt}}c@{\hspace{1pt}}c@{}}
\begin{overpic}[width=0.25\textwidth]{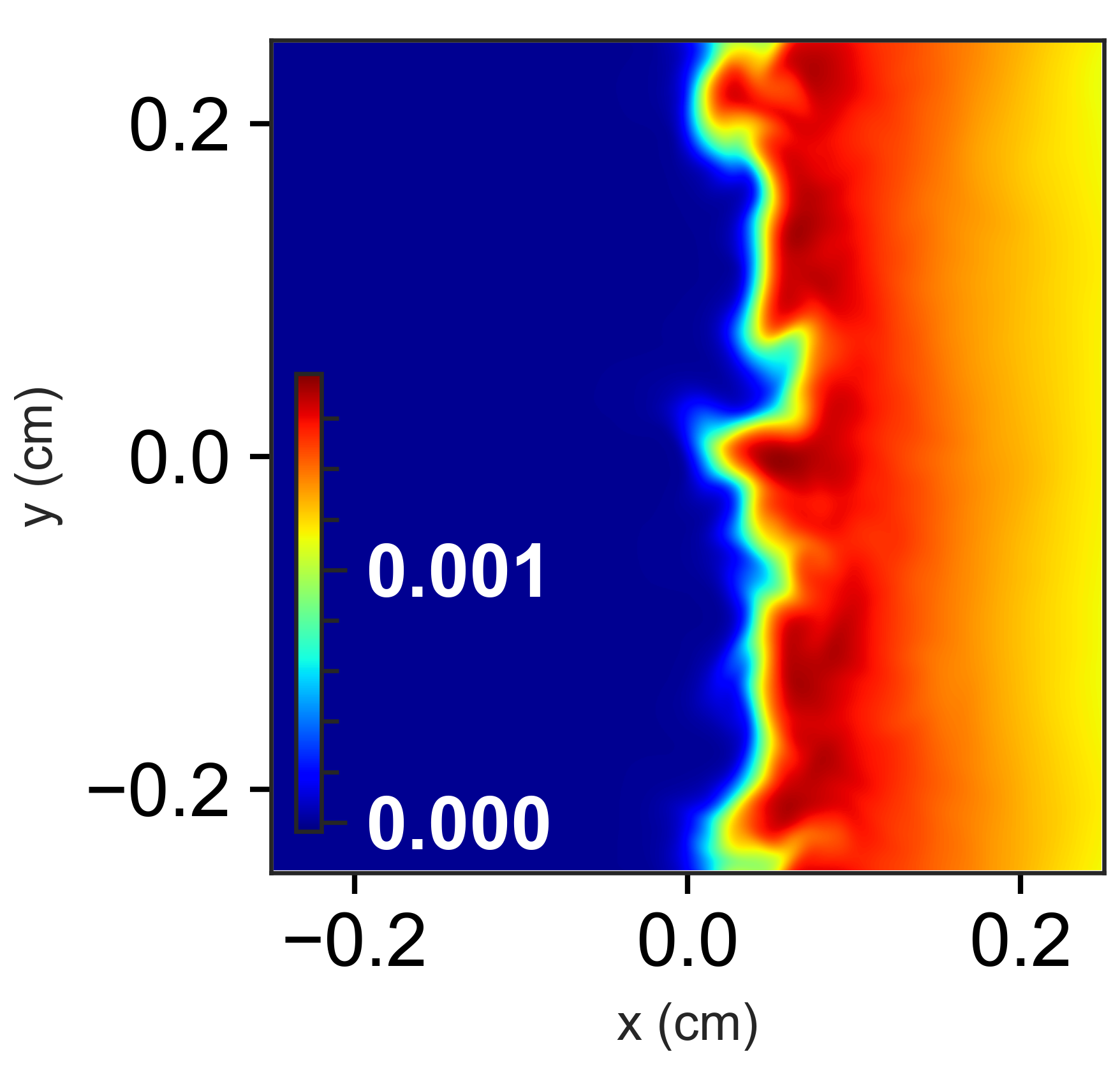}
\put(2,95){\footnotesize\textbf{(a)}}
\end{overpic} &
\begin{overpic}[width=0.25\textwidth]{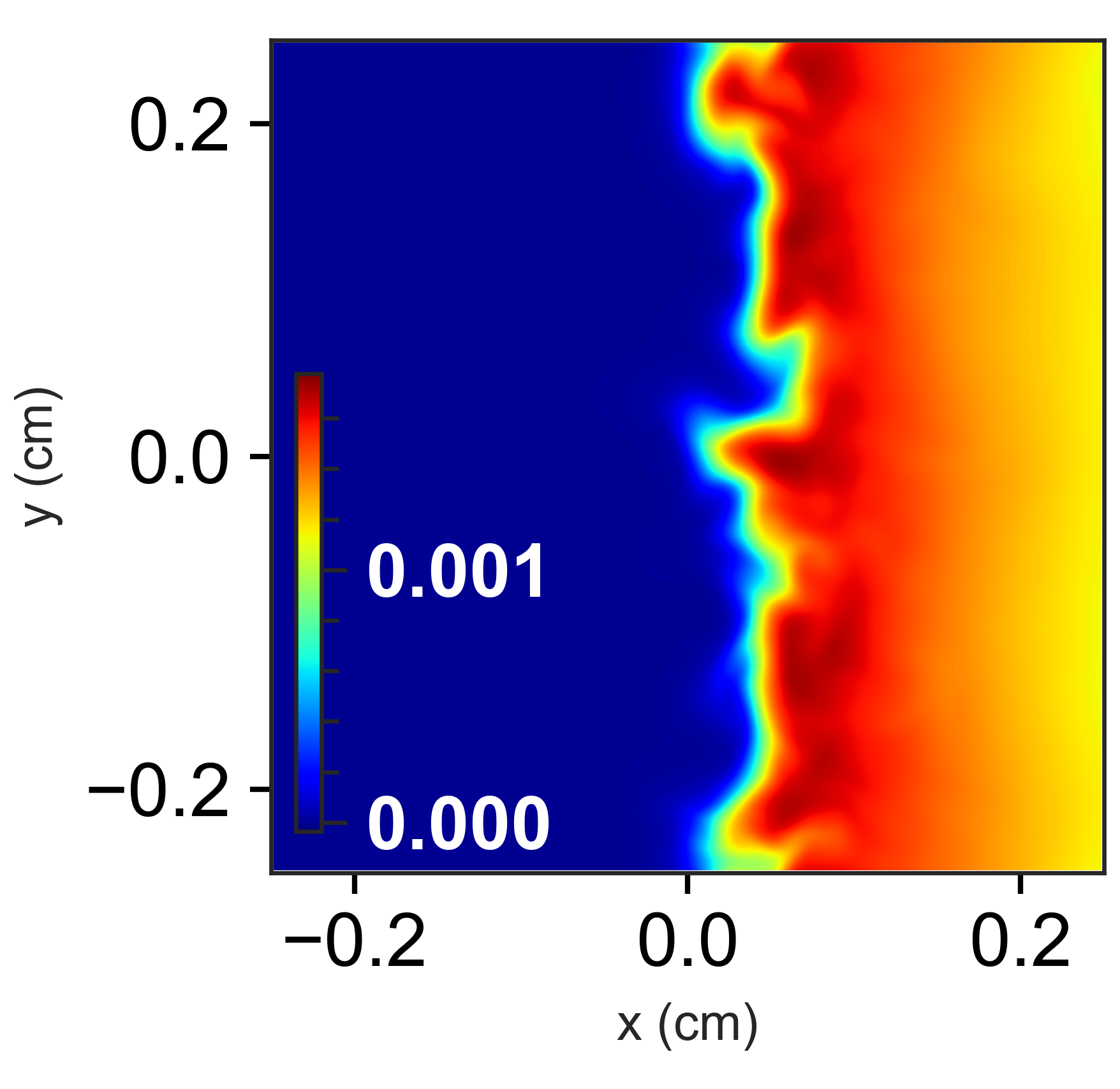}
\put(2,95){\footnotesize\textbf{(b)}}
\end{overpic} &
\begin{overpic}[width=0.25\textwidth]{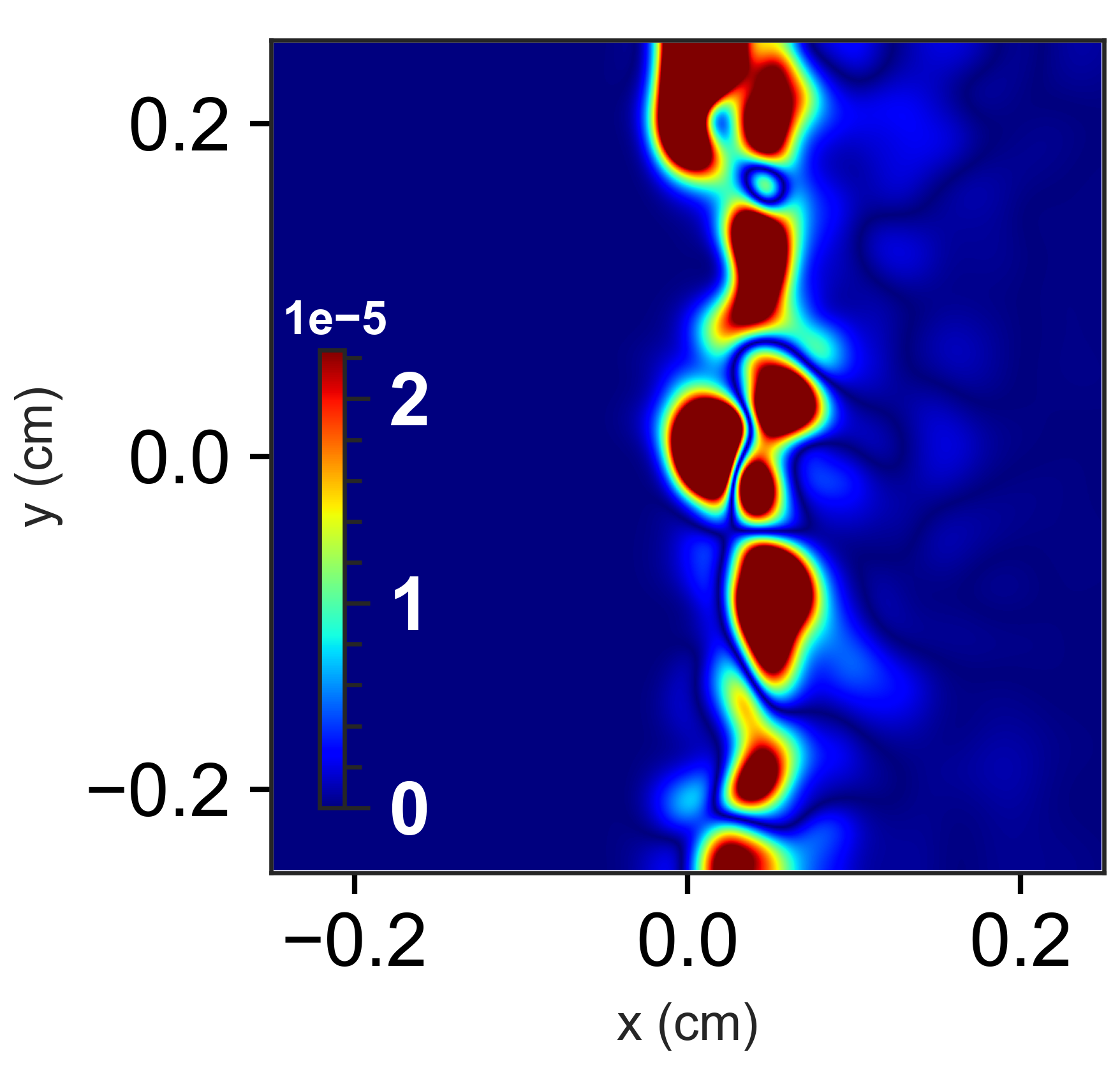}
\put(2,95){\footnotesize\textbf{(c)}}
\end{overpic}
\end{tabular}
\caption{\footnotesize $Y_{\rm OH}$ validation at $T_{\rm in} = 400$~K,
$\phi = 0.7$: (a)~full mechanism; (b)~ML model; (c)~absolute error.}
\label{fig:S_OH_400_p7}
\end{figure*}

\begin{figure*}[t]
\centering
\begin{tabular}{@{}c@{\hspace{1pt}}c@{\hspace{1pt}}c@{}}
\begin{overpic}[width=0.25\textwidth]{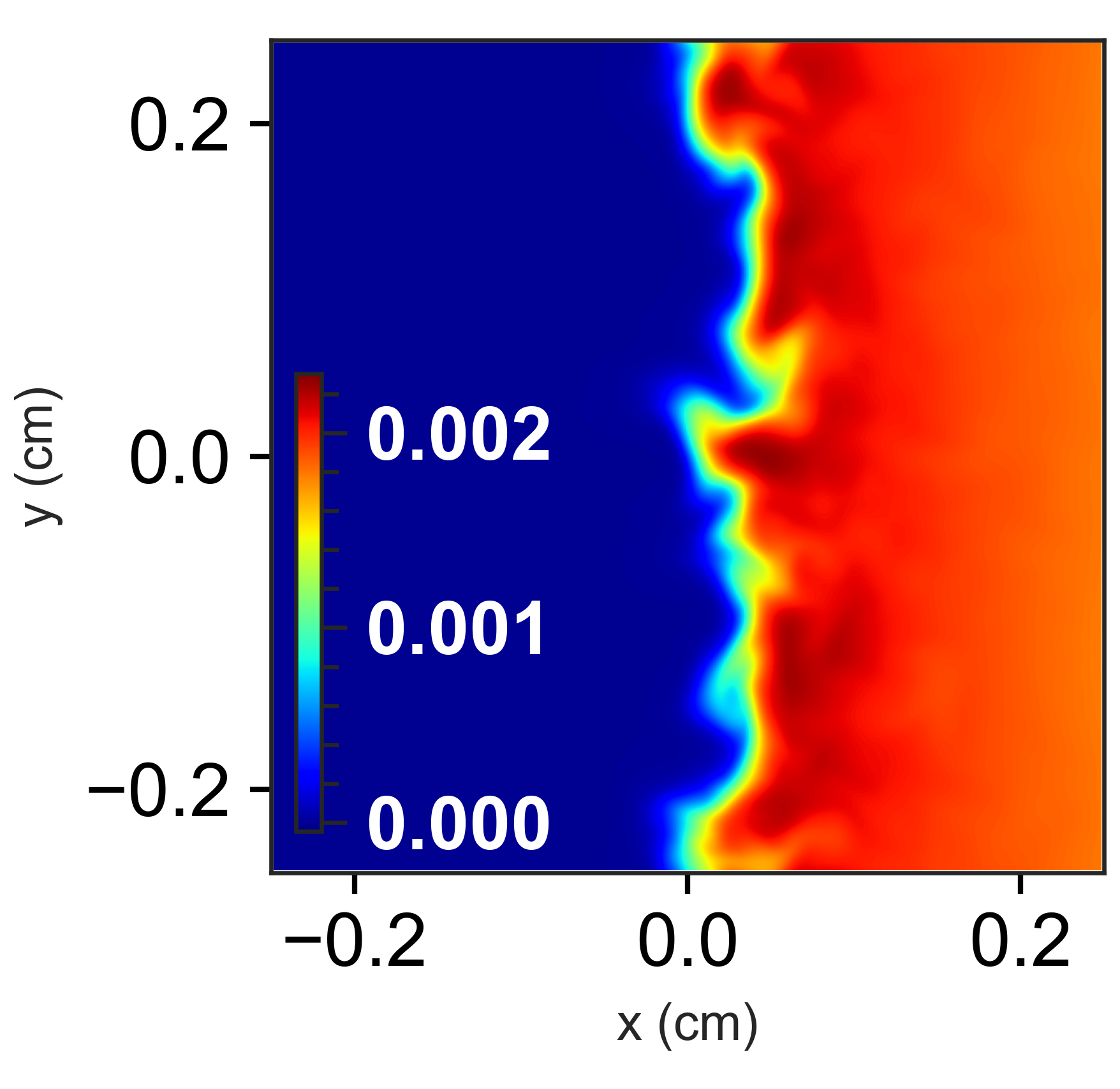}
\put(2,95){\footnotesize\textbf{(a)}}
\end{overpic} &
\begin{overpic}[width=0.25\textwidth]{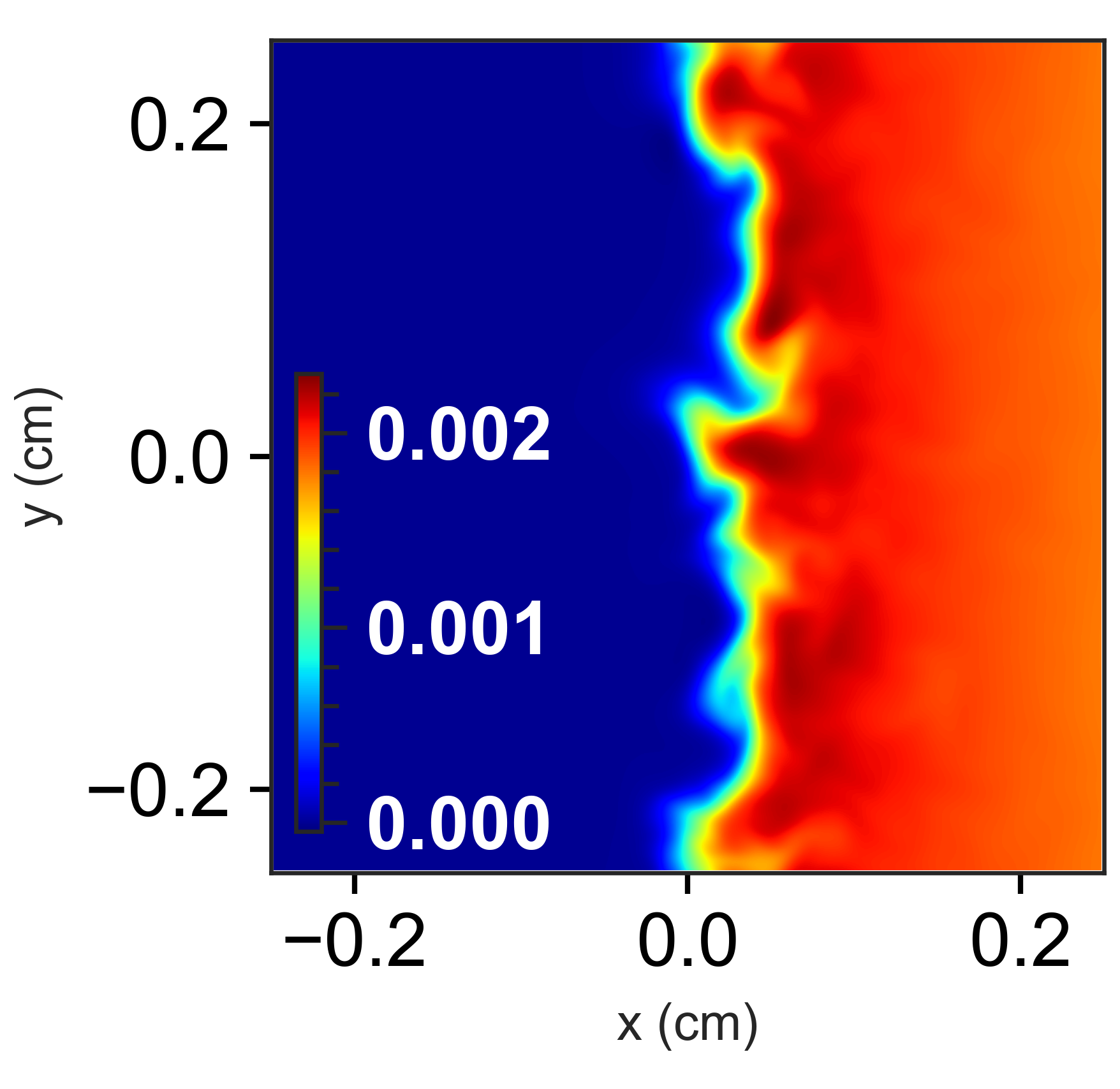}
\put(2,95){\footnotesize\textbf{(b)}}
\end{overpic} &
\begin{overpic}[width=0.25\textwidth]{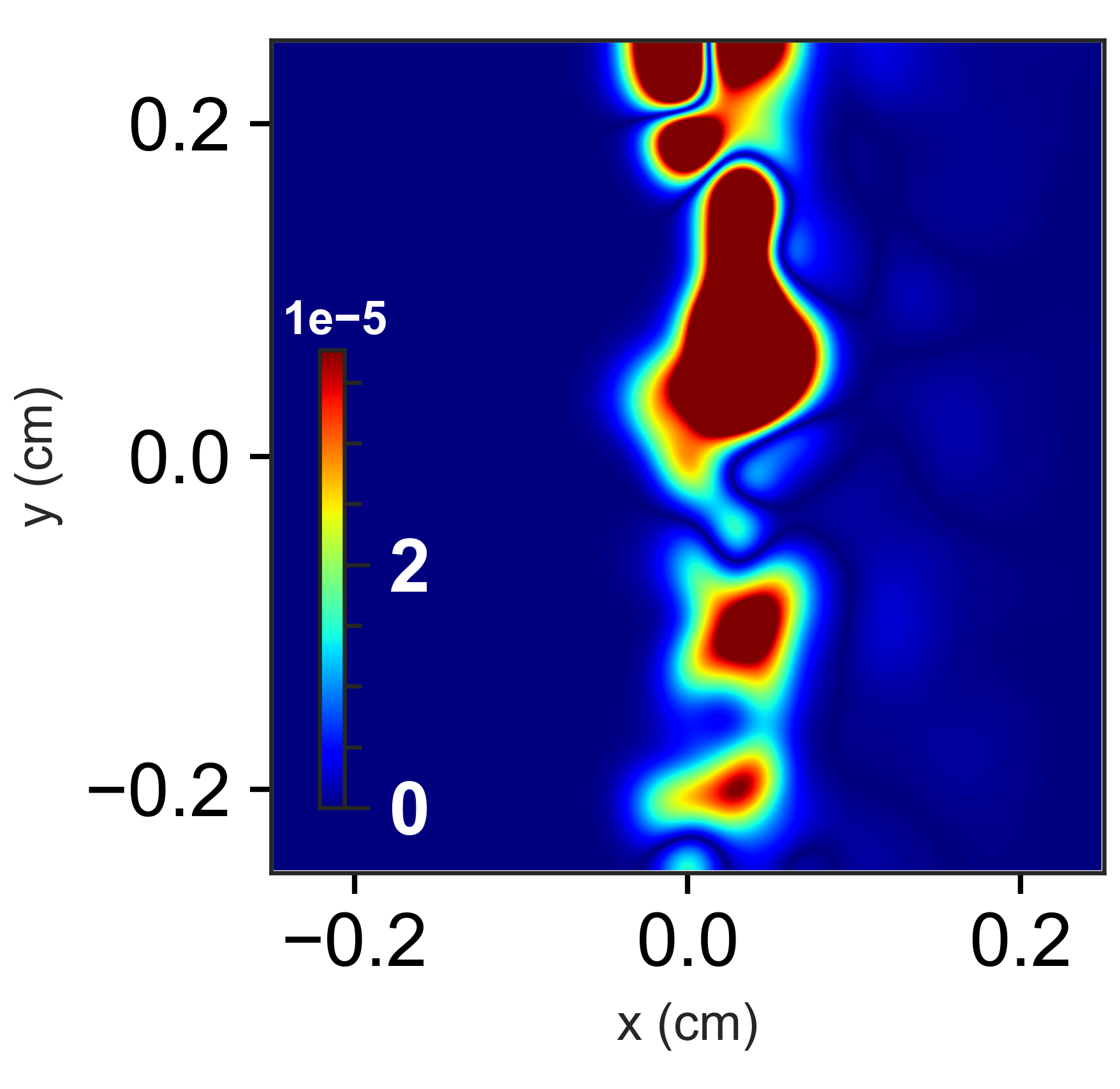}
\put(2,95){\footnotesize\textbf{(c)}}
\end{overpic}
\end{tabular}
\caption{\footnotesize $Y_{\rm OH}$ validation at $T_{\rm in} = 600$~K,
$\phi = 0.7$: (a)~full mechanism; (b)~ML model; (c)~absolute error.}
\label{fig:S_OH_600_p7}
\end{figure*}

\begin{figure*}[t]
\centering
\begin{tabular}{@{}c@{\hspace{1pt}}c@{\hspace{1pt}}c@{}}
\begin{overpic}[width=0.25\textwidth]{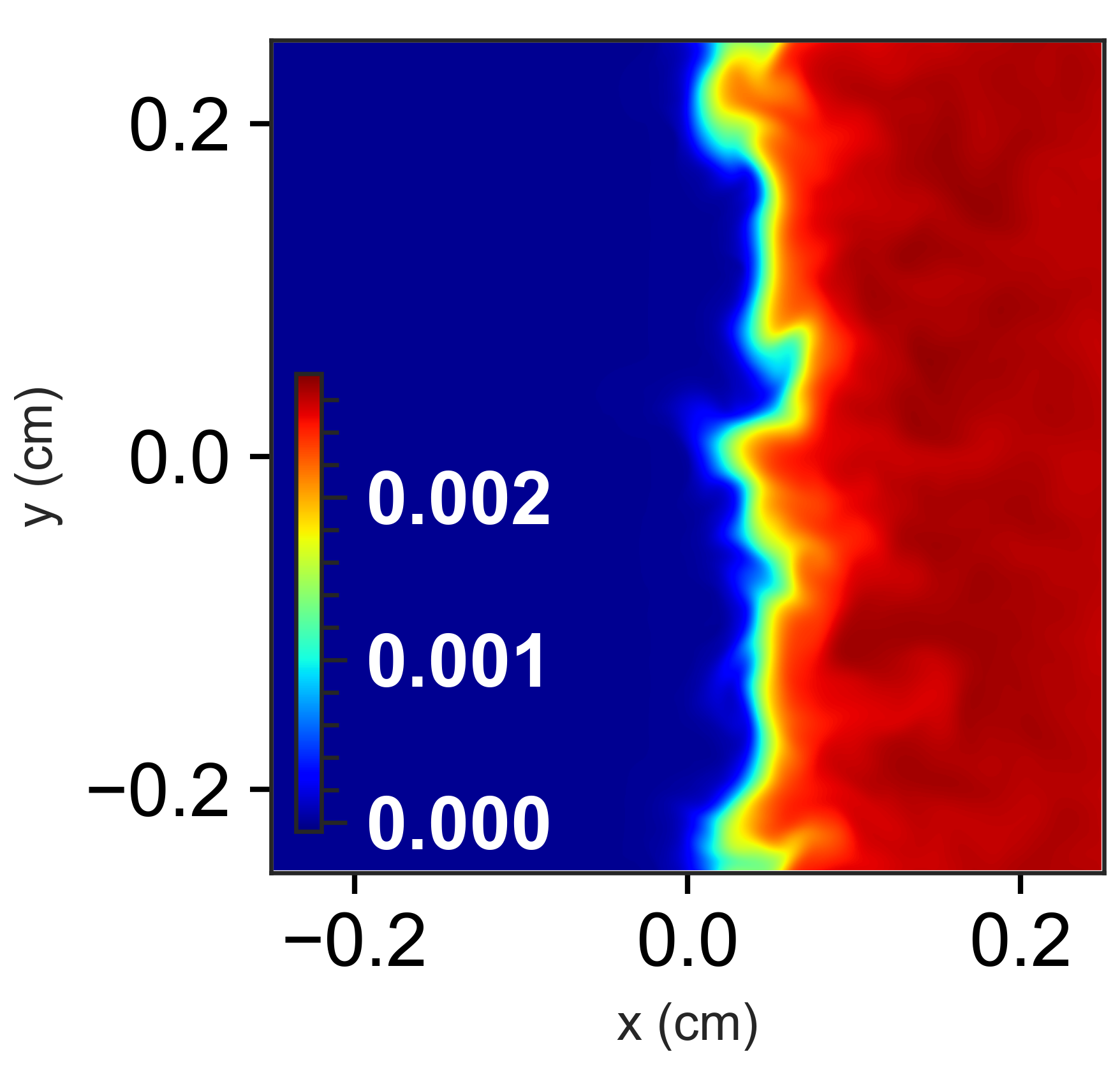}
\put(2,95){\footnotesize\textbf{(a)}}
\end{overpic} &
\begin{overpic}[width=0.25\textwidth]{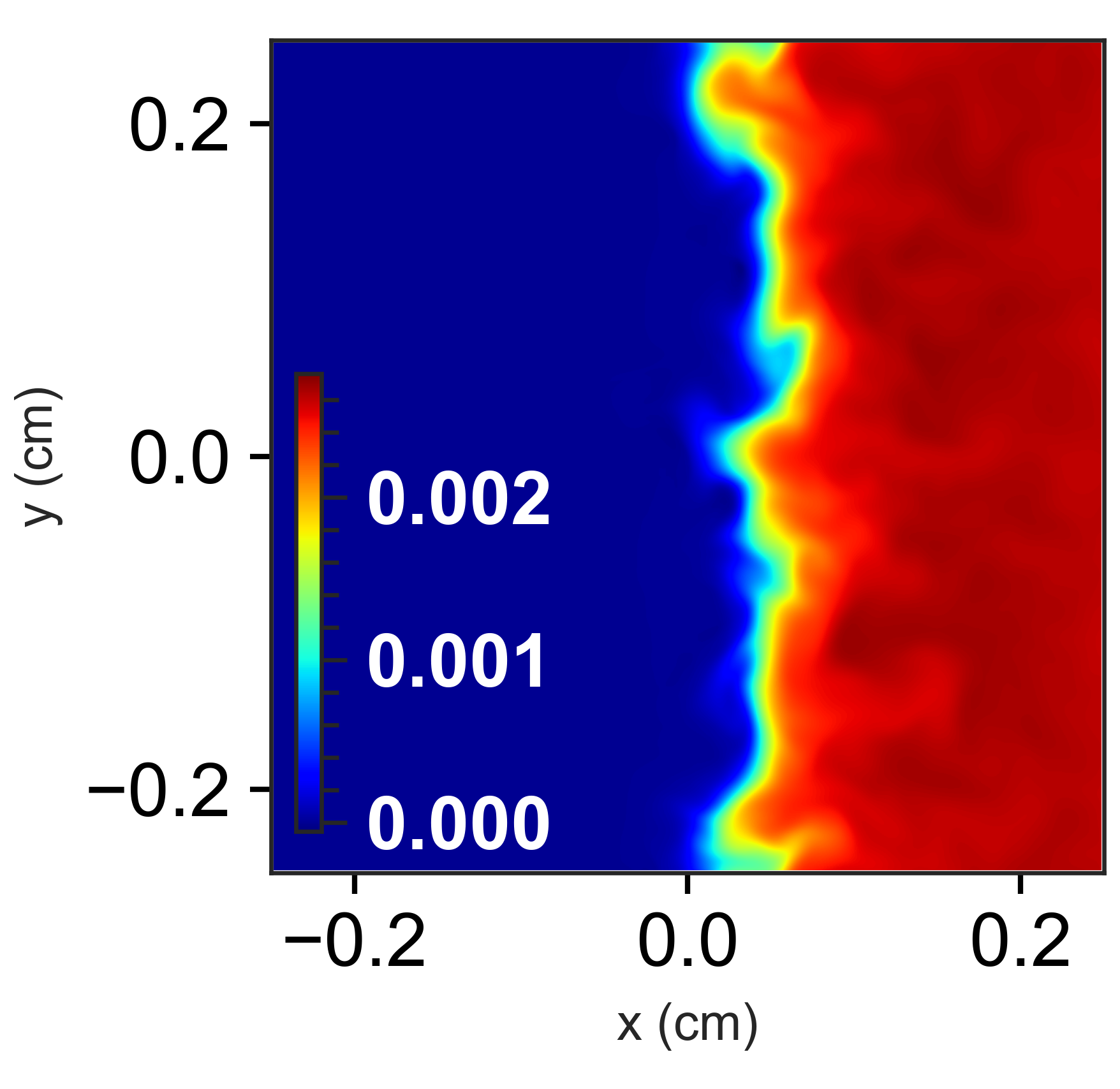}
\put(2,95){\footnotesize\textbf{(b)}}
\end{overpic} &
\begin{overpic}[width=0.25\textwidth]{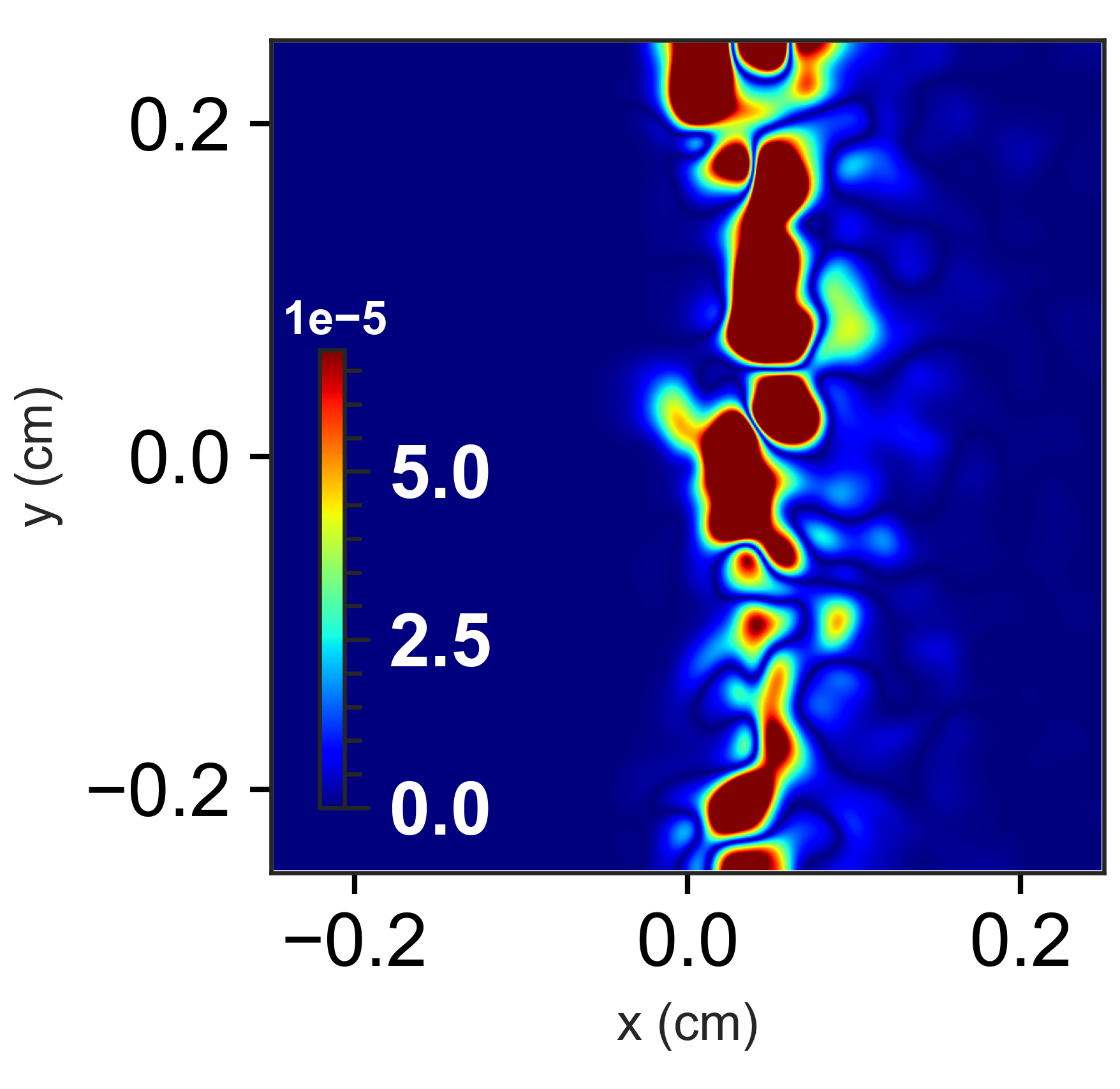}
\put(2,95){\footnotesize\textbf{(c)}}
\end{overpic}
\end{tabular}
\caption{\footnotesize $Y_{\rm OH}$ validation at $T_{\rm in} = 300$~K,
$\phi = 1.0$: (a)~full mechanism; (b)~ML model; (c)~absolute error.}
\label{fig:S_OH_300_1p0}
\end{figure*}

\begin{figure*}[t]
\centering
\begin{tabular}{@{}c@{\hspace{1pt}}c@{\hspace{1pt}}c@{}}
\begin{overpic}[width=0.25\textwidth]{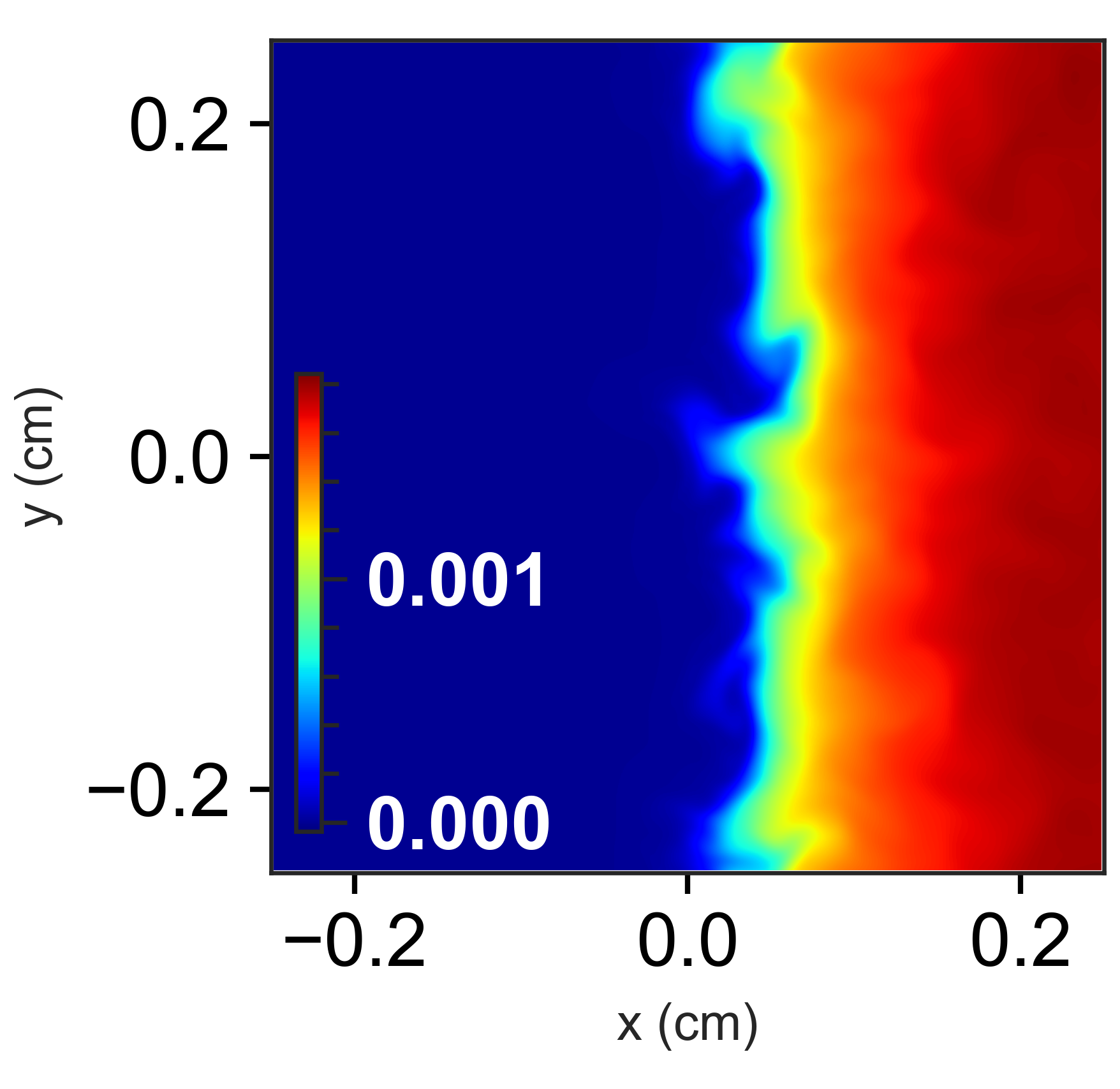}
\put(2,95){\footnotesize\textbf{(a)}}
\end{overpic} &
\begin{overpic}[width=0.25\textwidth]{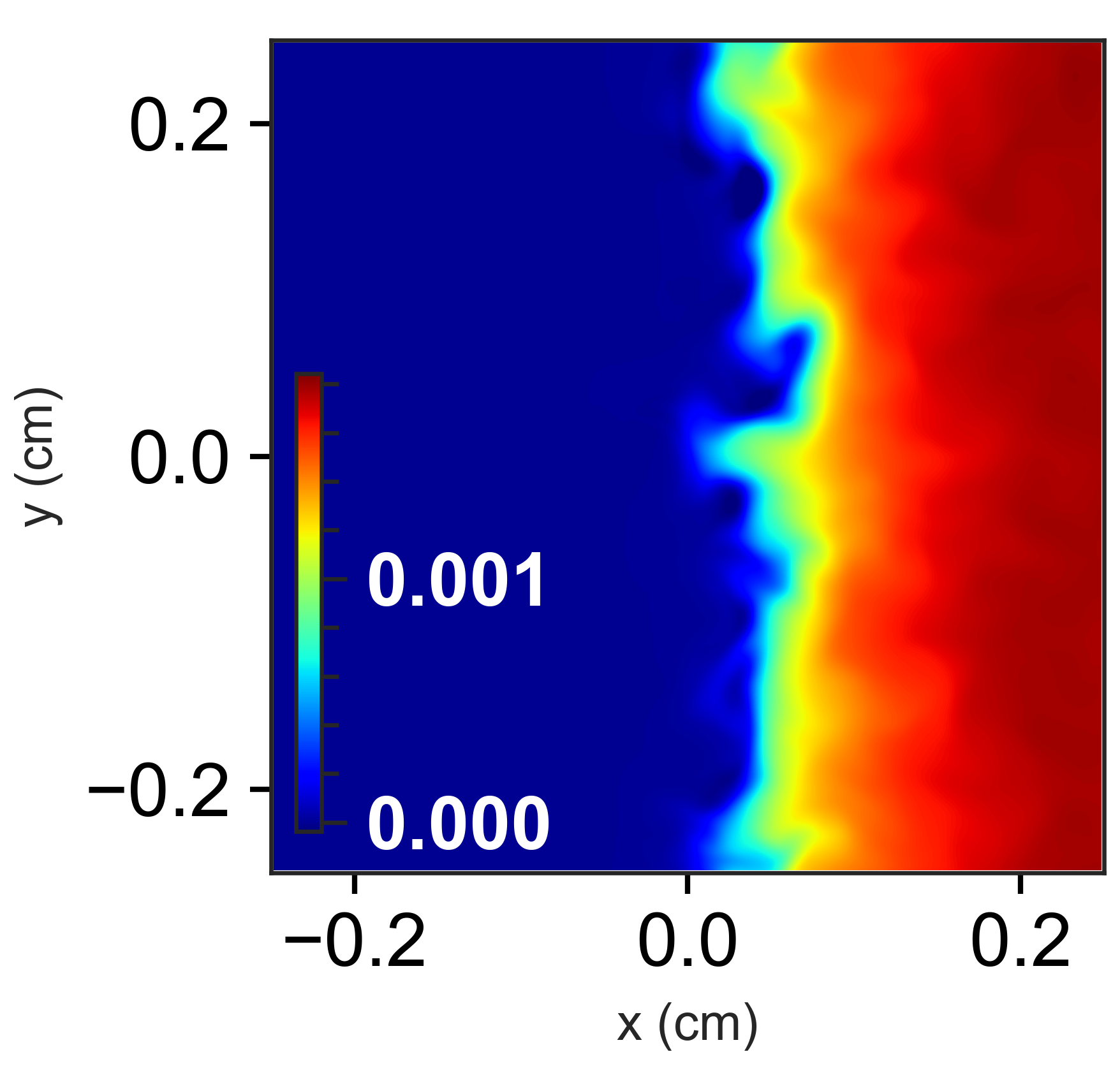}
\put(2,95){\footnotesize\textbf{(b)}}
\end{overpic} &
\begin{overpic}[width=0.25\textwidth]{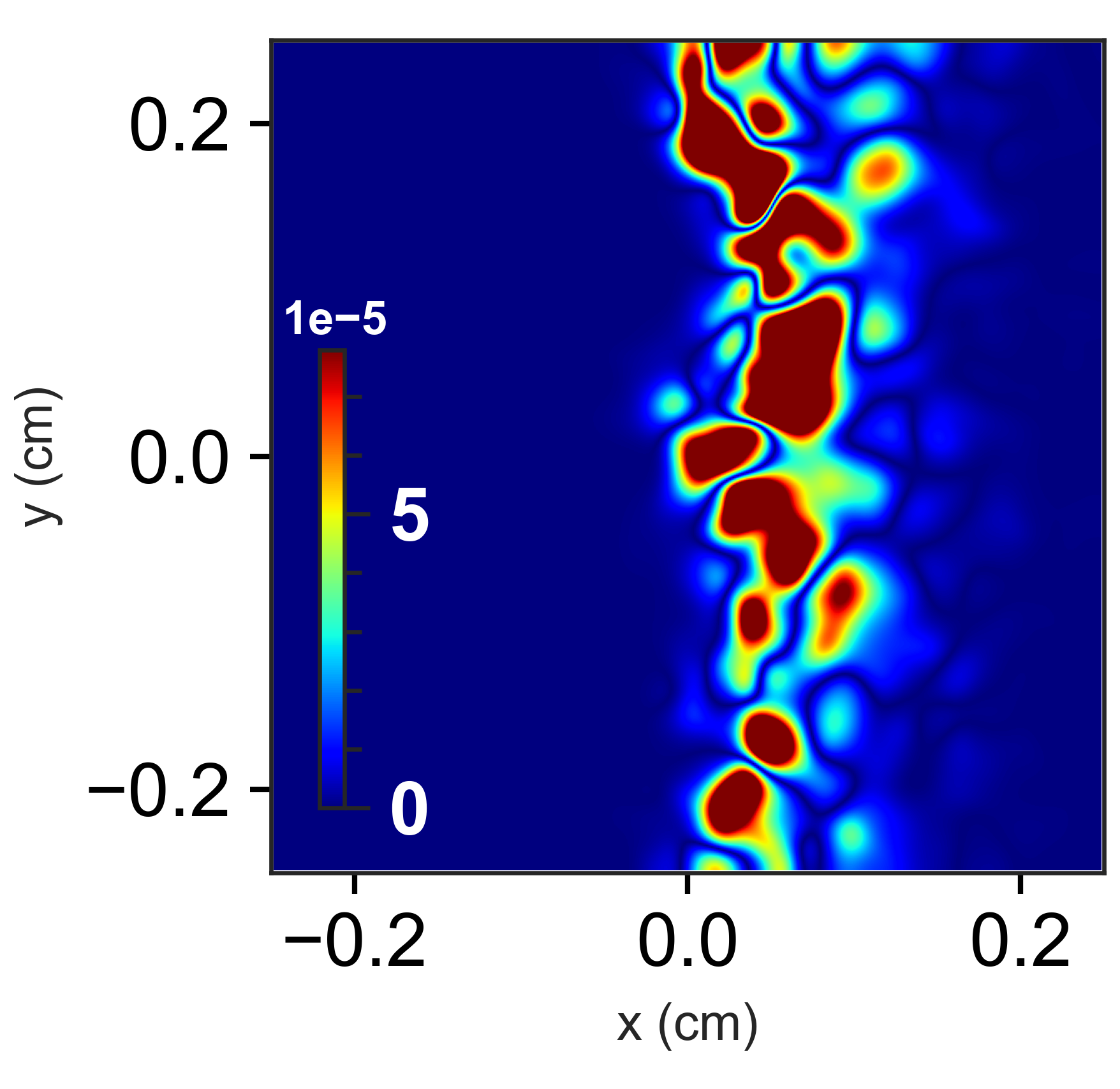}
\put(2,95){\footnotesize\textbf{(c)}}
\end{overpic}
\end{tabular}
\caption{\footnotesize $Y_{\rm OH}$ validation at $T_{\rm in} = 500$~K,
$\phi = 1.2$: (a)~full mechanism; (b)~ML model; (c)~absolute error.}
\label{fig:S_OH_500_1p2}
\end{figure*}

\section{Global fuel consumption rate\label{sec:fuel_consumption}}
\addvspace{1pt}

\begin{figure*}[h!]
\centering
\includegraphics[width=0.5\columnwidth]{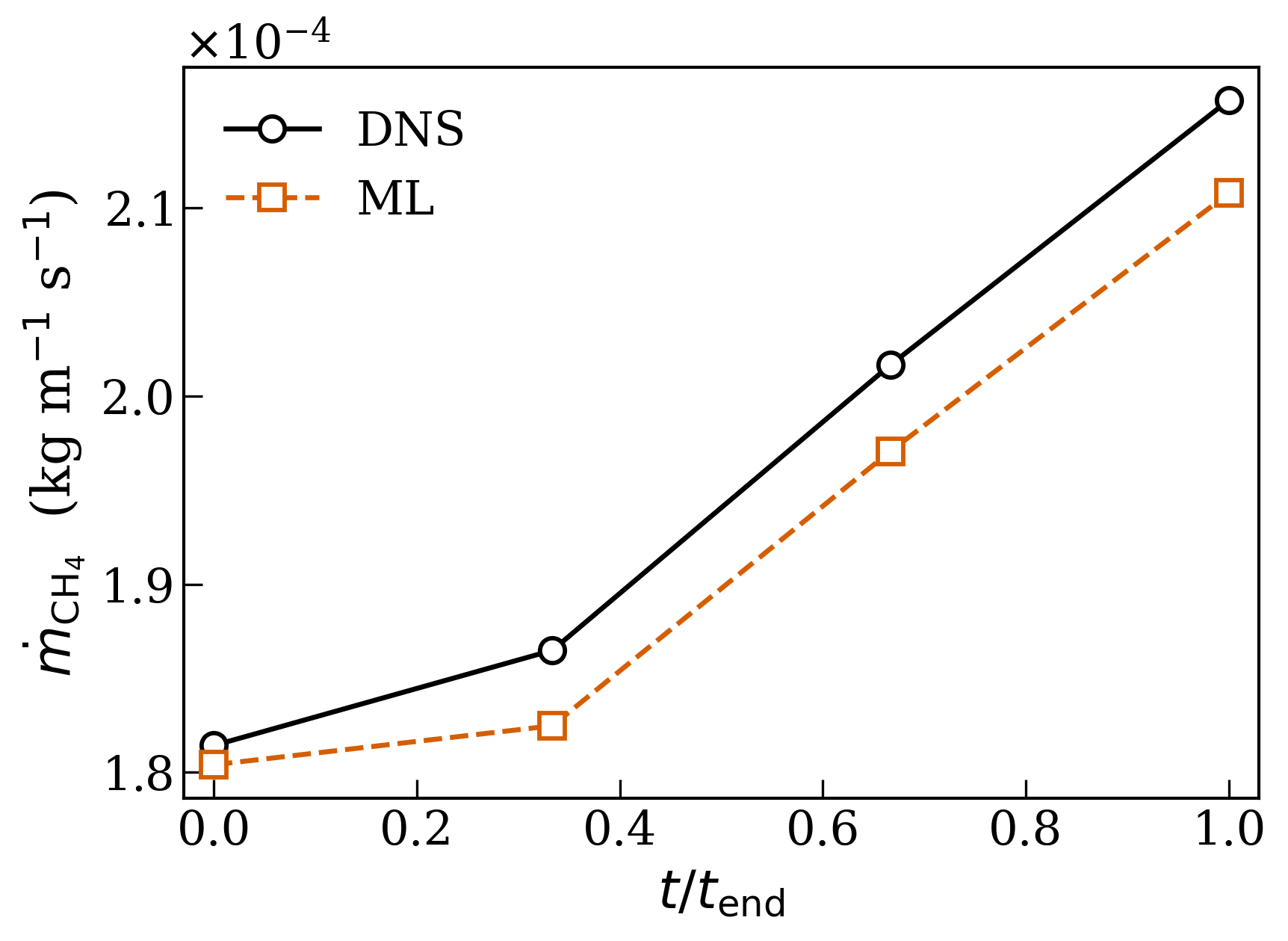}
\caption{\footnotesize Global $\mathrm{CH}_4$ consumption rate versus time for the test condition $T_{\rm in} = 300$~K,
$\phi = 0.7$.}
\label{fig:S_fuel_consumption}
\end{figure*}

Figure~\ref{fig:S_fuel_consumption} compares the global $\mathrm{CH}_4$
consumption rate $\dot{m}_{\mathrm{CH}_4}$, obtained by integrating the
local $\mathrm{CH}_4$ source term over the domain, between the ML model
and the detailed-chemistry DNS for the test condition $T_{\rm in} = 300$~K,
$\phi = 0.7$. At $t = 0$,
$\dot{m}_{\mathrm{CH}_4} = 1.815 \times 10^{-5}$~kg\,m$^{-1}$\,s$^{-1}$
for the DNS and $1.805 \times 10^{-5}$~kg\,m$^{-1}$\,s$^{-1}$ for the ML
model, an error of $0.6\%$. As the flame wrinkles, the error grows to
$2.2\%$ at $t = 50~\mu$s and stabilizes near $2.4\%$ through
$t = 150~\mu$s.

\end{document}